\pdfoutput=1
\RequirePackage[T1]{fontenc}
\documentclass[12pt,final]{article}
\usepackage{lmodern}
\usepackage{amsmath}
\usepackage{amssymb}
\usepackage{amsfonts}
\usepackage{bm}
\usepackage[dvipdfmx]{graphicx}
\usepackage{systeme}
\usepackage{enumerate}
\usepackage{comment}
\usepackage{here}
\usepackage{mathrsfs}
\usepackage{url}
\usepackage{braket}
\usepackage{cite}
\usepackage[svgnames]{xcolor}
\usepackage[colorlinks,citecolor=DarkGreen,linkcolor=FireBrick,urlcolor=FireBrick,breaklinks=true,linktoc=all]{hyperref}
\usepackage[top=30truemm,bottom=30truemm,left=25truemm,right=25truemm]{geometry}
\usepackage{showkeys}
\usepackage{authblk}
\usepackage{tikz}
\usetikzlibrary{shapes.geometric, positioning, arrows.meta}
\usepackage{ytableau}

\title{Gapless Foliated-Exotic Duality}

\author{Kantaro Ohmori, Shutaro Shimamura}
\affil{\textit{Department of Physics, The University of Tokyo, Bunkyo-ku, Tokyo 113-0033, Japan}}

\date{}

\renewcommand{\L}{\mathcal{L}}
\newcommand{\Z}{\mathbb{Z}}
\newcommand{\R}{\mathbb{R}}
\newcommand{\T}{\mathbb{T}}

\newcommand {\ali}[1] {\begin{align} #1 \end{align}}

\newcommand {\alis}[1] {\begin{align} \begin{split} #1 \end{split} \end{align}}

\usepackage{xcolor}

\usepackage[status=draft,inline,nomargin]{fixme}
\fxusetheme{color}
\FXRegisterAuthor{ko}{ako}{KO}

\FXRegisterAuthor{shu}{ashu}{\color{magenta}SHU}
    
\begin{document}
\maketitle

\vspace{1.5cm}

\begin{abstract}
 In this work, we construct a new foliated quantum field theory equivalent to the exotic $\phi$-theory -- a fractonic gapless scalar field theory described by tensor gauge fields and exhibiting $U(1) \times U(1)$ subsystem global symmetry. This subsystem symmetry has an 't Hooft anomaly, which is captured by a subsystem symmetry-protected topological (SSPT) phase in one dimension higher via the anomaly inflow mechanism. By analyzing both the anomaly inflow structure and the foliated-exotic duality in the SSPT phases, we establish the foliated-exotic duality in the $\phi$-theories. Furthermore, we also investigate the foliated-exotic duality in the $\hat\phi$-theory, which is dual to the $\phi$-theory, and construct the foliated $\hat\phi$-theory. These are the first examples of the foliated-exotic duality in gapless theories.
\end{abstract}

\thispagestyle{empty}
\clearpage
\addtocounter{page}{-1}

\newpage

\setlength{\parskip}{2mm}
\setlength{\abovedisplayskip}{10pt} 
\setlength{\belowdisplayskip}{10pt} 
\numberwithin{equation}{section}

\tableofcontents

\newpage

\section{Introduction}
\label{section1}

Subsystem global symmetry is one of the generalizations of global symmetries --- further extending the notion originally introduced by Gaiotto et al. \cite{Gaiotto:2014kfa} --- that appears in fracton phases (see reviews \cite{Nandkishore:2018sel,Pretko:2020cko,Gromov:2022cxa}).
Symmetry operators of subsystem symmetry are supported on specific submanifolds such as lines or planes.
In this sense, subsystem symmetry resembles higher-form global symmetry, another and the original generalization of global symmetries.
However, in contrast to higher-form global symmetries, these submanifolds cannot be deformed in arbitrary directions, meaning that subsystem symmetry is not fully topological.
Consequently, symmetry operators supported on each submanifold, which cannot be continuously deformed into one another, carry independent charges \cite{Seiberg:2019vrp}.
While this property is inconsistent with relativistic systems, subsystem symmetry does appear in non-relativistic systems, including lattice models \cite{Paramekanti:2002iup,Haah:2011drr,Vijay:2015mka,Vijay:2016phm} and quantum field theories (QFTs) lacking continuous rotational symmetry \cite{Pretko:2018jbi,You:2019ciz,Slagle:2017wrc,Seiberg:2020bhn,Seiberg:2020wsg,Seiberg:2020cxy}.
It leads to characteristic features such as restricted mobility of excitations, which are called fractons, and a sub-extensive ground state degeneracy \cite{Vijay:2016phm}, motivating various research.

Fractonic QFTs with subsystem symmetry have multiple formulations, and the most common one is referred to as exotic QFT, which employs tensor gauge fields \cite{Pretko:2018jbi,You:2019ciz,Slagle:2017wrc,Seiberg:2020bhn,Seiberg:2020wsg,Seiberg:2020cxy,Gorantla:2020xap,Gorantla:2021bda,Gorantla:2020jpy,Geng:2021cmq,Yamaguchi:2021qrx,Yamaguchi:2021xeq,Burnell:2021reh,Gorantla:2022eem,Gorantla:2022ssr,Honda:2022shd}.
The tensor gauge fields appearing here are in the representations of discrete spatial rotational symmetry, making the exotic QFT manifestly invariant under this symmetry.
Another formulation is called foliated QFT \cite{Slagle:2018swq,Slagle:2020ugk,Hsin:2021mjn,Geng:2021cmq, Ohmori:2022rzz, Hsin:2023ooo, Shimamura:2024kwf, Ebisu:2024mbb}.
Foliated QFT is QFT coupled with a lattice-like structure called \textit{foliation}, which is a decomposition of a space into a stack of infinitely many subspaces, such as planes or lines.
It involves foliated gauge fields, which can be regarded as gauge fields defined on these subspaces, as well as \textit{bulk} gauge fields that mediate between them.
Both exotic and foliated fractonic QFTs exhibit UV/IR mixing, where a certain low-energy physics is sensitive to microscopic details in the UV physics.
As a manifestation of this, the tensor and foliated gauge fields can have singularities and discontinuities in specific directions \cite{Seiberg:2020bhn,Seiberg:2020wsg,Seiberg:2020cxy,Slagle:2020ugk,Hsin:2021mjn}.

It was noted in \cite{Hsin:2021mjn} that there are pairs of exotic and foliated QFTs sharing the same subsystem symmetries and related quantities, and thus we expect such a pair to describe the same physics.
In \cite{Ohmori:2022rzz}, the authors of the present paper considered the exotic $BF$ theory \cite{Slagle:2017wrc,Seiberg:2020cxy} and the foliated $BF$ theory \cite{Slagle:2020ugk,Hsin:2021mjn}, each of which was considered as a continuous exotic or foliated QFT description of the X-cube model \cite{Vijay:2016phm}, and demonstrated the equivalence of the two theories.
This correspondence is referred to as foliated-exotic duality, and subsequently, correspondences have been established for other theories as well \cite{Spieler:2023wkz,Cao:2023rrb,Ebisu:2023idd,Ebisu:2024cke,Hsin:2024eyg}.

Beyond matching known pairs of foliated and exotic theories, one might search for a new dual description of a given exotic or foliated theory.
In \cite{Shimamura:2024kwf}, one of the present authors constructed a previously unknown foliated QFT description for the 3+1d subsystem symmetry-protected topological (SSPT) phase for $\Z_N \times \Z_N$ subsystem symmetry, whose exotic QFT description was known in \cite{Burnell:2021reh}.
This construction relies on the anomaly inflow mechanism \cite{Callan:1984sa}. In the context of fracton phases, it is known that an anomaly of a subsystem symmetry can be canceled by coupling it to an SSPT phase in one dimension higher (see \cite{You:2018oai,Devakul:2018fhz} for discussion in lattice models, and  \cite{Burnell:2021reh,Hsin:2021mjn,Okuda:2024azp} for discussion in continuum QFTs).\footnote{In the anomaly inflow mechanism for an ordinary global symmetry, an ’t Hooft anomaly in a $d$-dimensional QFT is captured by a classical field theory in $d+1$ dimensions, known as an invertible field theory \cite{Freed:2016rqq} or a symmetry-protected topological (SPT) phase \cite{Gu:2009dr,Chen:2010gda}.}
In \cite{Shimamura:2024kwf}, the foliated QFT description of the SSPT phase is constructed by taking the known exotic QFT description of the inflow mechanism and extending the foliated-exotic duality from the boundary theory into the SSPT phase.

Building on our previous work where we constructed a foliated SSPT phase from an anomalous boundary theory via the foliated-exotic duality, we now reverse the approach. 
In this paper, we outline a method to derive a boundary theory from an SSPT phase in one dimension higher, leading to a construction of a foliated $\phi$-theory. The foliated $\phi$-theory can be shown to be equivalent to the exotic $\phi$-theory \cite{Seiberg:2020bhn}, a gapless scalar field theory with $U(1) \times U(1)$ anomalous subsystem symmetry \cite{Gorantla:2021bda}, by tuning parameters and applying correspondences between the fields.
The resulting foliated theory contains a 1+1d compact scalar field $\Phi^k$ on each layer of the foliation in the $x^k$ direction $(k=1,2)$, coupled to a bulk 2+1d compact scalar field $\Phi$.
The main idea here is to identify the compact scalar field $\phi$ in the exotic $\phi$-theory with the bulk field $\Phi$, i.e., $\phi \simeq \Phi$.
As a correspondence between Lagrangians, 
the exotic $\phi$-theory is described by
\ali{
    \L_{\phi,\text{e}} = \frac{\mu_0}{2} (\partial_0 \phi )^2 + \frac{1}{2 \mu_{12}} (\partial_1 \partial_2 \phi )^2 \,, 
}
which is equivalent to the foliated $\phi$-theory described by
\alis{
    &\L_{\phi,\text{e} \rightarrow \text{f}}  = \frac{\mu_0}{2} ( \partial_0 \Phi )^2 + \frac{1}{4\mu_{12}} ( \partial_2 \Phi^1 )^2 + \frac{1}{4\mu_{12}} ( \partial_1 \Phi^2 )^2 \\
    & \qquad + \frac{i}{2\pi} \left[ -\hat{c}_{02} ( \partial_1 \Phi - \Phi^1 )  + \hat{c}_{01} ( \partial_2 \Phi - \Phi^2 ) \right]  \,, 
}
where $\hat{c}_{01}$ and $\hat{c}_{02}$ are Lagrange multipliers. 
Furthermore, we also investigate the structure of foliated-exotic duality for the $\hat\phi$-theory \cite{Seiberg:2020bhn,Spieler:2024fby}, which is dual to the $\phi$-theory. The exotic $\hat\phi$-theory is described by the Lagrangian
\ali{
    \L_{\hat{\phi},\text{e}} &= \frac{\hat{\mu}_0}{2} ( \partial_0  \hat{\phi}^{12} )^2 + \frac{1}{2 \hat{\mu}_{12}} ( \partial_1 \partial_2 \hat{\phi}^{12} )^2  \,,
}
where $\hat\phi^{12}$ is a compact scalar field. This exotic theory is equivalent to the foliated $\hat\phi$-theory described by
\alis{
    \L_{\hat{\phi},\text{e} \rightarrow \text{f}} &= \hat{\mu}_0 ( \partial_0 \hat{\Phi}^1 - \hat{\Phi}_0 )^2 + \hat{\mu}_0 ( \partial_0 \hat{\Phi}^2 - \hat{\Phi}_0 )^2  + \frac{1}{2\hat{\mu}_{12}}( \partial_1 \hat{\Phi}_2 - \partial_2 \hat{\Phi}_1 )^2 \\
    & + \frac{i}{2\pi} \left[ h_{01} ( \partial_2 \hat{\Phi}^1 - \hat{\Phi}_2 ) - h_{02} ( \partial_1 \hat{\Phi}^2 - \hat{\Phi}_1 ) \right] \,,
}
where $h_{01}$ and $h_{02}$ are Lagrange multipliers,  under the correspondence $\hat\phi^{12} \simeq \hat\Phi^1 - \hat\Phi^2$. The foliated $\hat\phi$-theory contains 1+1d compact scalar fields $\hat\Phi^k \ (k=1,2)$ and a bulk 2+1d one-form gauge field $\hat\Phi$. 
The foliated $\hat\phi$-theory can be regarded as resulting from taking $T$-duals of the foliated fields $\Phi^k\, (k=1,2)$ and the bulk field $\Phi$. These are the first examples of the foliated-exotic duality in gapless scalar field theories. 

The fractonic QFTs can be seen as layers of 1+1d compact scalar field theory coupled by the bulk gauge fields from a perspective of the foliated QFT description. This decomposes the fractonic QFT into more conventional building blocks, which might enable us to apply established tools from standard QFT to fractonic systems. For example, we might be able to provide a framework for constructing fermionic fractonic QFT \cite{Yamaguchi:2021qrx,Honda:2022shd,Katsura:2022xkg,
Kawakami:2025vox} by establishing a boson-fermion duality in foliated QFT, which is considered easier than in the case of exotic QFT.\footnote{The boson-fermion duality in lattice models with subsystem symmetry is studied in \cite{Cao:2022lig}.}

Here is a comment related to dipole symmetry.
In the exotic and foliated QFTs discussed in this paper, the tensor gauge fields are symmetric and \textit{hollow}, meaning that they satisfy $A_{ij} = A_{ji}$ and $A_{ii} = 0$, and the components of the foliated gauge fields $A^k_{kij...}$ are gauged.
The subsystem symmetry appearing in the theories is also referred to as dipole symmetry \cite{Seiberg:2020bhn}.
However, in recent studies on dipole symmetry \cite{Radzihovsky:2019jdo,Ebisu:2023idd,Ebisu:2024cke,Ebisu:2024eew,Huang:2023zhp,Han:2024nvu}, the theories typically involve tensor gauge fields where $A_{ii} \neq 0$ and the components of the foliated (or dipole) gauge fields $A^k_{kij...}$ are not gauged.
Regarding scalar field theories, a similar model is studied in the context of dipole symmetry \cite{Huang:2023zhp}.

\paragraph{Note added:} 
While finalizing this draft, we noticed the work \cite{ApruzziBedognaMancani}, which independently constructs the foliated $\phi$-theory based on the SymTFT for the subsystem symmetry.

\subsection*{Organization}
\label{section11}

\begin{figure}[H]
   \begin{center}
    \includegraphics[width=1\hsize]{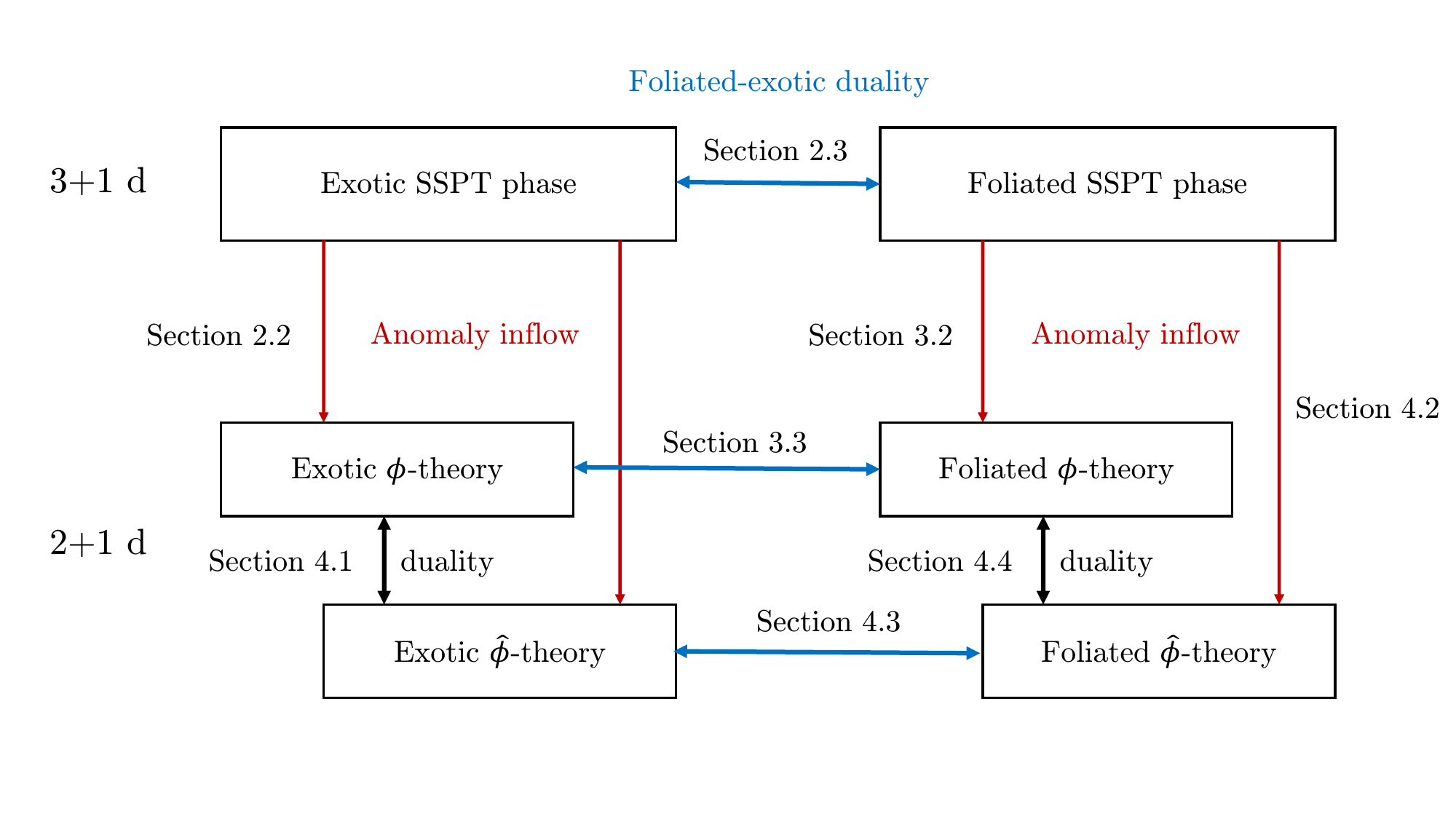} 
    \end{center}
    \vspace{-1cm}
      \caption{The structure of the anomaly inflow and dualities.}
      \label{fig1}
\end{figure}

The organization of the paper is as follows and is also summarized in Figure \ref{fig1}.

In Section \ref{section21}, we will review the exotic description of the SSPT phase for $U(1) \times U(1)$ subsystem symmetry in 3+1 dimensions \cite{Burnell:2021reh}.
This exotic SSPT phase cancels the 't Hooft anomaly of $U(1) \times U(1)$ subsystem symmetry in the exotic $\phi$-theory in 2+1 dimensions \cite{Seiberg:2020bhn}, which we will review in Section \ref{section22}.
In Section \ref{section23}, we establish correspondences between tensor gauge fields and foliated gauge fields, and then construct the foliated description of the SSPT phase, which is equivalent to the exotic SSPT phase for $U(1) \times U(1)$ subsystem symmetry.
This foliated-exotic duality is almost identical to the duality in the fractonic $BF$ theory with two flat foliations \cite{Spieler:2023wkz}.

Section \ref{section31} describes a way to construct an anomalous theory on a boundary from an SPT phase in one dimension higher.
As a relativistic example, we will construct a compact scalar field theory in 1+1 dimensions from the SPT phase for $U(1) \times U(1)$ global symmetry in 2+1 dimensions.
In Section \ref{section32}, we will construct a foliated $\phi$-theory from the foliated SSPT phase.
In Section \ref{section33}, we discuss the foliated-exotic duality between the exotic $\phi$-theory and a foliated $\phi$-theory.

In Section \ref{section4}, we investigate the relation between the foliated-exotic duality and the duality between the $\phi$-theory and the $\hat\phi$-theory.
In Section \ref{section41}, we will review the (self-)duality between the exotic $\phi$-theory and the exotic $\hat\phi$-theory \cite{Seiberg:2020bhn,Spieler:2024fby}.
In Section \ref{section42}, we will construct a foliated $\hat\phi$-theory from the foliated SSPT phase in the same way as in Section \ref{section32}.
Section \ref{section43} discusses the foliated-exotic duality between the exotic $\hat\phi$-theory and a foliated $\hat\phi$-theory.
 In Section \ref{section44}, we will describe the duality between the foliated $\phi$-theory and the foliated $\hat\phi$-theory, which is parallel to the discussion in Section \ref{section41}.

Further explanations of exotic QFT and tensor gauge fields can be found in Appendix \ref{appendix1}, and those of foliated QFT and foliated gauge fields are provided in Appendix \ref{appendix2}.

\section{Exotic and Foliated SSPT Phase for \texorpdfstring{$U(1) \times U(1)$}{U(1) × U(1)} Subsystem Symmetry}
\label{section2}

In this section, we first review the exotic description of the SSPT phase for $U(1) \times U(1)$ subsystem symmetry in 3+1 dimensions \cite{Burnell:2021reh} and an anomalous $\phi$-theory on (2+1)-dimensional boundary \cite{Seiberg:2020bhn}. This SSPT phase describes an 't Hooft anomaly of a $U(1) \times U(1)$ subsystem symmetry in the $\phi$-theory \cite{Gorantla:2021bda} via the anomaly inflow mechanism. Next, we will consider the field correspondences between tensor gauge fields and foliated gauge fields, and construct the foliated description of the SSPT phase.

\subsection{Exotic SSPT Phase in 3+1 Dimensions}
\label{section21}
We review the exotic SSPT phase for $U(1) \times U(1)$ subsystem symmetry in 3+1 dimensions \cite{Burnell:2021reh}. We take Euclidean spacetime to be a (3+1)-dimensional torus $\T^{3+1}$ of lengths $l^0$, $l^1$, $l^2$ and $l^3$, and the coordinate $(x^0, x^1, x^2 ,x^3)$ on it. First, we consider the exotic SSPT phase on the torus, and then we will restrict the space to the region $x^3 \geq 0$ and take the boundary $x^3 = 0$. The SSPT phase has the 90-degree rotational symmetry of $\Z_4$ for $(x^1,x^2)$ and the continuous rotational symmetry of $SO(2)$ for $(x^0,x^3)$. Related to that, the theory has $U(1)$ background tensor gauge fields in representations of $\Z_4 \times SO(2)$.\footnote{The irreducible representations of $\Z_4$ are one-dimensional ones $\bm{1}_n$ ($n = 0, \pm 1, 2$), where $n$ is the spin. See Appendix \ref{appendix1} for more details.} The background tensor gauge fields are $\bm{C}=(C_0, C_{12}, C_3)$ and $\bm{\hat{C}}=(\hat{C}^{12}_{0}, \hat{C}, \hat{C}^{12}_{3})$. For $\Z_4$ on subspace $(x^1,x^2)$, the tensor gauge fields $C_{12}$, $\hat{C}^{12}_{0}$ and $\hat{C}^{12}_{3}$ are in the representation $\bm{1}_2$, and $C_0$, $C_3$ and $\hat{C}$ are in the representation $\bm{1}_0$. For $SO(2)$ on subspace $(x^0,x^3)$, the fields $(C_0, C_3)$ and $(\hat{C}_0^{12}, \hat{C}_3^{12})$ are in the vector representation, and $C_{12}$ and $\hat{C}$ are in the trivial representation.  
The background gauge transformations of $\bm{C} = (C_0, C_{12}, C_3)$ and $\bm{\hat{C}} = (\hat{C}^{12}_{0}, \hat{C}, \hat{C}^{12}_{3})$ are 
\ali{
    C_0 &\sim C_0 + \partial_0 \Lambda \,, \label{bkgugec0} \\
    C_{12} &\sim C_{12} + \partial_1\partial_2 \Lambda \,, \label{bkgugec12} \\ 
    C_3 &\sim C_3 + \partial_3 \Lambda  \,, \label{bkgugec3}
}
and
\ali{
    \hat{C}^{12}_0 &\sim \hat{C}^{12}_0 + \partial_0  \hat{\Lambda}^{12} \,, \label{bkgugechat120}\\
    \hat{C} &\sim \hat{C} + \partial_1 \partial_2  \hat{\Lambda}^{12} \,, \label{bkgugechat}\\
    \hat{C}^{12}_3 &\sim \hat{C}^{12}_3 + \partial_3  \hat{\Lambda}^{12} \,, \label{bkgugechat123}
}
where $\Lambda$ and $\hat{\Lambda}^{12}$ are gauge parameters in $\bm{1}_0$ and $\bm{1}_2$ respectively.

The 3+1d SSPT phase is described by the exotic Lagrangian
\alis{
    &\L_{\text{SSPT,e}}\left[ \bm{C},\bm{\hat{C}} \right] \\ 
     & \quad = \frac{i}{2\pi} \left[ \hat{C} ( \partial_0 C_3 - \partial_3 C_0 ) - \hat{C}_0^{12} ( \partial_3 C_{12} - \partial_1 \partial_2 C_3 )  + \hat{C}_3^{12} ( \partial_0 C_{12} - \partial_1 \partial_2 C_0 )  \right] \,. \label{exoticsspt}
}
The Lagrangian is gauge invariant on spacetime without a boundary. However, if spacetime has a boundary, a gauge variation arises on the boundary, which will match a mixed 't Hooft anomaly of $U(1) \times U(1)$ subsystem symmetry. Then, we consider the spacetime $\T^{2+1} \times \R_{x^3 \geq 0}$ with the boundary $x^3 = 0$. The gauge variation of the Lagrangian is
\alis{
    \delta_g \L_{\text{SSPT,e}} 
      &= \frac{i}{2\pi} \left[ \partial_1 \partial_2 \hat{\Lambda}^{12} ( \partial_0 C_3 - \partial_3 C_0 ) - \partial_0 \hat{\Lambda}^{12} ( \partial_3 C_{12} - \partial_1 \partial_2 C_3 ) + \partial_3 \hat{\Lambda}^{12} ( \partial_0 C_{12} - \partial_1 \partial_2 C_0 )  \right] \\
     & = \partial_3 \left[ \frac{i}{2\pi} \hat{\Lambda}^{12}  ( \partial_0 C_{12} - \partial_1 \partial_2 C_0 ) \right] \,,
}
where we have dropped the total derivative terms with respect to $x^0$, $x^1$ and $x^2$, so the variation of the action is
\ali{
    \delta_g S_{\text{SSPT,e}} = - \int_{x^3 = 0} dx^0 dx^1 dx^2 \left[ \frac{i}{2\pi} \hat{\Lambda}^{12}  ( \partial_0 C_{12} - \partial_1 \partial_2 C_0 ) 
    \right] \,. \label{variexoticsspt}
}

\subsection{Exotic \texorpdfstring{$\phi$}{φ}-Theory in 2+1 Dimensions}
\label{section22}

Next, we review the exotic $\phi$-theory in 2+1 dimensions \cite{Seiberg:2020bhn}, which is the continuum QFT of the XY-plaquette model \cite{Paramekanti:2002iup}. We take Euclidean spacetime to be a (2+1)-dimensional torus $\T^{2+1}$ of lengths $l^0$, $l^1$ and $l^2$, and the coordinate $(x^0, x^1, x^2)$ on it. This theory has two types of $U(1)$ subsystem symmetries: the momentum dipole global symmetry and the winding dipole global symmetry. These symmetries have a mixed 't Hooft anomaly, so cannot be gauged simultaneously \cite{Gorantla:2021bda}. The anomaly is described by the SSPT phase in section \ref{section21}, so the $\phi$-theory is considered as a boundary theory of it \cite{Burnell:2021reh}.

The Lagrangian describing the exotic $\phi$-theory is
\ali{
    \L_{\phi,\text{e}} = \frac{\mu_0}{2} (\partial_0 \phi )^2 + \frac{1}{2 \mu_{12}} (\partial_1 \partial_2 \phi )^2 \,,  \label{exoticphi}
}
where $\mu_0$ and $\mu_{12}$ are parameters with mass dimension one. The field $\phi$ is a dynamical compact scalar field with the periodicity $\phi \sim \phi + 2\pi$, so we have a gauge transformation
\ali{
    \phi \sim \phi + 2\pi w^1 + 2\pi w^2 \,.
}
$w^1$ and $w^2$ are integer-valued gauge parameters and they can have step function discontinuities in the $x^1$ and $x^2$ directions respectively. Then, $\phi$ can also have the discontinuities and these configurations have contributions to the theory due to the term of the second-order derivative. This theory has the discrete spatial rotational symmetry $\Z_4$, and the field $\phi$ can  be considered as the tensor gauge field in the representation $\bm{1}_0$.

The equation of motion is
\ali{
    - \mu_0 \partial_0^2 \phi + \frac{1}{\mu_{12}} \partial^2_1 \partial_2^2 \phi = 0 \,. \label{phieom}
}

Let us discuss subsystem symmetry. The currents of the momentum dipole symmetry are
\ali{
    J_0 &= i \mu_0 \partial_0 \phi \,,  \label{momsym1} \\
    J_{12} &=  \frac{i}{\mu_{12}} \partial_1 \partial_2 \phi \,, \label{momsym2}
}
and the conservation law is
\ali{
    \partial_0 J_0 - \partial_1 \partial_2 J_{12} = 0 \,,
}
from the equation of motion \eqref{phieom}. The conserved charges are
\ali{
    Q^1(x^1) = \oint dx^2 J_0 \,, \\
    Q^2(x^2) = \oint dx^1 J_0 \,. 
}
By integrating them over the fixed width and exponentiating them, we have the symmetry operators
\ali{
    U_{\alpha}^1 \left[ [x_1^1, x_2^1] \times \mathcal{C}^2  \right] = \exp \left[ i \alpha \int_{x_1^1}^{x_2^1} dx^1 \oint_{\mathcal{C}^2} dx^2  \mu_0 \partial_0 \phi  \right] \,, \\
    U_{\alpha}^2 \left[ [x_1^2, x_2^2] \times \mathcal{C}^1 \right] = \exp \left[ i \alpha  \int_{x_1^2}^{x_2^2} dx^2 \oint_{\mathcal{C}^1} dx^1  \mu_0 \partial_0 \phi  \right] \,,
}
where $\alpha$ is $2\pi$-periodic and $\mathcal{C}^k$ is a one-dimensional loop along the $x^k$ direction. The momentum dipole symmetry acts on the charged operator
\ali{
    V_n [x] = e^{in\phi}  \,,
}
as
\ali{
    U_{\alpha}^1 \, V_n \, \left( U_{\alpha}^1 \right)^{-1} = e^{in\alpha} \,V_n \,, \quad x_1^1 < x^1 < x_2^1 \,, \\
    U_{\alpha}^2 \, V_n \, \left( U_{\alpha}^2 \right)^{-1} = e^{in\alpha} \,V_n \,, \quad x_1^2 < x^2 < x_2^2 \,.
}
For the field action, we have
\ali{
    \phi &\rightarrow \phi + \Lambda^1(x^1) + \Lambda^2(x^2) \,,
}
where $\Lambda^k(x^k)$ is a $2\pi$-periodic scalar that can have step function discontinuities in the $x^k$ direction.

On the other hand, the currents of the winding dipole symmetry are
\ali{
    \hat{J}^{12}_0 &= \frac{1}{2 \pi} \partial_1 \partial_2 \phi \,, \label{winsym1} \\
    \hat{J} &= \frac{1}{2 \pi} \partial_0 \phi \,,  \label{winsym2}
}
and the conservation law is
\ali{
    \partial_0 \hat{J}^{12}_0 - \partial_1 \partial_2 \hat{J} = 0 \,.
}
The conserved charges are
\ali{
    \hat{Q}^1(x^1) = \oint dx^2 \hat{J}^{12}_0 \,, \\
    \hat{Q}^2(x^2) = \oint dx^1 \hat{J}^{12}_0 \,. 
}
Then, we have the symmetry operators
\ali{
    \hat{U}_{\beta}^1 \left[ [x_1^1, x_2^1] \times \mathcal{C}^2 \right] = \exp \left[ i \beta  \int_{x_1^1}^{x_2^1} dx^1 \oint_{\mathcal{C}^2} dx^2  \frac{1}{2 \pi} \partial_1 \partial_2 \phi  \right] \,, \\
    \hat{U}_{\beta}^2 \left[  [x_1^2, x_2^2] \times \mathcal{C}^1 \right] = \exp \left[ i \beta  \int_{x_1^2}^{x_2^2} dx^2 \oint_{\mathcal{C}^1} dx^1   \frac{1}{2 \pi} \partial_1 \partial_2 \phi  \right] \,,
}
where $\beta$ is $2\pi$-periodic. The winding dipole symmetry acts on a charged operator, but it is written by the dual field $\hat{\phi}^{12}$ not $\phi$. We will consider the duality in Section \ref{section4}.

Let us see that the $U(1) \times U(1)$ subsystem symmetry has a mixed 't Hooft anomaly. To see this, we couple the subsystem symmetries to background gauge fields. Then, the symmetry action on the field becomes the local transformation:
\ali{
    \phi &\sim \phi + \Lambda \,.
}
where $\Lambda$ is a $2\pi$-periodic local parameter and it is absorbed into the gauge transformations of the background gauge fields.

The momentum dipole symmetry is coupled to the background tensor gauge field $\bm{C} = (C_0 ,C_{12})$ in the representation $(\bm{1}_0, \bm{1}_2)$. Their background gauge transformations are
\ali{
    C_0 &\sim C_0 + \partial_0 \Lambda \,,  \\
    C_{12} &\sim C_{12} + \partial_1\partial_2 \Lambda \,. 
}
The winding dipole symmetry is coupled to the background tensor gauge field $\bm{\hat{C}} =  (\hat{C}_0^{12} ,\hat{C})$ in the representation $(\bm{1}_2, \bm{1}_0)$. Their background gauge transformations are
\ali{
    \hat{C}^{12}_0 &\sim \hat{C}^{12}_0 + \partial_0  \hat{\Lambda}^{12} \,,  \\
    \hat{C} &\sim \hat{C} + \partial_1 \partial_2  \hat{\Lambda}^{12} \,.
}

Then, the Lagrangian of the $\phi$-theory including the background gauge fields is
\alis{
    \L_{\phi,\text{e}}\left[ \bm{C}, \bm{\hat{C}} \right] &= \frac{\mu_0}{2} (\partial_0 \phi - C_0 )^2 + \frac{1}{2 \mu_{12}} (\partial_1 \partial_2 \phi - C_{12})^2 \\
    & \quad + \frac{i}{2\pi} \hat{C}^{12}_0 ( \partial_1 \partial_2 \phi - C_{12}) + \frac{i}{2\pi} \hat{C} ( \partial_0 \phi - C_0 ) \,.  \label{exoticphic}
}
Under the background gauge transformation, it varies as
\alis{
    \delta_g \L_{\phi,\text{e}} &=  \frac{i}{2\pi} \left[ \partial_0 \hat{\Lambda}^{12} ( \partial_1 \partial_2 \phi - C_{12} ) + \partial_1 \partial_2 \hat{\Lambda}^{12} ( \partial_0 \phi - C_0 )   \right] \\
    &= \frac{i}{2\pi} \hat{\Lambda}^{12} ( \partial_0 C_{12} - \partial_1 \partial_2 C_0 ) \,. \label{variexoticphi}
}
Therefore, the partition function is not invariant and we cannot gauge the $U(1) \times U(1)$ subsystem symmetry, indicating a mixed ’t Hooft anomaly. This anomaly is canceled by combining the SSPT phase \eqref{exoticsspt} in one dimension higher. We put the $\phi$-theory on the boundary $x^3 = 0$, and set the boundary conditions
\ali{
    C_{\text{SSPT},0} |_{x^3 = 0} &= C_{\phi,0} \,, \\
    C_{\text{SSPT},12} |_{x^3 = 0} &= C_{\phi,12} \,, \\
    \hat{C}^{12}_{\text{SSPT},0} |_{x^3 = 0} &= \hat{C}^{12}_{\phi,0} \,, \\
    \hat{C}_{\text{SSPT}} |_{x^3 = 0} &= \hat{C}_{\phi} \,,
}
so then the variation of the SSPT phase \eqref{variexoticsspt} cancels the anomaly \eqref{variexoticphi}.

\subsection{Field Correspondences and Foliated SSPT Phase in 3+1 Dimensions}
\label{section23}

In Section \ref{section21}, we reviewed the exotic description of the SSPT phase with the spatial rotational symmetry $\Z_4$ in 3+1 dimensions. In this section, we use the foliated-exotic duality between the exotic and foliated SSPT phase, and construct the foliated description of the SSPT phase for $U(1) \times U(1)$ subsystem symmetry with two flat foliations. In this duality, the field correspondences are almost the same as those for the fractonic $BF$ theory with two flat foliations in 3+1 dimensions \cite{Spieler:2023wkz}. As for the foliated gauge fields, further details can be found in Appendix \ref{appendix2}.

The foliated SSPT phase has the background foliated gauge fields $C^k \wedge dx^k \  (k=1,2)$, $\hat{C}^k \ (k=1,2)$, $c$ and $\hat{c}_{12} \ dx^1 \wedge dx^2$. $C^k \wedge dx^k$ is a $U(1)$ type-$A$ foliated (1+1)-form gauge field and $c$ is a $U(1)$ type-$A$ bulk one-form gauge field. We assume correspondences between $\bm{C} = (C_0, C_{12}, C_3)$ and  $(C^k \wedge dx^k, c)$ such as
\ali{
    C_0 &\simeq c_0 \label{corr0-0} \,, \\
    C_{12} &\simeq C^1_2 + \partial_2 c_1 \,,  \label{corr12-1} \\
    C_3 &\simeq c_3 \label{corr3-3} \,.
}
In addition, to impose relations to the fields $C^k \wedge dx^k$ and $c$, we will add to the Lagrangian the term
\ali{
    \frac{i}{2\pi} \hat{c} \wedge \left. \left( \sum^{2}_{k=1} C^k \wedge dx^k + dc \right)\right|_{\hat{c}_{12} = 0} \,, \label{constterm}
}
where $\hat{c}_{ij} \ dx^i \wedge dx^j \ ((i,j) = (0,1), (0,2), (0,3), (2,3), (3,1))$ are dynamical two-form fields. On the other hand, $\hat{C}^k$ is a $U(1)$ type-$B$ foliated one-form gauge field and $\hat{c}_{12}$ is the $x^1 x^2$-component of a $U(1)$ type-$B$ bulk two-form gauge field. We also assume correspondences between $\bm{\hat{C}} = (\hat{C}_0^{12}, \hat{C}, \hat{C}_3^{12})$ and $(\hat{C}^k,\hat{c}_{12})$ such as
\ali{
    \hat{C}_0^{12} &\simeq \hat{C}_0^1 - \hat{C}_0^2 \,, \label{ssptcorr8} \\
    \hat{C} &\simeq \partial_1 \hat{C}_2^1 - \partial_2 \hat{C}_1^2 + \hat{c}_{12}  \,, \\
    \hat{C}_3^{12} &\simeq \hat{C}_3^1 - \hat{C}_3^2 \,. \label{ssptcorr10}
}

The background gauge transformations of the foliated gauge fields are
\ali{
    C^k \wedge dx^k &\sim C^k \wedge dx^k +d \lambda^k \wedge dx^k  \, , \\
    \hat{C}^k &\sim \hat{C}^k +d \hat{\lambda}^k - \hat{\kappa} \,, \\
     c &\sim c + d\lambda - \sum^{2}_{k=1} \lambda^k dx^k\,,  \\
     \hat{c}_{12} &\sim \hat{c}_{12} + \partial_1 \hat{\kappa}_2 - \partial_2 \hat{\kappa}_1 \,, 
}
where $\lambda^k \ dx^k$ is a type-$A$ foliated (0+1)-form gauge parameter, $\hat{\lambda}^k$ is a type-$B$ foliated zero-form gauge parameter, $\lambda$ is a type-$A$ bulk zero-form gauge parameter, and $\hat{\kappa}$ is a type-$B$ bulk one-form gauge parameter. From the background gauge transformations of the tensor gauge field \eqref{bkgugec0}--\eqref{bkgugec3} and \eqref{bkgugechat120}--\eqref{bkgugechat123}, we have correspondences between the background gauge parameters
\ali{
    \Lambda &\simeq \lambda \,, \label{ssptcorr11}\\
    \hat{\Lambda}^{12} &\simeq \hat{\lambda}^1 - \hat{\lambda}^2 \,. \label{ssptcorr12}
}
Note that the gauge parameters $\lambda^k\ dx^k$ and $\hat{\kappa}$ do not appear on the exotic side, and then the degrees of freedom of the fields are the same on both sides.\footnote{In more detail, the background gauge parameters also have their own gauge transformation. For example, $\hat{\Lambda}^{12}$ transforms as $\hat{\Lambda}^{12} \sim \hat{\Lambda}^{12} + 2\pi \hat{P}^1 - 2\pi \hat{P}^2$,  $\hat{\kappa}$ transforms as $\hat{\kappa} \sim \hat{\kappa} + d \hat{\nu}$, and $\hat{\lambda}^k$ transforms as $\hat{\lambda}^k \sim \hat{\lambda}^k + 2\pi \hat{p}^k  + \hat{\nu}$, where $\hat{\nu}$ is a type-$B$ bulk zero-form gauge parameter and $\hat{P}^k$ and $\hat{p}^k$ are integer-valued gauge parameters that can have step function discontinuities in the $x^k$ direction. Then, as for the form $\hat{\lambda}^1 - \hat{\lambda}^2$, the parameter $\hat{\nu}$ cancels out and the gauge transformation is $2\pi \hat{p}^1 - 2\pi \hat{p}^2$, which is consistent with $\hat{\Lambda}^{12} \simeq \hat{\lambda}^1 - \hat{\lambda}^2$ under the correspondence $\hat{P}^k \simeq \hat{p}^k$.}

Using the correspondences \eqref{corr0-0}--\eqref{corr3-3} and \eqref{ssptcorr8}--\eqref{ssptcorr10}, we can construct the foliated description of the SSPT phase for $U(1) \times U(1)$ subsystem symmetry. The Lagrangian can be written as
\alis{
    \L_{\text{SSPT,e}} &\simeq  \frac{i}{2\pi} \left[ (\partial_1 \hat{C}^1_2 - \partial_2 \hat{C}^2_1 + \hat{c}_{12}) ( \partial_0 c_3 - \partial_3 c_0 ) - ( \hat{C}^1_0 - \hat{C}^2_0 ) \left\{ \partial_3 ( C^1_2 + \partial_2 c_1 ) - \partial_1 \partial_2 c_3   \right\} \right. \\
    & \left. \quad  + ( \hat{C}^1_3 - \hat{C}^2_3 ) \left\{ \partial_0  ( C^1_2 + \partial_2 c_1 )   - \partial_1 \partial_2 c_0 \right\}  \right] \,. \label{ssptetof1}
}
Integrating by parts, dropping the $x^0$-, $x^1$- and $x^2$-derivative terms, and then adding the term for constraints \eqref{constterm}, we have
\alis{
    &\L_{\text{SSPT,f}} \\
    & \quad = \frac{i}{2\pi} \left[ - \hat{C}^1_2  \partial_1 ( \partial_0 c_3 - \partial_3 c_0 ) + \hat{C}^2_1  \partial_2 ( \partial_0 c_3 - \partial_3 c_0 ) - ( \hat{C}^1_0 - \hat{C}^2_0 ) \left\{ \partial_3 ( C^1_2 + \partial_2 c_1 ) - \partial_1 \partial_2 c_3   \right\} \right. \\
    & \left. \quad \quad + ( \hat{C}^1_3 - \hat{C}^2_3 ) \left\{ \partial_0  ( C^1_2 + \partial_2 c_1 )   - \partial_1 \partial_2 c_0 \right\} + \hat{c}_{12} ( \partial_0 c_3 - \partial_3 c_0 )  \right] \\
    & \quad \quad + \frac{i}{2\pi} \left[ \hat{c}_{01} ( - C^2_3 + \partial_2 c_3 - \partial_3 c_2 ) + \hat{c}_{02} ( C^1_3 + \partial_3 c_1 - \partial_1 c_3 )  \right. \\
    & \left. \quad \quad \quad + \hat{c}_{03} ( C^2_1 - C^1_2 + \partial_1 c_2 - \partial_2 c_1 )  + \hat{c}_{23} ( C^1_0 + \partial_0 c_1 - \partial_1 c_0 ) + \hat{c}_{31} ( C^2_0 + \partial_0 c_2 - \partial_2 c_0 ) \right]\,. \label{ssptlag2}
}
Next, let us consider changes of the dynamical variables\footnote{Since we choose the correspondence \eqref{corr12-1}, we consider $\hat{c}_{03} \rightarrow \hat{c}_{03} + \partial_0 \hat{C}^2_3 - \partial_3 \hat{C}^2_0$ to replace $C^1_2 + \partial_2 c_1$ with $C^2_1 + \partial_1 c_2$. Instead of that, we can adopt the correspondence $C_{12} \simeq C^2_1 + \partial_1 c_2$ and then we have to use the change of variable $\hat{c}_{03} \rightarrow \hat{c}_{03} + \partial_0 \hat{C}^1_3 - \partial_3 \hat{C}^1_0$.} as
\ali{
    \hat{c}_{ij} \rightarrow \hat{c}_{ij} + \partial_i \hat{C}^k_j - \partial_j \hat{C}^k_i \,, \ ( (i,j,k) = (0,1,2), (0,2,1), (0,3,2), (2,3,1), (3,1,2) ) \,.
}
The term $\partial_i \hat{C}^k_j - \partial_j \hat{C}^k_i$ has the background gauge transformation 
\ali{
    \partial_i \hat{C}^k_j - \partial_j \hat{C}^k_i \sim \partial_i \hat{C}^k_j - \partial_j \hat{C}^k_i - \partial_i \hat{\kappa}_j + \partial_j \hat{\kappa}_i \,,
}
so the new $\hat{c}_{ij}$ also has the background gauge transformation
\ali{
    \hat{c}_{ij} &\sim \hat{c}_{ij} + \partial_i \hat{\kappa}_j - \partial_j \hat{\kappa}_i \,, \quad ((i,j) = (0,1), (0,2), (0,3), (2,3), (3,1) ) \,. \label{dychatbkg}
}
Under the changes of variables we have additional terms
\alis{
    \frac{i}{2\pi} & \big[ - \hat{C}^1_2 \left\{ -\partial_1 ( \partial_0 c_3 - \partial_3 c_0 ) + ( \partial_0 C^1_3 - \partial_3 C^1_0 )  \right\} + \hat{C}^2_1 \left\{ -\partial_2 ( \partial_0 c_3 - \partial_3 c_0 ) + ( \partial_0 C^2_3 - \partial_3 C^2_0 )  \right\} 
     \\
    & \quad  - \hat{C}^1_0 \left\{ \partial_1 \partial_2 c_3 - \partial_2 ( C^1_3 + \partial_3 c_1 ) \right\} + \hat{C}^2_0 \left\{ -\partial_3 (C^1_2 + \partial_2 c_1 ) + \partial_1 \partial_2 c_3 + ( \partial_3 C^2_1 - \partial_1 C^2_3 ) \right\} \\
    & \quad + \hat{C}^1_3 \left\{ \partial_1 \partial_2 c_0 - \partial_2 ( C^1_0 + \partial_0 c_1 ) \right\} - \hat{C}^2_3 \left\{ -\partial_0 (C^1_2 + \partial_2 c_1 ) + \partial_1 \partial_2 c_0  +  (\partial_0 C^2_1 - \partial_1  C^2_0 )  \right\}  \\
    & \quad + \partial_3 \left\{ -\hat{C}^2_0 ( C^2_1 - C^1_2 + \partial_1 c_2 - \partial_2 c_1 ) - \hat{C}^1_2 ( C^1_0 + \partial_0 c_1 - \partial_1 c_0 ) + \hat{C}^2_1 ( C^2_0 + \partial_0 c_2 - \partial_2 c_0 ) \right\} \big] \,, \label{ssptaddterm}
}
where we have dropped the total derivative terms with respect to $x^0$, $x^1$ and $x^2$. The total derivative term with respect to $x^3$ has an effect when the spacetime has a boundary $x^3 = 0$ and will be considered in the boundary theory, so we drop it here. Adding \eqref{ssptaddterm} to \eqref{ssptlag2}, we can derive the foliated Lagrangian of the SSPT phase\footnote{We use $\L$ and $L$ as letters for Lagrangians. $\L$ is often referred to as Lagrangian density and $L$ is $\L \, d^dx$ in $d$-dimensional spacetime, so the action is written as $S = \int \L \, d^dx = \int L$.}
\alis{
    &L_{\text{SSPT,f}} \left[ C^k \wedge dx^k, c, \hat{C}^k, \hat{c}_{12}\ dx^1 \wedge dx^2 \right] \\
    & \quad = \frac{i}{2\pi} \left[  \sum_{k = 1}^2 \hat{C}^k \wedge d C^k \wedge dx^k + \hat{c} \wedge \left( \sum_{k=1}^2 C^k \wedge dx^k + dc \right) \right]  \,, \label{folissptlag}
}
where we combine the background bulk gauge field $\hat{c}^{12} \ dx^1 \wedge dx^2$ and the dynamical fields $\hat{c}_{ij} \ dx^i \wedge dx^j \ ((i,j) = (0,1), (0,2), (0,3), (2,3), (3,1))$ into $\hat{c}$.

We can also derive the foliated description \eqref{folissptlag} by integrating out the dynamical fields $\hat{c}_{ij} \ ((i,j) = (0,1), (0,2), (0,3), (2,3), (3,1))$ as Lagrange multipliers. From the term \eqref{constterm}, have a constraint
\ali{
    \left[ \sum^{2}_{k=1} C^k \wedge dx^k + dc \right]_{ij} = 0 \,,\quad (i,j) = (2,3), (3,1), (1,2), (0,1), (0,2)  \,, 
}
or in components,
\ali{
    -C^2_3 + \partial_2 c_3 - \partial_3 c_2 &= 0 \,, \label{constraint1} \\
    C^1_3 + \partial_3 c_1 - \partial_1 c_3 &= 0 \,, \\
    C^2_1 - C^1_2 + \partial_1 c_2 - \partial_2 c_1 &= 0  \,, \\
    C^1_0 + \partial_0 c_1 - \partial_1 c_0 &= 0 \,, \\
    C^2_0 + \partial_0 c_2 - \partial_2 c_0 &= 0 \,. \label{constraint5}
}
Combining these constraints with the correspondences \eqref{corr0-0}--\eqref{corr3-3}, we can derive a dictionary 
\ali{
    C_0 &\simeq c_0 \,, \label{ssptdic1} \\
    C_{12} &\simeq C^1_2 + \partial_2 c_1 = C^2_1 + \partial_1 c_2  \,,  \\
    C_3 &\simeq c_3 \,, \\
    \partial_1 C_0 &\simeq C^1_0 + \partial_0 c_1 \,, \\
    \partial_2 C_0 &\simeq C^2_0 + \partial_0 c_2 \,, \\
    \partial_1 C_3 &\simeq C^1_3 + \partial_3 c_1 \,, \\
    \partial_2 C_3 &\simeq C^2_3 + \partial_3 c_2 \,, \label{ssptdic7} \\
    \hat{C}_0^{12} &\simeq \hat{C}_0^1 - \hat{C}_0^2 \,,  \\
    \hat{C} &\simeq \partial_1 \hat{C}_2^1 - \partial_2 \hat{C}_1^2 + \hat{c}_{12}  \,, \\
    \hat{C}_3^{12} &\simeq \hat{C}_3^1 - \hat{C}_3^2 \,. \label{ssptdic10}
}
Using these correspondences \eqref{ssptdic1}--\eqref{ssptdic10}, integrating by parts and dropping the $x^0$-, $x^1$- and $x^2$-derivative terms, we have
\alis{
    \L_{\text{SSPT,e}} \, d^4 x &\simeq \frac{i}{2\pi} \left[ - \hat{C}^1_2 ( \partial_0 C^1_3 - \partial_3 C^1_0 ) + \hat{C}^2_1 ( \partial_0 C^2_3 - \partial_3 C^2_0 ) + \hat{c}_{12} (\partial_0 c_3 - \partial_3 c_0 )
     \right. \\
     & \quad - \hat{C}^1_0 ( \partial_3 C^1_2 - \partial_2 C^1_3 ) + \hat{C}^2_0 ( \partial_3 C^2_3 - \partial_1 C^2_3 ) \\
    & \left. \quad  + \hat{C}^1_3 ( \partial_0 C^1_2 - \partial_2 C^1_0 ) - \hat{C}^2_3 ( \partial_0 C^2_1 - \partial_1 C^2_0 )  \right] \, d^4 x \\
    & = \frac{i}{2\pi} \left[ \sum^2_{k=1} \hat{C}^k  \wedge d C^k \wedge dx^k + \hat{c}_{12} ( \partial_0 c_3 - \partial_3 c_0 ) \, d^4 x \right] \,,
}
where $d^4 x = dx^0 \wedge dx^1 \wedge dx^2 \wedge dx^3$. By recovering the term \eqref{constterm}, we can derive the foliated SSPT phase \eqref{folissptlag}.\footnote{In this method, we cannot detect the background gauge transformation of the dynamical fields \eqref{dychatbkg} systematically.}

Conversely, to obtain the exotic SSPT phase \eqref{exoticsspt} from the foliated SSPT phase \eqref{folissptlag}, we can integrate out the dynamical fields $\hat{c}_{ij} \ dx^i \wedge dx^j \ ((i,j) = (0,1), (0,2), (0,3), (2,3), (3,1))$ and use the dictionary \eqref{ssptdic1}--\eqref{ssptdic10}.

As in the case of the exotic SSPT phase, we put the foliated SSPT phase on the spacetime $\T^{2+1} \times \R_{x^3 \geq 0}$ with the boundary $x^3 = 0$. Then under the background gauge transformations, the gauge variation is
\alis{
    \delta_g L_{\text{SSPT,f}} &=  \frac{i}{2\pi} \left[  \sum^2_{k=1} ( d\hat{\lambda}^k - \hat{\kappa} ) \wedge d C^k \wedge dx^k + d\hat{\kappa} \wedge \left( \sum^2_{k=1}  C^k \wedge dx^k + dc   \right) \right] \\
    &= d \left[ \frac{i}{2\pi} \sum^2_{k=1} \hat{\lambda}^k \ d C^k \wedge dx^k + \frac{i}{2\pi} \hat{\kappa} \wedge \left( \sum^2_{k=1} C^k \wedge dx^k + dc \right) \right] \,.
}
The variation of the action is
\ali{
    \delta_g S_{\text{SSPT,f}} = \int_{x^3 = 0}  \left[ \frac{i}{2\pi} \sum^2_{k=1} \hat{\lambda}^k \ d C^k \wedge dx^k + \frac{i}{2\pi} \hat{\kappa} \wedge \left( \sum^2_{k=1} C^k \wedge dx^k + dc \right) \right] \,. \label{varifolisspt}
}
Using the correspondences \eqref{ssptdic1}--\eqref{ssptdic10} and \eqref{ssptcorr12}, we can derive the variation of the exotic SSPT phase \eqref{variexoticsspt} from the foliated one \eqref{varifolisspt}. Indeed, we can see
\alis{
    \delta_g S_{\text{SSPT,f}} &\simeq \int_{x^3 = 0}   \frac{i}{2\pi} \left[ \hat{\lambda}^1 ( \partial_1 \partial_2 C_0 - \partial_0 C_{12} ) + \hat{\lambda}^2 ( \partial_0 C_{12} - \partial_1 \partial_2 C_0 ) \right] dx^0 \wedge dx^1 \wedge dx^2  \\
    &\simeq \int_{x^3 = 0}  dx^0 dx^1 dx^2 \left[ - \frac{i}{2\pi}  \hat{\Lambda}^{12} ( \partial_0 C_{12} - \partial_1 \partial_2 C_0 )  \right] =  \delta_g S_{\text{SSPT,e}} \,.
}

\section{Boundary Theory for Foliated SSPT Phase}
\label{section3}

In the previous section, we have reviewed the anomaly inflow mechanism between the exotic SSPT phase and the exotic $\phi$-theory, and constructed the foliated SSPT phase for $U(1) \times U(1)$ subsystem symmetry using the foliated-exotic duality. In this section, we construct a foliated $\phi$-theory that has an 't Hooft anomaly of the $U(1) \times U(1)$ subsystem symmetry from the foliated SSPT phase \eqref{folissptlag}. This foliated $\phi$-theory is considered as a boundary theory of the foliated SSPT phase. Furthermore, by tuning the parameters in the foliated Lagrangian, we will derive the foliated $\phi$-theory that is equivalent to the exotic $\phi$-theory \eqref{exoticphi}, and construct the foliated-exotic duality in the gapless theory.

\subsection{Construction of Boundary Theory}
\label{section31}

First, we show a way to construct a boundary theory from an SPT phase in a relativistic case. We consider the SPT phase for a relativistic $U(1) \times U(1)$ symmetry in 2+1 dimensions, and construct the compact scalar field theory in 1+1 dimensions, which is a boundary theory of it.

As in Section \ref{section2}, we take Euclid spacetime to be a (2+1)-dimensional torus $\T^{2+1}$. The SPT phase for a $U(1) \times U(1)$ global symmetry is described by the Lagrangian
\ali{
    L_{\text{SPT}}\left[ C, \hat{C} \right] = \frac{i}{2\pi} \hat{C} \wedge d C \,, \label{spt1}
}
or
\ali{
    L'_{\text{SPT}} \left[ C, \hat{C} \right] = \frac{i}{2\pi} d \hat{C} \wedge  C \,, \label{spt2}
}
where $C$ and $\hat{C}$ are $U(1)$ one-form background gauge fields. These two Lagrangians differ only by a total derivative term, so they are the same on the spacetime without a boundary. The background gauge transformations of $C$ and $\hat{C}$ are
\ali{
    C &\sim C + d\lambda \,, \label{sptgauge1} \\
    \hat{C} &\sim \hat{C} + d\hat{\lambda} \,, \label{sptgauge2}
}
where $\lambda$ and $\hat{\lambda}$ are zero-form background gauge parameters. Under these gauge transformations, the Lagrangian is invariant if the spacetime does not have a boundary. However, when the spacetime has a boundary, we have a variation on the boundary. Let us consider the SPT phase \eqref{spt1} in the spacetime $\T^{1+1} \times \R_{x^2 \geq 0}$ with the boundary $x^2 = 0$. Then, under the gauge transformations, we have
\alis{
    \delta_g L_{\text{SPT}} &= \frac{i}{2\pi} d \hat{\lambda} \wedge d C \\
    &= d \left[ \frac{i}{2\pi}  \hat{\lambda} \ d C \right] \,,
}
and the variation of action is
\ali{
    \delta_g S_{\text{SPT}} = \int_{x^2 = 0} \frac{i}{2\pi} \hat{\lambda} \ d C \,. \label{varispt}
}
From the anomaly inflow mechanism, this variation should match an 't Hooft anomaly of some boundary theory. In this situation, we have one way to construct a boundary theory.

First, we choose a background gauge field $C$ and introduce a one-form lower dynamical gauge field that cancels the background gauge transformation \eqref{sptgauge1}. Since the gauge field $C$ is one-form, we introduce a zero-form scalar field $\phi$, whose dynamical gauge transformation is
\ali{
    \phi \sim \phi + 2\pi w \,,
}
where $w$ is a locally constant function valued in an integer. This gauge equivalence makes $\phi$ a compact scalar: $\phi \sim \phi + 2\pi$.\footnote{The compact scalar field $\phi$ in Section \ref{section31} is different from the compact scalar field $\phi$ in the exotic $\phi$-theory. } We set the background gauge transformation of $\phi$ as
\ali{
    \phi \sim \phi + \lambda \,,
}
and then, the term $C - d\phi$ becomes invariant under the background gauge transformations. Next, we consider the Lagrangian \eqref{spt2} on $\T^{1+1} \times \R_{x^2 \geq 0}$ in which $\hat{C}$ has a exterior derivative. Then, $L'_{\text{SPT}}$ is not gauge invariant due to \eqref{sptgauge1}, so we substitute $C - d\phi$ for $C$ in the $L'_{\text{SPT}}$.\footnote{We can choose the gauge field $\hat{C}$ and introduce a compact scalar $\hat{\phi}$. Then, we can derive the dual theory of $\hat{\phi}$. We will see it in Section \ref{section4}.} Then, we have
\alis{
    L''_{\text{SPT}} &= \frac{i}{2\pi} d \hat{C} \wedge ( C  - d\phi ) \\
    &= d \left[ \frac{i}{2\pi} \hat{C} \wedge ( C  - d\phi ) \right] + \frac{i}{2\pi} \hat{C} \wedge d C  \,. \label{spt''}
}
The boundary term in the action is
\ali{
    \int_{x^2 = 0} -\frac{i}{2\pi} \hat{C} \wedge ( C  - d\phi ) \,,
}
on the boundary $x^2 = 0$, and the other term is the SPT phase \eqref{spt1}.\footnote{Instead, we can use $L''_{\text{SPT}} = d \left[ - \frac{i}{2\pi} \hat{C} \wedge  d\phi \right] + \frac{i}{2\pi} d \hat{C} \wedge C $. Then, the boundary term is $\int_{x^3 = 0} \frac{i}{2\pi} \hat{C} \wedge  d\phi$ and the Lagrangian of the SPT phase is \eqref{spt2}.} Finally, we add a gauge-invariant quadratic term to the boundary, and the result is
\ali{
    L_{\phi, \text{r}} \left[ C, \hat{C} \right] = \frac{R^2}{2} ( d\phi - C ) \wedge \ast ( d\phi - C ) + \frac{i}{2\pi} \hat{C} \wedge ( d\phi - C )  \,, \label{phicc1}
}
or
\ali{
   \L_{\phi, \text{r}} \left[ C, \hat{C} \right]  = \frac{R^2}{2} \left\{ (\partial_0 \phi - C_0 ) ^2 + (\partial_1 \phi - C_1 )^2  \right\} + \frac{i}{2\pi} \left\{ \hat{C}_0 ( \partial_1 \phi - C_1 ) - \hat{C}_1 ( \partial_0 \phi - C_0 )  \right\} \,,
}
where $R$ is a parameter with mass dimension zero. Since the Lagrangian \eqref{spt''} is gauge invariant, the gauge variation of the action $S_{\phi, \text{r}} = \int L_{\phi, \text{r}}$ matches the variation of the SPT phase \eqref{varispt}. Therefore, this scalar field theory is an anomaly theory coupled to background gauge fields, whose 't Hooft anomaly is described by the SPT phase \eqref{spt1}.

We can identify the currents of the global symmetries. Coupling between background gauge fields and currents of global symmetry is in the form $i C \wedge \ast J$. To derive the conservation law, we just have to consider the gauge transformation of $C$ only, and set $d \lambda \wedge \ast J$ to be zero. Then, we have the conservation law $d \ast J = 0 $.\footnote{From the Noether theorem, we can derive the conservation law by considering a local transformation of $\phi$ in the theory without the background gauge field $C$, which is equivalent to considering only the gauge transformation of $C$ in the theory coupled to $C$.} Since coupling in \eqref{phicc1} is
\ali{
    i C \wedge \ast \left( i R^2 d\phi \right) + i \hat{C} \wedge \ast \left( - \frac{1}{2\pi} \ast d\phi \right) \,,
}
the currents are
\ali{
    J &= i R^2 d\phi \,, \\
    \hat{J} &= - \frac{1}{2\pi} \ast d\phi \,,
}
and the conservation laws are
\ali{
    d \ast J &= 0 \,, \\
    d \ast \hat{J} &= 0 \,.
}
The global symmetry with $J$ is called the $U(1)$ momentum global symmetry, and the global symmetry with $\hat{J}$ is called the $U(1)$ winding global symmetry. In 1+1 dimensions, both of the global symmetries are zero-form symmetry.

By setting the background gauge fields and parameters to zero, we can derive the compact scalar field theory
\ali{
    L_{\phi, \text{r}}  = \frac{R^2}{2}  d\phi \wedge \ast  d\phi \,,
}
or
\ali{
    \L_{\phi, \text{r}}   = \frac{R^2}{2} \left\{ (\partial_0 \phi) ^2 + (\partial_1 \phi )^2  \right\} \,.
}

Applying this method to the exotic SSPT phase \eqref{exoticsspt} , we can also derive the boundary exotic $\phi$-theory \eqref{exoticphic} and \eqref{exoticphi}. In this situation, the exotic $\phi$-theory does not have a continuous rotational symmetry, so we can introduce different parameters $\mu_0$ and $\mu_{12}$ as the coefficients of the kinetic terms.

\subsection{Construction of Foliated \texorpdfstring{$\phi$}{φ}-Theory}
\label{section32}

We construct a foliated $\phi$-theory in 2+1 dimensions by using the method in Section \ref{section31}. We take the foliated SSPT phase \eqref{folissptlag}, and choose the background gauge fields $C^k \wedge dx^k$ and $c$. Then, we have to consider the Lagrangian integrated by parts
\alis{
    &L'_{\text{SSPT,f}} \left[ C^k \wedge dx^k, c, \hat{C}^k, \hat{c}_{12}\ dx^1 \wedge dx^2 \right] \\
    & \quad = \frac{i}{2\pi} \left[  \sum_{k = 1}^2 ( d \hat{C}^k + \hat{c} ) \wedge C^k \wedge dx^k   - d\hat{c} \wedge c  \right]  \,,
}
on the spacetime $\T^{2+1} \times \R_{x^3 \geq 0}$. Next, we introduce a $U(1)$ type-$A$ foliated (0+1)-form gauge field $\Phi^k\, dx^k\ (k = 1,2)$ and a $U(1)$ type-$A$ bulk zero-form gauge field $\Phi$, whose dynamical gauge transformations are
\ali{
    \Phi^k \, dx^k &\sim \Phi^k\, dx^k + 2\pi d W^k \,, \label{dyngaugephik} \\
    \Phi &\sim \Phi + 2\pi W^1 + 2\pi W^2 \,,
}
where $W^k$ is an integer-valued gauge parameters and it can have step function discontinuities in the $x^k$ direction. Since the background gauge transformations of the foliated gauge fields are
\ali{
    C^k \wedge dx^k &\sim C^k \wedge dx^k +d \lambda^k \wedge dx^k  \, , \label{bkggaugetr1} \\
    \hat{C}^k &\sim \hat{C}^k +d \hat{\lambda}^k - \hat{\kappa} \,, \label{bkggaugetr2}\\
     c &\sim c + d\lambda - \sum^{2}_{k=1} \lambda^k dx^k\,, \label{bkggaugetr3} \\
     \hat{c} &\sim \hat{c} + d \hat{\kappa} \,, \label{bkggaugetr4} 
}
$d\hat{C}^k + \hat{c}$ and $d\hat{c}$ are gauge invariant, but the parts of $C^k \wedge dx^k$ and $c$ are not on the spacetime with boundary. As in the case of the relativistic theory in Section \ref{section31}, we set the background gauge transformations of $\Phi^k dx^k$ and $\Phi$ as
\ali{
    \Phi^k\, dx^k &\sim \Phi^k\, dx^k  + \lambda^k\, dx^k  \,, \label{bkggaugephik} \\
    \Phi &\sim \Phi + \lambda \,. \label{bkggaugephi}
}
and substitute the gauge-invariant combinations $ (C^k - d\Phi^k ) \wedge dx^k $ and $c - d\Phi + \sum_{k = 1}^2 \Phi^k\, dx^k$ for $C^k \wedge dx^k$ and $c$ in $L'_{\text{SSPT,f}}$ respectively. Then, we have
\alis{
    &L''_{\text{SSPT,f}}  \\
    & \quad = \frac{i}{2\pi} \left[  \sum_{k = 1}^2 ( d \hat{C}^k + \hat{c} ) \wedge ( C^k - d\Phi^k ) \wedge dx^k - d\hat{c} \wedge \left( c - d\Phi + \sum_{k = 1}^2 \Phi^k\, dx^k \right) \right]   \\
    & \quad = \frac{i}{2\pi} \left[ \sum_{k = 1}^2  \hat{C}^k \wedge d C^k \wedge dx^k + \hat{c} \wedge \left( \sum_{k = 1}^2 C^k \wedge dx^k + dc \right) \right] \\
    &\qquad +  d \left[ \frac{i}{2\pi}\sum_{k = 1}^2  \hat{C}^k \wedge ( C^k - d\Phi^k ) \wedge dx^k - \frac{i}{2\pi} \hat{c} \wedge \left( c - d\Phi +  \sum_{k = 1}^2 \Phi^k\, dx^k  \right)  \right] \,.
}
The boundary term in the action is
\alis{
    &\int_{x^3 = 0} \frac{i}{2\pi} \left[ - \sum_{k = 1}^2  \hat{C}^k \wedge ( d\Phi^k - C^k ) \wedge dx^k + \hat{c} \wedge \left( d\Phi - \sum_{k = 1}^2 \Phi^k\, dx^k - c \right) \right] \\
    & = \int_{x^3 = 0} \frac{i}{2\pi} \left[ \hat{C}^1_0 ( 
 \partial_2 \Phi^1 - C^1_2 ) - \hat{C}^1_2 ( 
 \partial_0 \Phi^1 - C^1_0 ) - \hat{C}^2_0 ( 
 \partial_1 \Phi^2 - C^2_1 ) + \hat{C}^2_1 ( 
 \partial_0 \Phi^2 - C^2_0 )  \right. \\
 & \qquad \qquad \left.  + \hat{c}_{12} ( \partial_0 \Phi - c_0 )  - \hat{c}_{02} ( \partial_1 \Phi - \Phi^1 - c_1  ) + \hat{c}_{01} ( \partial_2 \Phi - \Phi^2 - c_2  )  \right] d^3 x 
}
where $d^3 x = dx^0 \wedge dx^1 \wedge dx^2$. Note that the gauge fields $\hat{c}_{01}$ and $\hat{c}_{02}$ on the boundary are dynamical, so integrating out these fields, we have the constraints
\ali{
    (  C^1_3 + \partial_3 c_1 - \partial_1 c_3 ) - ( \partial_1 \Phi - \Phi^1 - c_1  ) \delta(x^3) &= 0 \,, \\
    ( - C^2_3 + \partial_2 c_3 - \partial_3 c_2 ) +  ( \partial_2 \Phi - \Phi^2 - c_2  ) \delta(x^3) &= 0 \,.
}
Since we have the constraints \eqref{constraint1}--\eqref{constraint5} in the (3+1)-dimensional spacetime, we have
\ali{
    \partial_1 \Phi - \Phi^1 - c_1  &= 0 \,, \label{constraintphi1} \\
    \partial_2 \Phi - \Phi^2 - c_2  &= 0 \label{constraintphi2} \,,
}
on the boundary.\footnote{The first terms $C^1_3 + \partial_3 c_1 - \partial_1 c_3$ and $- C^2_3 + \partial_2 c_3 - \partial_3 c_2$ do not have contribution proportional to $\delta(x^3)$ because the theory does not have a foliation in the $x^3$ direction.} In addition, the background gauge fields are the ones in the (3+1)-dimensional spacetime restricted to the (2+1)-dimensional boundary, we also have constraints on the background gauge fields $C^k \wedge dx^k$ and $c$ as
\ali{
    C^2_1 - C^1_2 + \partial_1 c_2 - \partial_2 c_1 &= 0 \label{constc12} \,, \\
    C^1_0 + \partial_0 c_1 - \partial_1 c_0 &= 0 \,, \label{constc01} \\
    C^2_0 + \partial_0 c_2 - \partial_2 c_0 &= 0 \,. \label{constc02}
}
Therefore, we add the term 
\alis{
    & \frac{i}{2\pi} \hat{\chi} \wedge \left( \sum^{2}_{k=1} C^k \wedge dx^k + dc \right) \\
    &= \frac{i}{2\pi} \left[ \hat{\chi}_0 ( C^2_1 - C^1_2 + \partial_1 c_2 - \partial_2 c_1 ) \right. \\
    & \quad \left. - \hat{\chi}_1 ( C^2_0 + \partial_0 c_2 - \partial_2 c_0) + \hat{\chi}_2 ( C^1_0 + \partial_0 c_1 - \partial_1 c_0 ) \right] d^3 x \,,
}
on the boundary, where $\hat{\chi}$ is a dynamical one-form field. Finally, we add background gauge-invariant and $\Z_4$ rotational invariant quadratic terms,\footnote{Quadratic terms $ ( \partial_1 \Phi - \Phi^1 - c_1 )^2$ and $ ( \partial_2 \Phi - \Phi^2 - c_2 )^2$ are absorbed to changes of the variables $\hat{c}_{01}$ and $\hat{c}_{02}$ on the boundary.} and we have
\alis{
    &\L_{\phi,\text{f}} \left[ C^k \wedge dx^k, c, \hat{C}^k, \hat{c}_{12}\ dx^1 \wedge dx^2 \right] \\
    & \quad = \frac{\mu_0}{2} ( \partial_0 \Phi - c_0 )^2 + \frac{1}{4\mu_{12}} ( \partial_2 \Phi^1 - C^1_2 )^2 + \frac{1}{4\mu_{12}} ( \partial_1 \Phi^2 - C^2_1 )^2 \\
    & \qquad + \frac{1}{2\mu_{012}} ( \partial_0 \Phi^1 - C^1_0 )^2  + \frac{1}{2\mu_{012}} ( \partial_0 \Phi^2 - C^2_0 )^2 \\
    & \qquad + \frac{i}{2\pi} \left[ \hat{C}^1_0 ( 
    \partial_2 \Phi^1 - C^1_2 ) - \hat{C}^1_2 ( 
    \partial_0 \Phi^1 - C^1_0 ) - \hat{C}^2_0 ( 
    \partial_1 \Phi^2 - C^2_1 )  \right. \\
    & \qquad \quad \left. + \hat{C}^2_1 ( 
    \partial_0 \Phi^2 - C^2_0 ) + \hat{c}_{12} ( \partial_0 \Phi - c_0 ) \right] \\
    & \qquad + \frac{i}{2\pi} \left[ - \hat{c}_{02} ( \partial_1 \Phi - \Phi^1 - c_1  ) + \hat{c}_{01} ( \partial_2 \Phi - \Phi^2 - c_2  ) \right] \\
    & \qquad + \frac{i}{2\pi} \left[ \hat{\chi}_0 ( C^2_1 - C^1_2 + \partial_1 c_2 - \partial_2 c_1 ) \right. \\
    & \qquad \quad \left. - \hat{\chi}_1 ( C^2_0 + \partial_0 c_2 - \partial_2 c_0) + \hat{\chi}_2 ( C^1_0 + \partial_0 c_1 - \partial_1 c_0 ) \right] \,, \label{genefoliatedphi}
}
where $\mu_0$, $\mu_{12}$ and $\mu_{012}$ are parameters with mass dimension one. We can check the $\Z_4$ rotational invariance of the Lagrangian using the transformation rules in Appendix \ref{appendix23}.\footnote{We can change the term $\frac{1}{4\mu_{12}} ( \partial_2 \Phi^1 - C^1_2 )^2 + \frac{1}{4\mu_{12}} ( \partial_1 \Phi^2 - C^2_1 )^2$ to $\frac{1}{2\mu^1_2} ( \partial_2 \Phi^1 - C^1_2 )^2 + \frac{1}{2\mu^2_1} ( \partial_1 \Phi^2 - C^2_1 )^2$, where $\mu^1_2$ and $\mu^2_1$ are parameters satisfying $1/\mu_{12} = 1/\mu^1_2 + 1/\mu^2_1$. Using the equations of motion \eqref{constraintphi1}, \eqref{constraintphi2} and \eqref{constc12}, this term becomes $\Z_4$ rotational invariant and the parameters $\mu^1_2$ and $\mu^2_1$ appear only in the form $1/\mu_{12} = 1/\mu^1_2 + 1/\mu^2_1$.} We will consider global symmetry in Section \ref{section33}. 

We can also write the foliated Lagrangian in differential form by tuning the metric:
\alis{
    &L_{\phi,\text{f}} \left[ C^k \wedge dx^k, c, \hat{C}^k, \hat{c}_{12}\ dx^1 \wedge dx^2 \right] \\
    & \quad = \frac{1}{2} \left( d \Phi - \sum_{k = 1}^2 \Phi^k \, dx^k - c \right) \wedge \ast \left( d \Phi - \sum_{k = 1}^2 \Phi^k \, dx^k - c \right) \\
    & \qquad + \frac{1}{2} \sum_{k=1}^2  \left( d \Phi^k \wedge dx^k - C^k \wedge dx^k  \right) \wedge \ast \left( d \Phi^k \wedge dx^k - C^k \wedge dx^k  \right)  \\
    & \qquad + \frac{i}{2\pi} \left[ - \sum_{k = 1}^2  \hat{C}^k \wedge ( d\Phi^k - C^k ) \wedge dx^k + \hat{c} \wedge \left( d\Phi - \sum_{k = 1}^2 \Phi^k\, dx^k - c \right) \right]  \\
    & \qquad + \frac{i}{2\pi} \hat{\chi} \wedge \left( \sum^{2}_{k=1} C^k \wedge dx^k + dc \right) \,. \label{foliphideff}
}

\subsection{Foliated-Exotic Duality in \texorpdfstring{$\phi$}{φ}-Theory}
\label{section33}

In this section, we compare the exotic $\phi$-theory \eqref{exoticphic} with the foliated $\phi$-theory \eqref{genefoliatedphi} in 2+1 dimensions. We will establish the field correspondences in the $\phi$-theory, and by tuning the parameters, we will see that the foliated $\phi$-theory can be equivalent to the exotic $\phi$-theory, which is a gapless example of the foliated-exotic duality.

In Section \ref{section22}, we reviewed the exotic $\phi$-theory coupled with the background tensor gauge fields
\alis{
    \L_{\phi,\text{e}}\left[ \bm{C}, \bm{\hat{C}} \right] &= \frac{\mu_0}{2} (\partial_0 \phi - C_0 )^2 + \frac{1}{2 \mu_{12}} (\partial_1 \partial_2 \phi - C_{12})^2 \\
    & \quad + \frac{i}{2\pi} \hat{C}^{12}_0 ( \partial_1 \partial_2 \phi - C_{12}) + \frac{i}{2\pi} \hat{C} ( \partial_0 \phi - C_0 ) \,.  \label{exoticphic2}
}
First, we assume correspondences between the scalar field $\phi$ in the exotic $\phi$-theory and the type-$A$ bulk zero-form gauge field $\Phi$ in a foliated $\phi$-theory as
\ali{
    \phi \simeq \Phi \,. \label{corrphi}
}
Since the dynamical gauge transformations of $\phi$ and $\Phi$ are
\ali{
    \phi &\sim \phi + 2 \pi w^1 + 2 \pi w^2 \,, \\
    \Phi &\sim \Phi + 2 \pi W^1 + 2 \pi W^2 \,,
}
the correspondences between the dynamical gauge parameters are
\ali{
    w^k \simeq W^k \,, \quad (k = 1,2) \,. 
}
In addition, to impose relations to the fields $\Phi$ and 
type-$A$ foliated (0+1)-form gauge fields $\Phi^k \, dx^k \ ( k = 1,2 )$, we add to the Lagrangian the term
\ali{
    \frac{i}{2\pi} \left[ -\hat{c}_{02} ( \partial_1 \Phi - \Phi^1 - c_1 )  + \hat{c}_{01} ( \partial_2 \Phi - \Phi^2 - c_2 ) \right] \,,
}
where $\hat{c}_{01}$ and $\hat{c}_{02}$ are dynamical gauge fields. The dynamical gauge transformation of $\Phi^k\ dx^k$ is \eqref{dyngaugephik}, and the background gauge transformations of $\Phi^k\ dx^k$ and $\Phi$ are \eqref{bkggaugephik} and \eqref{bkggaugephi}.

For the background gauge fields, the background gauge fields in a foliated theory are $U(1)$ type-$A$ foliated (1+1)-form gauge fields $C^k \wedge dx^k \ (k = 1,2)$, a $U(1)$ type-$A$ bulk one-form gauge field $c$, $U(1)$ type-$B$ foliated one-form gauge fields $\hat{C}^k \ (k = 1,2)$ and the $x^1x^2$-component of a $U(1)$ type-$B$ bulk two-form gauge field $\hat{c}_{12}$. Then, we assume correspondences
\ali{
    C_0 &\simeq c_0 \label{bcorr0-0} \,, \\
    C_{12} &\simeq C^1_2 + \partial_2 c_1 \,,  \label{bcorr12-1} 
}
and
\ali{
    \hat{C}_0^{12} &\simeq \hat{C}_0^1 - \hat{C}_0^2 \,, \label{bcorr012} \\
    \hat{C} &\simeq \partial_1 \hat{C}_2^1 - \partial_2 \hat{C}_1^2 + \hat{c}_{12}  \,. \label{bcorr1221}
}
Their background gauge transformations are \eqref{bkggaugetr1}--\eqref{bkggaugetr4}. As in the previous section, we add to the Lagrangian the term
\alis{
    & \frac{i}{2\pi} \hat{\chi} \wedge \left( \sum^{2}_{k=1} C^k \wedge dx^k + dc \right) \\
    &= \frac{i}{2\pi} \left[ \hat{\chi}_0 ( C^2_1 - C^1_2 + \partial_1 c_2 - \partial_2 c_1 ) \right. \\
    & \quad \left. - \hat{\chi}_1 ( C^2_0 + \partial_0 c_2 - \partial_2 c_0) + \hat{\chi}_2 ( C^1_0 + \partial_0 c_1 - \partial_1 c_0 ) \right] d^3 x \,.
}

Using these correspondences, the Lagrangian of the $\phi$-theory can be written as
\alis{
    \L_{\phi,\text{e}} &\simeq \frac{\mu_0}{2} (\partial_0 \Phi - c_0 )^2 + \frac{1}{2 \mu_{12}} (\partial_1 \partial_2 \Phi - C^1_2 - \partial_2 c_1 )^2 \\
    & \quad + \frac{i}{2\pi} ( \hat{C}^{1}_0 - \hat{C}^2_0 ) ( \partial_1 \partial_2 \Phi - C^1_2 - \partial_2 c_1 ) \\
    & \quad - \frac{i}{2\pi} \hat{C}_2^1 ( \partial_0 \partial_1 \Phi - \partial_1 c_0 ) + \frac{i}{2\pi} \hat{C}_1^2 ( \partial_0 \partial_2 \Phi - \partial_2 c_0 )  + \frac{i}{2\pi} \hat{c}_{12} ( \partial_0 \Phi - c_0 ) \\
    & \quad + \frac{i}{2\pi} \left[ -\hat{c}_{02} ( \partial_1 \Phi - \Phi^1 - c_1 )  + \hat{c}_{01} ( \partial_2 \Phi - \Phi^2 - c_2 ) \right] \\
    & \quad + \frac{i}{2\pi} \left[ \hat{\chi}_0 ( C^2_1 - C^1_2 + \partial_1 c_2 - \partial_2 c_1 ) \right. \\
    & \qquad \left. - \hat{\chi}_1 ( C^2_0 + \partial_0 c_2 - \partial_2 c_0) + \hat{\chi}_2 ( C^1_0 + \partial_0 c_1 - \partial_1 c_0 ) \right] \,,
}
where we have integrated by parts and dropped the total derivative terms.  

Next, we consider changes of the dynamical variables
\ali{
    \hat{c}_{02} &\rightarrow \hat{c}_{02} + \frac{i\pi}{2\mu_{12}}  \partial_2 \left\{ \partial_2 ( \partial_1 \Phi + \Phi^1 + c_1 ) - 2 ( C^1_2 +  \partial_2 c_1 )   \right\} \\
    \hat{c}_{01} &\rightarrow \hat{c}_{01} - \frac{i\pi}{2\mu_{12}}  \partial_1 \left\{ \partial_1 ( \partial_2 \Phi + \Phi^2 + c_2 ) - 2 ( C^1_2 +  \partial_2 c_1 )   \right\} \\
    \hat{\chi}_0 &\rightarrow \hat{\chi}_0 - \frac{i\pi}{2\mu_{12}} \left\{ C^2_1 + \partial_1 c_2 + C^1_2 + \partial_2 c_1 - 2 \partial_1 ( \Phi^2 + c_2 ) \right\} \,.
}
The additional terms are background gauge invariant, so the background gauge transformations of the dynamical fields are not added. Then, we can see 
\alis{
    &\frac{1}{2 \mu_{12}} (\partial_1 \partial_2 \Phi - C^1_2 - \partial_2 c_1 )^2 \\
    & \quad \rightarrow \frac{1}{4 \mu_{12}} ( \partial_2 \Phi^1 - C^1_2 )^2 + \frac{1}{4 \mu_{12}} ( \partial_1 \Phi^2 - C^2_1 )^2 \,.
}
Furthermore, we consider changes of the dynamical variables
\ali{
    \hat{c}_{02} &\rightarrow \hat{c}_{02} + \partial_0 \hat{C}^1_2 - \partial_2 \hat{C}^1_0  \,, \\
    \hat{c}_{01} &\rightarrow \hat{c}_{01} + \partial_0 \hat{C}^2_1 - \partial_1 \hat{C}^2_0  \,,
}
where the new $\hat{c}_{01}$ and $\hat{c}_{02}$ get the background gauge transformations
\ali{
    \hat{c}_{01} \sim \hat{c}_{01} + \partial_0 \hat{\kappa}_1 - \partial_1 \hat{\kappa}_0 \,, \\ 
    \hat{c}_{02} \sim \hat{c}_{02} + \partial_0 \hat{\kappa}_2 - \partial_2 \hat{\kappa}_0 \,.
}
In addition, we add the boundary term of \eqref{ssptaddterm}:
\ali{
     \frac{i}{2\pi} \left[ \hat{C}^2_0 ( C^2_1 - C^1_2 + \partial_1 c_2 - \partial_2 c_1 ) - \hat{C}^2_1 ( C^2_0 + \partial_0 c_2 - \partial_2 c_0 ) + \hat{C}^1_2 ( C^1_0 + \partial_0 c_1 - \partial_1 c_0 )   \right] \,.
}
Then, we can see
\alis{
    & \frac{i}{2\pi} ( \hat{C}^{1}_0 - \hat{C}^2_0 ) ( \partial_1 \partial_2 \Phi - C^1_2 - \partial_2 c_1 ) \\
    & \quad - \frac{i}{2\pi} \hat{C}_2^1 ( \partial_0 \partial_1 \Phi - \partial_1 c_0 ) + \frac{i}{2\pi} \hat{C}_1^2 ( \partial_0 \partial_2 \Phi - \partial_2 c_0 ) \\
    & \rightarrow  \frac{i}{2\pi} \left[ \hat{C}^{1}_0 ( \partial_2  \Phi^1 - C^1_2 ) - \hat{C}^{2}_0 ( \partial_1  \Phi^2 - C^2_1 ) \right. \\
    & \qquad\quad \left. -  \hat{C}_2^1 ( \partial_0  \Phi^1 - C^1_0 ) + \hat{C}_1^2 ( \partial_0 \Phi^2 - C^2_0 ) \right] \,.
}
After all, we derive the foliated $\phi$-theory that is equivalent to the exotic $\phi$-theory \eqref{exoticphic2}:
\alis{
    &\L_{\phi,\text{e} \rightarrow \text{f}} \left[ C^k \wedge dx^k, c, \hat{C}^k, \hat{c}_{12}\ dx^1 \wedge dx^2 \right] \\
    & \quad = \frac{\mu_0}{2} ( \partial_0 \Phi - c_0 )^2 + \frac{1}{4\mu_{12}} ( \partial_2 \Phi^1 - C^1_2 )^2 + \frac{1}{4\mu_{12}} ( \partial_1 \Phi^2 - C^2_1 )^2 \\
    & \qquad + \frac{i}{2\pi} \left[ \hat{C}^1_0 ( 
    \partial_2 \Phi^1 - C^1_2 ) - \hat{C}^1_2 ( 
    \partial_0 \Phi^1 - C^1_0 ) - \hat{C}^2_0 ( 
    \partial_1 \Phi^2 - C^2_1 )  \right. \\
    & \qquad \quad \left. + \hat{C}^2_1 ( 
    \partial_0 \Phi^2 - C^2_0 ) + \hat{c}_{12} ( \partial_0 \Phi - c_0 ) \right] \\
    & \qquad + \frac{i}{2\pi} \left[ -\hat{c}_{02} ( \partial_1 \Phi - \Phi^1 - c_1 )  + \hat{c}_{01} ( \partial_2 \Phi - \Phi^2 - c_2 ) \right] \\
    & \qquad + \frac{i}{2\pi} \left[ \hat{\chi}_0 ( C^2_1 - C^1_2 + \partial_1 c_2 - \partial_2 c_1 ) \right. \\
    & \qquad \quad \left. - \hat{\chi}_1 ( C^2_0 + \partial_0 c_2 - \partial_2 c_0) + \hat{\chi}_2 ( C^1_0 + \partial_0 c_1 - \partial_1 c_0 ) \right] \,. \label{etofoliatedphi}
}
If we set the background gauge fields to zero, we can derive the foliated $\phi$-theory without the background gauge fields
\alis{
    &\L_{\phi,\text{e} \rightarrow \text{f}}  = \frac{\mu_0}{2} ( \partial_0 \Phi )^2 + \frac{1}{4\mu_{12}} ( \partial_2 \Phi^1 )^2 + \frac{1}{4\mu_{12}} ( \partial_1 \Phi^2 )^2 \\
    & \qquad + \frac{i}{2\pi} \left[ -\hat{c}_{02} ( \partial_1 \Phi - \Phi^1 )  + \hat{c}_{01} ( \partial_2 \Phi - \Phi^2 ) \right]  \,. \label{etofoliatedphiczero}
}

We can also derive the foliated description \eqref{etofoliatedphi} by integrating out the dynamical fields $\hat{c}_{01}$, $\hat{c}_{02}$ and $\hat{\chi}_i \ (i = 0,1,2)$ as Lagrange multipliers. Then, we have the constraints \eqref{constraintphi1}, \eqref{constraintphi2} and \eqref{constc12}--\eqref{constc02}, and we can use a dictionary
\ali{
    \phi &\simeq \Phi \,,  \label{phidic1} \\
    \partial_1 \phi &\simeq \Phi^1 + c_1 \,, \\
    \partial_2 \phi &\simeq \Phi^2 + c_2 \,, \\
    C_0 &\simeq c_0 \,, \\
    C_{12} &\simeq C^1_2 + \partial_2 c_1 =  C^2_1 + \partial_1 c_2 \,, \\
    \partial_1 C_0 &\simeq C^1_0 + \partial_0 c_1 \,, \\
    \partial_2 C_0 &\simeq C^2_0 + \partial_0 c_2 \,, \\ 
    \hat{C}_0^{12} &\simeq \hat{C}_0^1 - \hat{C}_0^2 \,,  \\
    \hat{C} &\simeq \partial_1 \hat{C}_2^1 - \partial_2 \hat{C}_1^2 + \hat{c}_{12}  \,, \label{phidic9}
}
instead of the correspondences \eqref{corrphi}, \eqref{bcorr0-0} and \eqref{bcorr12-1}. Using these correspondences and recovering the terms of $\hat{c}_{01}$, $\hat{c}_{02}$ and $\hat{\chi}_i \ (i = 0,1,2)$, we can derive the foliated Lagrangian \eqref{etofoliatedphi} again.

Comparing the foliated Lagrangian \eqref{etofoliatedphi} with \eqref{genefoliatedphi}, the difference is the terms
\ali{
    \frac{1}{2\mu_{012}} ( \partial_0 \Phi^1 - C^1_0 )^2  + \frac{1}{2\mu_{012}} ( \partial_0 \Phi^2 - C^2_0 )^2 \,,
}
so we have only to take the limit $\mu_{012} \rightarrow \infty$ to get \eqref{etofoliatedphi} from \eqref{genefoliatedphi}. In the exotic description, these terms correspond to
\ali{
    \frac{1}{2\mu_{012}} \left\{ \partial_1 ( \partial_0 \phi - C_0 ) \right\}^2  + \frac{1}{2\mu_{012}} \left\{ \partial_2 ( \partial_0 \phi - C_0 ) \right\}^2 \,.
}
Note that, due to the UV/IR mixing, these terms have contribution to the energy of the same order as the leading term \cite{Seiberg:2020bhn}.

Conversely, to obtain the exotic $\phi$-theory \eqref{exoticphic2} from the foliated $\phi$-theory \eqref{etofoliatedphi}, we can integrate out the dynamical fields $\hat{c}_{01}$, $\hat{c}_{02}$ and $\hat{\chi}_i \ (i = 0,1,2)$ and use the dictionary \eqref{phidic1}--\eqref{phidic9}.

Let us consider subsystem symmetry. From the background gauge transformations of $C^k \wedge dx^k$ and $c$, which are \eqref{bkggaugetr1} and \eqref{bkggaugetr3}, in the Lagrangian with the background gauge fields \eqref{etofoliatedphi}, we have
\begin{gather}
    \partial_0 \left( i \mu_0 \partial_0 \Phi \right) + \partial_1 \left( \frac{1}{2\pi} \hat{c}_{02} \right) - \partial_2 \left( \frac{1}{2\pi} \hat{c}_{01} \right) = 0 \,, \\
    \partial_2 \left( \frac{i}{2\mu_{12}} \partial_2 \Phi^1 \right) + \frac{1}{2\pi} \hat{c}_{02} = 0 \,, \\
    \partial_1 \left( \frac{i}{2\mu_{12}} \partial_1 \Phi^2 \right) - \frac{1}{2\pi} \hat{c}_{01} = 0 \,,
\end{gather}
where $\hat{\chi}$ is canceled out. These equations can also be obtained as the equations of motion of $\Phi$, $\Phi^1$ and $\Phi^2$ in the theory where the background gauge fields are zero \eqref{etofoliatedphiczero}.
Combining these equations, we derive a conservation law
\ali{
    \partial_0 J_0 - \partial_1 \partial_2 J_{12}  = 0 \,,
}
where the currents are\footnote{In the theory \eqref{genefoliatedphi}, the $x^0$ component of the current $J_0$ is modified to $J_0 = i \mu_0 \partial_0 \Phi -  \frac{i}{\mu_{012}} \partial_0 \partial_1 \Phi^1 - \frac{i}{\mu_{012}} \partial_0 \partial_2 \Phi^2$.}
\ali{
    J_0 &= i \mu_0 \partial_0 \Phi \,, \\
    J_{12} &= \frac{i}{2\mu_{12}} ( \partial_2 \Phi^1 + \partial_1 \Phi^2 ) \,.
}
Using the constraints derived by integrating out the dynamical fields $\hat{c}_{01}$ and $\hat{c}_{02}$
\ali{
    \partial_1 \Phi - \Phi^1 &= 0 \,, \label{constphinoc1} \\
    \partial_2 \Phi - \Phi^2 &= 0 \,, \label{constphinoc2}
}
the $x^1 x^2$-component of the current $J_{12}$ becomes
\ali{
    J_{12} = \frac{i}{\mu_{12}} \partial_1 \partial_2 \Phi \,.
}
These currents are equivalent to the currents \eqref{momsym1} and \eqref{momsym2} under the correspondence \eqref{corrphi}, and thus generate the momentum dipole symmetry. We also have conservation laws related to winding symmetry. From the background gauge transformations of $\hat{C}^k$ and $\hat{c}$, which are \eqref{bkggaugetr2} and \eqref{bkggaugetr4}, we have relations
\begin{gather}
    \partial_0 \left( \frac{1}{2\pi} \partial_2 \Phi^1  \right) - \partial_2 \left( \frac{1}{2\pi} \partial_0 \Phi^1  \right) = 0 \,, \\
    \partial_0 \left( -\frac{1}{2\pi} \partial_1 \Phi^2  \right) + \partial_1 \left( \frac{1}{2\pi} \partial_0 \Phi^2  \right) = 0 \,, \\
    \frac{1}{2\pi} \partial_2 \Phi^1 - \frac{1}{2\pi}\partial_1 \Phi^2 + \partial_2 \left\{ \frac{1}{2\pi} ( \partial_1 \Phi - \Phi^1 ) \right\} - \partial_1 \left\{ \frac{1}{2\pi} ( \partial_2 \Phi - \Phi^2 ) \right\} = 0 \,, \\
    \frac{1}{2\pi} \partial_0 \Phi^2 - \partial_2 \left( \frac{1}{2\pi} \partial_0 \Phi \right) + \partial_0 \left\{ \frac{1}{2\pi} ( \partial_2 \Phi - \Phi^2 ) \right\} = 0 \,, \\
    -\frac{1}{2\pi} \partial_0 \Phi^1 + \partial_1 \left( \frac{1}{2\pi} \partial_0 \Phi \right) - \partial_0 \left\{ \frac{1}{2\pi} ( \partial_1 \Phi - \Phi^1 ) \right\} = 0 \,,
\end{gather}
which are locally trivial. Under the constraints \eqref{constphinoc1} and \eqref{constphinoc2}, these relations become a conservation law
\ali{
    \partial_0 \hat{J}^{12}_0 - \partial_1 \partial_2 \hat{J} = 0 \,, 
}
where the currents are
\ali{
    \hat{J}^{12}_0 &= \frac{1}{2\pi} \partial_1 \partial_2 \Phi \,, \\
    \hat{J} &=  \frac{1}{2\pi} \partial_0 \Phi \,.
}
These currents are equivalent to the currents \eqref{winsym1} and \eqref{winsym2} under the correspondence \eqref{corrphi}, and thus generate the winding dipole symmetry.

\section{Duality and Foliated \texorpdfstring{$\hat{\phi}$}{φhat}-Theory}
\label{section4}

The exotic $\phi$-theory in 2+1 dimensions has the momentum dipole symmetry and the winding dipole symmetry. While the momentum dipole symmetry acts on the scalar field $\phi$, the winding dipole symmetry acts on the dual field $\hat{\phi}^{12}$ in the $\hat\phi$-theory \cite{Seiberg:2020bhn,Spieler:2024fby}. In this section, we consider the duality between the $\phi$-theory and the $\hat{\phi}$-theory, which is a self-duality in the $\phi$-theory, and construct the foliated description of the $\hat{\phi}$-theory.

\subsection{Exotic \texorpdfstring{$\hat{\phi}$}{φhat}-Theory}
\label{section41}

In this section, we review the duality between the exotic $\phi$-theory and the exotic $\hat{\phi}$-theory\cite{Seiberg:2020bhn}, which is similar to the $T$-duality in the compact scalar field theory.

We consider the exotic Lagrangian including the background tensor gauge fields
\alis{
    \L_{\phi,\text{e}}\left[ \bm{C}, \bm{\hat{C}} \right] &= \frac{\mu_0}{2} ( E_0 )^2 + \frac{1}{2 \mu_{12}} ( B_{12} )^2  \\
    & \quad +\frac{i}{2\pi} \hat{E}^{12}_0(\partial_1 \partial_2 \phi - C_{12} - B_{12}) + \frac{i}{2\pi} \hat{B} (\partial_0 \phi - C_0 - E_0 )  \\
    & \quad + \frac{i}{2\pi} \hat{C}^{12}_0 ( \partial_1 \partial_2 \phi - C_{12}) + \frac{i}{2\pi} \hat{C} ( \partial_0 \phi - C_0 ) \,, \label{exoticphic-eb}
}
where $E_0$, $B_{12}$, $\hat{E}^{12}_0$ and $\hat{B}$ are dynamical tensor gauge fields in $\bm{1}_0$, $\bm{1}_2$, $\bm{1}_2$ and $\bm{1}_0$ respectively. Integrating out $\hat{E}^{12}_0$ and $\hat{B}$ leads to the equations
\ali{
    B_{12} &= \partial_1 \partial_2 \phi - C_{12} \,, \label{e12re} \\
    E_0 &= \partial_0 \phi - C_0 \,, \label{b0re}
}
and then, we recover the Lagrangian of the exotic $\phi$-theory \eqref{exoticphic}. On the other hand, integrating out $E_0$ and $B_{12}$ leads to the equations
\ali{
    E_0 &= \frac{i}{2\pi \mu_0} \hat{B} \,, \label{b0ehat} \\
    B_{12} &= \frac{i\mu_{12}}{2\pi} \hat{E}^{12}_0 \,, \label{e12bhat012}
}
so we can write the Lagrangian as
\alis{
    \L_{\phi,\text{e}}\left[ C_0, C_{12}, \hat{C}^{12}_0, \hat{C} \right] &= \frac{\mu_{12}}{8\pi^2} ( \hat{E}^{12}_0 )^2 + \frac{1}{8 \pi^2 \mu_0} ( \hat{B} )^2  \\
    & \quad +\frac{i}{2\pi} (\hat{E}^{12}_0 + \hat{C}^{12}_0 )(\partial_1 \partial_2 \phi - C_{12}) + \frac{i}{2\pi} (\hat{B} + \hat{C} ) (\partial_0 \phi - C_0)  \,. \label{exoticphic-eb2}
}
Next, we integrate out $\phi$ to have the equation
\ali{
    \partial_1 \partial_2 (\hat{E}^{12}_0 + \hat{C}^{12}_0) = \partial_0 (\hat{B} + \hat{C}) \,,
}
which we can solve locally in terms of a tensor gauge field $\hat{\phi}^{12}$ in $\bm{1}_2$:
\ali{
    \hat{E}^{12}_0 &= \partial_0  \hat{\phi}^{12} - \hat{C}^{12}_0 \,, \label{bhat120re} \\
    \hat{B} &= \partial_1 \partial_2 \hat{\phi}^{12} - \hat{C} \,. \label{ehatre}
}
The dual field $\hat{\phi}^{12}$ is a dynamical compact scalar field with the periodicity $\hat{\phi}^{12} \sim \hat{\phi}^{12} + 2\pi$, and we have gauge transformation of $\hat{\phi}^{12}$ as
\ali{
    \hat{\phi}^{12} \sim \hat{\phi}^{12}  + 2\pi \hat{w}^1 - 2\pi \hat{w}^2 \,.
}
$\hat{w}^1$ and $\hat{w}^2$ are integer-valued gauge parameters and they can have step function discontinuities in the $x^1$ and $x^2$ directions respectively. Then, we derive the Lagrangian of the exotic $\hat{\phi}$-theory including the background tensor gauge fields
\alis{
    \L_{\hat{\phi},\text{e}}\left[ \bm{C}, \bm{\hat{C}} \right] &= \frac{\hat{\mu}_0}{2} ( \partial_0  \hat{\phi}^{12} - \hat{C}^{12}_0 )^2 + \frac{1}{2 \hat{\mu}_{12}} ( \partial_1 \partial_2 \hat{\phi}^{12} - \hat{C} )^2  \\
    & \quad - \frac{i}{2\pi} C_{0} \partial_1 \partial_2 \hat{\phi}^{12} - \frac{i}{2\pi} C_{12} \partial_0  \hat{\phi}^{12}   \,. \label{exoticphihatc1}
}
where $\hat{\mu}_0 = \mu_{12}/(4\pi^2)$ and $\hat{\mu}_{12} = 4\pi^2 \mu_0$. We can obtain relations between $\phi$ and $\hat{\phi}^{12}$ from \eqref{e12re}, \eqref{b0re}, \eqref{b0ehat}, \eqref{e12bhat012}, \eqref{bhat120re} and \eqref{ehatre}:
\ali{
    \partial_0 \phi - C_0 &=  \frac{i}{2\pi \mu_0} (\partial_1 \partial_2 \hat{\phi}^{12} - \hat{C} ) \,, \label{phiphihatcre1} \\
    \partial_1 \partial_2 \phi - C_{12} &= \frac{i\mu_{12}}{2\pi} ( \partial_0 \hat{\phi}^{12} - \hat{C}_0^{12} ) \,. \label{phiphihatcre2}
}

We can consider the duality between the exotic $\phi$-theory and the exotic $\hat{\phi}$-theory as a self-duality. The exotic $\hat{\phi}$-theory is also a boundary theory of the exotic SSPT phase \eqref{exoticsspt} in the spacetime $\T^{2+1} \times \R_{x^3 \geq 0}$. Then, we add the terms
\ali{
    \frac{i}{2\pi} ( C_0 \hat{C} + C_{12} \hat{C}^{12}_0 ) \,, 
}
on the boundary and the terms
\ali{
    \partial_3 \left[ \frac{i}{2\pi} ( C_0 \hat{C} + C_{12} \hat{C}^{12}_0 )  \right] \,,
}
in the SSPT phase in 3+1 dimensions, the sum of which is zero. The boundary $\hat{\phi}$-theory becomes
\alis{
    \L'_{\hat{\phi},\text{e}}\left[ \bm{C}, \bm{\hat{C}} \right] &= \frac{\hat{\mu}_0}{2} ( \partial_0  \hat{\phi}^{12} - \hat{C}^{12}_0 )^2 + \frac{1}{2 \hat{\mu}_{12}} ( \partial_1 \partial_2 \hat{\phi}^{12} - \hat{C} )^2  \\
    & \quad - \frac{i}{2\pi} C_{0} ( \partial_1 \partial_2 \hat{\phi}^{12} - \hat{C} ) - \frac{i}{2\pi} C_{12} ( \partial_0  \hat{\phi}^{12} - \hat{C}^{12}_0 )   \,,
}
and the SSPT phase \eqref{exoticsspt} becomes
\alis{
    &\L'_{\text{SSPT,e}}\left[ \bm{C}, \bm{\hat{C}} \right] \\
    &\quad = \frac{i}{2\pi} \left[ - C_{12} ( \partial_0 \hat{C}^{12}_3 - \partial_3 \hat{C}^{12}_0 ) + C_0 (  \partial_3 \hat{C} - \partial_1 \partial_2 \hat{C}^{12}_3 ) - C_3 ( \partial_0 \hat{C} - \partial_1 \partial_2 \hat{C}^{12}_0 )  \right] \,. \label{exoticsspt2}
}
Therefore, considering transformations
\ali{
    \phi &\rightarrow \hat{\phi}^{12} \,, \\
    C_0 &\rightarrow \hat{C}^{12}_0 \,, \\
    C_{12} &\rightarrow \hat{C} \,, \\
    C_3 &\rightarrow \hat{C}^{12}_3 \,, \\
    \hat{C}^{12}_0 &\rightarrow -C_0 \,, \\
    \hat{C} &\rightarrow -C_{12} \,, \\
    \hat{C}^{12}_3 &\rightarrow -C_3 \,, 
}
and replacing the $\Z_4$ rotations with the $\Z_4$ rotations multiplied by the charge conjugation, under which all the fields are multiplied by $-1$, we can derive the original theory \eqref{exoticphic} and \eqref{exoticsspt}, when the parameters satisfy $\mu_0 = \hat{\mu_0}$ and $\mu_{12} = \hat{\mu}_{12}$, i.e., $4\pi^2 \mu_0 = \mu_{12}$ \cite{Seiberg:2020bhn,Spieler:2024fby}.

Let us consider the subsystem symmetries. We consider the exotic $\hat{\phi}$-theory without the background tensor gauge fields
\ali{
    \L_{\hat{\phi},\text{e}} &= \frac{\hat{\mu}_0}{2} ( \partial_0  \hat{\phi}^{12}  )^2 + \frac{1}{2 \hat{\mu}_{12}} ( \partial_1 \partial_2 \hat{\phi}^{12} )^2  \,.
}
Then, the relations \eqref{phiphihatcre1} and \eqref{phiphihatcre2} become
\ali{
    \partial_0 \phi  &=  \frac{i}{2\pi \mu_0} \partial_1 \partial_2 \hat{\phi}^{12}  \,, \label{phiphihatre1} \\
    \partial_1 \partial_2 \phi  &= \frac{i\mu_{12}}{2\pi}  \partial_0 \hat{\phi}^{12}  \,. \label{phiphihatre2}
}
Thus, the currents of the momentum dipole symmetry \eqref{momsym1} and \eqref{momsym2} in the exotic $\phi$-theory are
\ali{
    J_0 &= i\mu_0 \partial_0 \phi = -\frac{1}{2\pi} \partial_1 \partial_2 \hat{\phi}^{12} \,, \label{hatwinsym1} \\
    J_{12} &=  \frac{i}{\mu_{12}} \partial_1 \partial_2 \phi =  -\frac{1}{2\pi} \partial_0 \hat{\phi}^{12} \,, \label{hatwinsym2}
}
which are the currents of the winding dipole symmetry in the exotic $\hat{\phi}$-theory.\footnote{The minus signs are compatible with the coupling terms with $C_0$ and $C_{12}$ in \eqref{exoticphihatc1}.} On the other hand, the currents of the winding dipole symmetry \eqref{winsym1} and \eqref{winsym2} in the exotic $\phi$-theory are
\ali{
    \hat{J}^{12}_0 &= \frac{1}{2 \pi} \partial_1 \partial_2 \phi = i \hat{\mu}_0 \partial_0 \hat{\phi}^{12} \,, \label{hatmomsym1} \\
    \hat{J} &= \frac{1}{2 \pi} \partial_0 \phi = \frac{i}{\hat{\mu}_{12}} \partial_1 \partial_2 \hat{\phi}^{12} \,, \label{hatmomsym2}
}
which are the currents of the momentum dipole symmetry in the exotic $\hat{\phi}$-theory. Then, the charged operator is
\ali{
    \hat{V}_n [x] = e^{in\hat{\phi}^{12}} \,, 
}
For the field action, the momentum dipole symmetry acts on $\hat{\phi}^{12}$ as
\ali{
    \hat{\phi}^{12} &\rightarrow \hat{\phi}^{12} + \hat{\Lambda}^1(x^1) - \hat{\Lambda}^2(x^2) \,,
}
where $\Lambda^k(x^k)$ is a $2\pi$-periodic scalar that can have step function discontinuities in the $x^k$ direction.

\subsection{Construction of Foliated \texorpdfstring{$\hat{\phi}$}{φhat}-Theory}
\label{section42}

As in the case of Section \ref{section32}, we can construct a foliated $\hat{\phi}$-theory in 2+1 dimensions by using the method in Section \ref{section31}. We take the foliated SSPT phase \eqref{folissptlag}, and choose the gauge fields $\hat{C}^k$ and $\hat{c}$. Then, we have to consider the Lagrangian
\alis{
    &L_{\text{SSPT,f}} \left[ C^k \wedge dx^k, c, \hat{C}^k, \hat{c}_{12}\ dx^1 \wedge dx^2 \right] \\
    & \quad = \frac{i}{2\pi} \left[  \sum_{k = 1}^2 \hat{C}^k \wedge d C^k \wedge dx^k + \hat{c} \wedge \left( \sum_{k=1}^2 C^k \wedge dx^k + dc \right) \right]  \,, 
}
on the spacetime $\T^{2+1} \times \R_{x^3 \geq 0}$. We introduce a $U(1)$ type-$B$ foliated zero-form gauge field $\hat{\Phi}^k\ (k = 1,2)$ and a $U(1)$ type-$B$ bulk one-form gauge field $\hat{\Phi}$, whose dynamical gauge transformations are
\ali{
    \hat{\Phi}^k  &\sim \hat{\Phi}^k + 2\pi \hat{W}^k + \hat{\xi} \,, \\
    \hat{\Phi} &\sim \hat{\Phi} + d \hat{\xi} \,,
}
where $\hat{W}^k$ is an integer-valued gauge parameter and it can have step function discontinuities in the $x^k$ direction, and $\hat{\xi}$ is a type-$B$ bulk zero-form gauge parameter. We set the background gauge transformations of $\hat{\Phi}^k$ and $\hat{\Phi}$ as
\ali{
    \hat{\Phi}^k  &\sim \hat{\Phi}^k + \hat{\lambda}^k \,, \label{bkggaugephihatk} \\
    \hat{\Phi} &\sim \hat{\Phi} + \hat{\kappa} \,,
}
and substitute the form $(\hat{C}^k - d\hat{\Phi}^k + \hat{\Phi})\wedge dx^k$ and $\hat{c} - d\hat{\Phi}$, which are the gauge-invariant combinations from \eqref{bkggaugetr1}--\eqref{bkggaugetr4}, for $\hat{C}^k \wedge dx^k$ and $\hat{c}$ in $L_{\text{SSPT,f}}$ respectively. Then, we have
\alis{
    &L'''_{\text{SSPT,f}}  \\
    & \quad = \frac{i}{2\pi} \left[  \sum_{k = 1}^2 (\hat{C}^k - d\hat{\Phi}^k + \hat{\Phi}) \wedge d C^k \wedge dx^k + (\hat{c} - d\hat{\Phi}) \wedge \left( \sum_{k=1}^2 C^k \wedge dx^k + dc \right) \right] \\
    & \quad = \frac{i}{2\pi} \left[  \sum_{k = 1}^2  \hat{C}^k \wedge d C^k \wedge dx^k + \hat{c} \wedge \left(  \sum_{k = 1}^2 C^k \wedge dx^k + dc \right)  \right] \\
    & \qquad + d \left[  \frac{i}{2\pi} \sum_{k = 1}^2 ( d\hat{\Phi}^k - \hat{\Phi}) \wedge  C^k \wedge dx^k - \frac{i}{2\pi} d \hat{\Phi} \wedge c  \right]\,, 
}
The boundary term in the action is
\alis{
    &\int_{x^3 = 0} \frac{i}{2\pi} \left[  \sum_{k = 1}^2 (d\hat{\Phi}^k - \hat{\Phi}) \wedge  C^k \wedge dx^k - d \hat{\Phi} \wedge c  \right] \\
    & = \int_{x^3 = 0} \frac{i}{2\pi} \left[ C^1_0 ( \partial_2 \hat{\Phi}^1 - \hat{\Phi}_2 ) - C^2_0 ( \partial_1 \hat{\Phi}^2 - \hat{\Phi}_1 ) - C^1_2 ( \partial_0 \hat{\Phi}^1 - \hat{\Phi}_0 ) + C^2_1 ( \partial_0 \hat{\Phi}^2 - \hat{\Phi}_0 )       \right. \\
    & \qquad \qquad \left. - c_0 ( \partial_1 \hat{\Phi}_2 - \partial_2 \hat{\Phi}_1 )  + c_1 ( \partial_0 \hat{\Phi}_2 - \partial_2 \hat{\Phi}_0 ) - c_2 ( \partial_0 \hat{\Phi}_1 - \partial_1 \hat{\Phi}_0 )    \right] d^3 x 
}
As in the case of Section \ref{section32}, we add the term 
\alis{
    & \frac{i}{2\pi} \hat{\chi} \wedge \left( \sum^{2}_{k=1} C^k \wedge dx^k + dc \right) \\
    &= \frac{i}{2\pi} \left[ \hat{\chi}_0 ( C^2_1 - C^1_2 + \partial_1 c_2 - \partial_2 c_1 ) \right. \\
    & \quad \left. - \hat{\chi}_1 ( C^2_0 + \partial_0 c_2 - \partial_2 c_0) + \hat{\chi}_2 ( C^1_0 + \partial_0 c_1 - \partial_1 c_0 ) \right] d^3 x \,,
}
on the boundary, where $\hat{\chi}$ is a dynamical one-form field. Finally, we add background gauge-invariant and $\Z_4$ rotational invariant quadratic terms, and we have
\alis{
    &\L_{\hat{\phi},\text{f}} \left[ C^k \wedge dx^k, c, \hat{C}^k, \hat{c}_{12}\ dx^1 \wedge dx^2 \right] \\
    & \quad = \hat{\mu}_0 ( \partial_0 \hat{\Phi}^1 - \hat{\Phi}_0 -\hat{C}^1_0 )^2 + \hat{\mu}_0 ( \partial_0 \hat{\Phi}^2 - \hat{\Phi}_0 -\hat{C}^2_0 )^2 \\ 
    & \qquad +  \frac{\hat{\mu}^{12}}{2}( \partial_2 \hat{\Phi}^1 - \hat{\Phi}_2 - \hat{C}^1_2 )^2 +   \frac{\hat{\mu}^{12}}{2}( \partial_1 \hat{\Phi}^2 - \hat{\Phi}_1 -\hat{C}^2_1 )^2 \\
    & \qquad + \frac{1}{2\hat{\mu}_{12}}( \partial_1 \hat{\Phi}_2 - \partial_2 \hat{\Phi}_1 - \hat{c}_{12} )^2 \\
    & \qquad + \frac{1}{2\hat{\mu}_{012}}( \partial_0 \hat{\Phi}_1 - \partial_1 \hat{\Phi}_0 - \hat{c}_{01} )^2 + \frac{1}{2\hat{\mu}_{012}}( \partial_0 \hat{\Phi}_2 - \partial_2 \hat{\Phi}_0 - \hat{c}_{02} )^2 \\
    & \qquad + \frac{i}{2\pi} \left[ C^1_0 ( \partial_2 \hat{\Phi}^1 - \hat{\Phi}_2 ) - C^2_0 ( \partial_1 \hat{\Phi}^2 - \hat{\Phi}_1 ) \right. \\
    & \qquad  \qquad - C^1_2 ( \partial_0 \hat{\Phi}^1 - \hat{\Phi}_0 )  + C^2_1 ( \partial_0 \hat{\Phi}^2 - \hat{\Phi}_0 )  \\
    & \qquad \qquad \left. - c_0 ( \partial_1 \hat{\Phi}_2 - \partial_2 \hat{\Phi}_1 )  + c_1 ( \partial_0 \hat{\Phi}_2 - \partial_2 \hat{\Phi}_0 ) - c_2 ( \partial_0 \hat{\Phi}_1 - \partial_1 \hat{\Phi}_0 )    \right] \\
    & \qquad + \frac{i}{2\pi} \left[ \hat{\chi}_0 ( C^2_1 - C^1_2 + \partial_1 c_2 - \partial_2 c_1 ) \right. \\
    & \qquad \quad \left. - \hat{\chi}_1 ( C^2_0 + \partial_0 c_2 - \partial_2 c_0) + \hat{\chi}_2 ( C^1_0 + \partial_0 c_1 - \partial_1 c_0 ) \right] \,, \label{genefoliatedphihat1}
}
where $\hat{\mu}_0$, $\hat{\mu}^{12}$ and $\hat{\mu}_{12}$  are parameters with mass dimension one.

We can also write the foliated Lagrangian in differential form by tuning the metric:
\alis{
    &L_{\hat{\phi},\text{f}} \left[ C^k \wedge dx^k, c, \hat{C}^k, \hat{c}_{12}\ dx^1 \wedge dx^2 \right] \\
    & \quad = \frac{1}{2} \sum_{k=1}^2 \left\{ \left(  d \hat{\Phi}^k - \hat{\Phi} - \hat{C}^k \right) \wedge dx^k \right\} \wedge \ast \left\{ \left( d \hat{\Phi}^k - \hat{\Phi} - \hat{C}^k \right) \wedge dx^k \right\} \\
    & \qquad + \frac{1}{2} \left( d \hat{\Phi} -\hat{c}  \right) \wedge \ast \left( d \hat{\Phi} -\hat{c}  \right)  \\
    & \qquad + \frac{i}{2\pi} \left[  \sum_{k = 1}^2  \left( d\hat{\Phi}^k - \hat{\Phi} \right)  \wedge C^k \wedge dx^k -  d\hat{\Phi} \wedge c   \right]  \\
    & \qquad + \frac{i}{2\pi} \hat{\chi} \wedge \left( \sum^{2}_{k=1} C^k \wedge dx^k + dc \right) \,. \label{foliatedphihatdeff}
}

We can integrate out the dynamical gauge fields $\hat{c}_{01}$ and $\hat{c}_{02}$ on the boundary, and then we have
\ali{
    \hat{c}_{01} &=  \partial_0 \hat{\Phi}_1 - \partial_1 \hat{\Phi}_0 \,, \\
    \hat{c}_{02} &= \partial_0 \hat{\Phi}_2 - \partial_2 \hat{\Phi}_0 \,.
}
Furthermore, we consider a change of the dynamical variable
\ali{
    \hat{\chi} \rightarrow \hat{\chi} + \hat{\Phi} \,.
}
Then, the field $\hat{\chi}$ has a dynamical gauge transformation
\ali{
    \hat{\chi} \sim \hat{\chi} - d\hat{\xi} \,,
}
and a background gauge transformation
\ali{
    \hat{\chi} \sim \hat{\chi} - \hat{\kappa} \,.
}
Finally, we have the foliated Lagrangian
\alis{
    &\L_{\hat{\phi},\text{f}} \left[ C^k \wedge dx^k, c, \hat{C}^k, \hat{c}_{12}\ dx^1 \wedge dx^2 \right] \\
    & \quad = \hat{\mu}_0 ( \partial_0 \hat{\Phi}^1 - \hat{\Phi}_0 -\hat{C}^1_0 )^2 + \hat{\mu}_0 ( \partial_0 \hat{\Phi}^2 - \hat{\Phi}_0 -\hat{C}^2_0 )^2 \\ 
    & \qquad +  \frac{\hat{\mu}^{12}}{2}( \partial_2 \hat{\Phi}^1 - \hat{\Phi}_2 -\hat{C}^1_2 )^2 +   \frac{\hat{\mu}^{12}}{2}( \partial_1 \hat{\Phi}^2 - \hat{\Phi}_1 -\hat{C}^2_1 )^2  \\
    & \qquad + \frac{1}{2\hat{\mu}_{12}}( \partial_1 \hat{\Phi}_2 - \partial_2 \hat{\Phi}_1 - \hat{c}_{12} )^2 \\
    & \qquad + \frac{i}{2\pi} \left( C^1_0  \partial_2 \hat{\Phi}^1  - C^2_0  \partial_1 \hat{\Phi}^2 - C^1_2 \partial_0 \hat{\Phi}^1  + C^2_1  \partial_0 \hat{\Phi}^2 \right)  \\
    & \qquad + \frac{i}{2\pi} \left[ \hat{\chi}_0 ( C^2_1 - C^1_2 + \partial_1 c_2 - \partial_2 c_1 ) \right. \\
    & \qquad \quad \left. - \hat{\chi}_1 ( C^2_0 + \partial_0 c_2 - \partial_2 c_0) + \hat{\chi}_2 ( C^1_0 + \partial_0 c_1 - \partial_1 c_0 ) \right] \,, \label{genefoliatedphihat2}
}

\subsection{Foliated-Exotic Duality in \texorpdfstring{$\hat{\phi}$}{φhat}-Theory}
\label{section43}

In this section, we establish the field correspondences in the $\hat{\phi}$-theory and obtain the foliated $\hat{\phi}$-theory, which is equivalent to the exotic $\hat{\phi}$-theory \eqref{exoticphihatc1}. We can also derive the foliated $\hat{\phi}$-theory by tuning the parameters in the general foliated $\hat{\phi}$-theory \eqref{genefoliatedphihat2}.

In section \ref{section41}, we have derived the exotic $\hat{\phi}$-theory coupled with the background tensor gauge fields
\alis{
    \L_{\hat{\phi},\text{e}}\left[ \bm{C}, \bm{\hat{C}}  \right] &= \frac{\hat{\mu}_0}{2} ( \partial_0  \hat{\phi}^{12} - \hat{C}^{12}_0 )^2 + \frac{1}{2 \hat{\mu}_{12}} ( \partial_1 \partial_2 \hat{\phi}^{12} - \hat{C} )^2  \\
    & \quad - \frac{i}{2\pi} C_{0} \partial_1 \partial_2 \hat{\phi}^{12} - \frac{i}{2\pi} C_{12} \partial_0  \hat{\phi}^{12}   \,. \label{exoticphihatc2} 
}

First, we assume correspondences between the tensor gauge field $\hat{\phi}^{12}$ in the exotic $\hat{\phi}$-theory and the type-$B$ foliated zero-form gauge fields $\hat{\Phi}^k\ (k=1,2)$ in a foliated $\hat{\phi}$-theory as
\ali{
    \hat{\phi}^{12} \simeq \hat{\Phi}^1 - \hat{\Phi}^2 \,. \label{corrphihat}
}
Since the dynamical gauge transformations of $\hat{\phi}^{12}$ and $\hat{\Phi}^k$ are
\ali{
    \hat{\phi}^{12} &\sim \hat{\phi}^{12} + 2 \pi \hat{w}^1 - 2 \pi \hat{w}^2 \,, \\
    \hat{\Phi}^k &\sim \hat{\Phi}^k + 2 \pi \hat{W}^k + \hat{\xi} \,,
}
the correspondences between the dynamical gauge parameters are
\ali{
    \hat{w}^k \simeq \hat{W}^k \,, \quad (k = 1,2) \,. 
}
The background gauge transformation of $\hat{\Phi}^k$ is \eqref{bkggaugephihatk}.

As in the case of Section \ref{section33}, the background gauge fields in a foliated theory are $U(1)$ type-$A$ foliated (1+1)-form gauge fields $C^k \wedge dx^k \ (k = 1,2)$, a $U(1)$ type-$A$ bulk one-form gauge field $c$, $U(1)$ type-$B$ foliated one-form gauge fields $\hat{C}^k \ (k = 1,2)$ and the $x^1x^2$-component of a $U(1)$ type-$B$ bulk two form gauge field $\hat{c}_{12}$. We assume correspondences
\ali{
    C_0 &\simeq c_0 \label{bcorr0-0,2} \,, \\
    C_{12} &\simeq C^1_2 + \partial_2 c_1 \,,  \label{bcorr12-1,2} 
}
and
\ali{
    \hat{C}_0^{12} &\simeq \hat{C}_0^1 - \hat{C}_0^2 \,, \label{bcorr012,2} \\
    \hat{C} &\simeq \partial_1 \hat{C}_2^1 - \partial_2 \hat{C}_1^2 + \hat{c}_{12}  \,. \label{bcorr1221,2}
}
Their background gauge transformations are \eqref{bkggaugetr1}--\eqref{bkggaugetr4}. We also add to the Lagrangian the term
\alis{
    & \frac{i}{2\pi} \hat{\chi} \wedge \left( \sum^{2}_{k=1} C^k \wedge dx^k + dc \right) \\
    &= \frac{i}{2\pi} \left[ \hat{\chi}_0 ( C^2_1 - C^1_2 + \partial_1 c_2 - \partial_2 c_1 ) \right. \\
    & \quad \left. - \hat{\chi}_1 ( C^2_0 + \partial_0 c_2 - \partial_2 c_0) + \hat{\chi}_2 ( C^1_0 + \partial_0 c_1 - \partial_1 c_0 ) \right] d^3 x \,.
}

Using these correspondences, the Lagrangian of the $\hat{\phi}$-theory can be written as
\alis{
    \L_{\hat{\phi},\text{e}} &\simeq \frac{\hat{\mu}_0}{2} \left[ ( \partial_0 \hat{\Phi}^1 - \hat{C}^1_0 ) - ( \partial_0 \hat{\Phi}^2 - \hat{C}^2_0 ) \right]^2 \\
    & \quad + \frac{1}{2\hat{\mu}_{12}} \left[ \partial_1 ( \partial_2 \hat{\Phi}^1 - \hat{C}^1_2 ) - \partial_2 ( \partial_1 \hat{\Phi}^2 - \hat{C}^2_1 ) - \hat{c}_{12} \right]^2 \\
    & \quad + \frac{i}{2\pi} ( \partial_1 c_0 \partial_2 \hat{\Phi}^1 - \partial_2 c_0 \partial_1 \hat{\Phi}^2 )  - \frac{i}{2\pi} ( C^1_2 + \partial_2 c_1 ) \partial_0 ( \hat{\Phi}^1 - \hat{\Phi}^2 ) \\
    & \quad + \frac{i}{2\pi} \left[ \hat{\chi}_0 ( C^2_1 - C^1_2 + \partial_1 c_2 - \partial_2 c_1 ) \right. \\
    & \qquad \left. - \hat{\chi}_1 ( C^2_0 + \partial_0 c_2 - \partial_2 c_0) + \hat{\chi}_2 ( C^1_0 + \partial_0 c_1 - \partial_1 c_0 ) \right] \,,
}
where we have integrated by parts and dropped the total derivative terms. As in the case of Section \ref{section33}, we have to add the boundary term of \eqref{ssptaddterm}:
\ali{
     \frac{i}{2\pi} \left[ \hat{C}^2_0 ( C^2_1 - C^1_2 + \partial_1 c_2 - \partial_2 c_1 ) - \hat{C}^2_1 ( C^2_0 + \partial_0 c_2 - \partial_2 c_0 ) + \hat{C}^1_2 ( C^1_0 + \partial_0 c_1 - \partial_1 c_0 )   \right] \,.
}

Next, we consider changes of the dynamical variables
\ali{
    \hat{\chi}_0 &\rightarrow \hat{\chi}_0 + \partial_0 \hat{\Phi}^2 - \hat{C}^2_0 \,, \\
    \hat{\chi}_1 &\rightarrow \hat{\chi}_1 + \partial_1 \hat{\Phi}^2 + \hat{C}^2_1 \,, \\
    \hat{\chi}_2 &\rightarrow \hat{\chi}_2 + \partial_2 \hat{\Phi}^1 - \hat{C}^1_2 \,, 
}
and then, the field $\hat{\chi}$ has a dynamical gauge transformation
\ali{
    \hat{\chi} \sim \hat{\chi} - d\hat{\xi} \,,
}
and a background gauge transformation
\ali{
    \hat{\chi} \sim \hat{\chi} - \hat{\kappa} \,,
}
as in the previous section. The result is
\alis{
    & \frac{i}{2\pi} ( \partial_1 c_0 \partial_2 \hat{\Phi}^1 - \partial_2 c_0 \partial_1 \hat{\Phi}^2 )  - \frac{i}{2\pi} ( C^1_2 + \partial_2 c_1 ) \partial_0 ( \hat{\Phi}^1 - \hat{\Phi}^2 ) \\
    & \quad + \frac{i}{2\pi} \left[ \hat{C}^2_0 ( C^2_1 - C^1_2 +  \partial_1 c_2 - \partial_2 c_1 ) - \hat{C}^2_1 ( C^2_0 + \partial_0 c_2 - \partial_2 c_0 ) + \hat{C}^1_2 ( C^1_0 + \partial_0 c_1 - \partial_1 c_0 )   \right] \\
    & \rightarrow \frac{i}{2\pi} \left[  ( C^1_0 + \partial_0 c_1 ) \partial_2 \hat{\Phi}^1 - ( C^2_0 + \partial_0 c_2 ) \partial_1 \hat{\Phi}^2 - ( C^1_2 + \partial_2 c_1 ) \partial_0 \hat{\Phi}^1 + ( C^2_1 + \partial_1 c_2 ) \partial_0 \hat{\Phi}^2 \right] \\
    &\quad =  \frac{i}{2\pi} \left(  C^1_0  \partial_2 \hat{\Phi}^1 -  C^2_0  \partial_1 \hat{\Phi}^2 -  C^1_2  \partial_0 \hat{\Phi}^1 +  C^2_1  \partial_0 \hat{\Phi}^2 \right) \,,
}
where we have integrated by parts and dropped the total derivative terms. Finally, we obtain the foliated $\hat{\phi}$-theory, which is equivalent to the exotic $\hat{\phi}$-theory \eqref{exoticphihatc2}:
\alis{
    &\L_{\hat{\phi},\text{e} \rightarrow \text{f}} \left[ C^k \wedge dx^k, c, \hat{C}^k, \hat{c}_{12}\ dx^1 \wedge dx^2 \right] \\
    & = \frac{\hat{\mu}_0}{2} \left[ ( \partial_0 \hat{\Phi}^1 - \hat{C}^1_0 ) - ( \partial_0 \hat{\Phi}^2 - \hat{C}^2_0 ) \right]^2 \\
    & \quad + \frac{1}{2\hat{\mu}_{12}} \left[ \partial_1 ( \partial_2 \hat{\Phi}^1 - \hat{C}^1_2 ) - \partial_2 ( \partial_1 \hat{\Phi}^2 - \hat{C}^2_1 ) - \hat{c}_{12} \right]^2 \\
    & \quad + \frac{i}{2\pi} \left(  C^1_0  \partial_2 \hat{\Phi}^1 -  C^2_0  \partial_1 \hat{\Phi}^2 -  C^1_2  \partial_0 \hat{\Phi}^1 +  C^2_1  \partial_0 \hat{\Phi}^2 \right) \\
    & \quad + \frac{i}{2\pi} \left[ \hat{\chi}_0 ( C^2_1 - C^1_2 + \partial_1 c_2 - \partial_2 c_1 ) \right. \\
    & \qquad \left. - \hat{\chi}_1 ( C^2_0 + \partial_0 c_2 - \partial_2 c_0) + \hat{\chi}_2 ( C^1_0 + \partial_0 c_1 - \partial_1 c_0 ) \right] \,. \label{foliphihatc}
}

We can also derive the foliated description \eqref{foliphihatc} by integrating out the dynamical fields $\hat{\chi}_i\ (i = 0,1,2)$ as Lagrange multipliers. We have the constraints \eqref{constc12}--\eqref{constc02}, and we can use a dictionary
\ali{
    \hat{\phi}^{12} &\simeq \hat{\Phi}^1 - \hat{\Phi}^2 \,, \label{phihatdic1} \\ 
    C_0 &\simeq c_0 \,, \\
    C_{12} &\simeq C^1_2 + \partial_2 c_1 =  C^2_1 + \partial_1 c_2 \,, \\
    \partial_1 C_0 &\simeq C^1_0 + \partial_0 c_1 \,, \\
    \partial_2 C_0 &\simeq C^2_0 + \partial_0 c_2 \,, \\
    \hat{C}_0^{12} &\simeq \hat{C}_0^1 - \hat{C}_0^2 \,,  \\
    \hat{C} &\simeq \partial_1 \hat{C}_2^1 - \partial_2 \hat{C}_1^2 + \hat{c}_{12} \,, \label{phihatdic7}
}
instead of the correspondences \eqref{corrphihat}, \eqref{bcorr0-0,2} and \eqref{bcorr12-1,2}. Using these correspondences and recovering the terms of $\hat{\chi}_i \ (i = 0,1,2)$, we can derive the foliated Lagrangian \eqref{foliphihatc} again.

Conversely, to obtain the exotic $\hat{\phi}$-theory \eqref{exoticphihatc2} from the foliated $\phi$-theory \eqref{foliphihatc}, we can integrate out the dynamical fields $\hat{\chi}_i \ (i = 0,1,2)$ and use the dictionary \eqref{phihatdic1}--\eqref{phihatdic7}.

We can also derive the foliated Lagrangian \eqref{foliphihatc} by tuning the parameters in the general foliated Lagrangian \eqref{genefoliatedphihat2}. First, the terms
\ali{
    \frac{\hat{\mu}^{12}}{2}( \partial_2 \hat{\Phi}^1 - \hat{\Phi}_2 -\hat{C}^1_2 )^2 +   \frac{\hat{\mu}^{12}}{2}( \partial_1 \hat{\Phi}^2 - \hat{\Phi}_1 -\hat{C}^2_1 )^2 \label{term12}
}
can be written as
\ali{
     \frac{1}{8\pi^2 \hat{\mu}^{12}} (h_{01})^2 + \frac{i}{2\pi} h_{01} ( \partial_2 \hat{\Phi}^1 - \hat{\Phi}_2 -\hat{C}^1_2 ) + \frac{1}{8\pi^2 \hat{\mu}^{12}} (h_{02})^2 - \frac{i}{2\pi} h_{02} ( \partial_1 \hat{\Phi}^2 - \hat{\Phi}_1 -\hat{C}^2_1 ) 
}
by using dynamical fields $h_{01}$ and $h_{02}$. Integrating out $h_{01}$ and $h_{02}$, we have the equations of motion
\ali{
     h_{01}  &=  2\pi i \hat{\mu}^{12} (\partial_2 \hat{\Phi}^1 - \hat{\Phi}_2 -\hat{C}^1_2 ) \,, \\
     h_{02}  &=  - 2\pi i \hat{\mu}^{12} (\partial_1 \hat{\Phi}^2 - \hat{\Phi}_1 -\hat{C}^2_1 ) \,.
}
Using them, we can recover the terms \eqref{term12}. Then, let us take the limit $\hat{\mu}^{12} \rightarrow \infty$, which leads to the terms
\ali{
    \frac{i}{2\pi} h_{01} ( \partial_2 \hat{\Phi}^1 - \hat{\Phi}_2 -\hat{C}^1_2 ) - \frac{i}{2\pi} h_{02} ( \partial_1 \hat{\Phi}^2 - \hat{\Phi}_1 -\hat{C}^2_1 )  \,.
}
Then, the foliated Lagrangian becomes
\alis{
    &\L_{\hat{\phi},\text{f}} \left[ C^k \wedge dx^k, c, \hat{C}^k, \hat{c}_{12}\ dx^1 \wedge dx^2 \right] \\
    & \quad = \hat{\mu}_0 ( \partial_0 \hat{\Phi}^1 - \hat{\Phi}_0 -\hat{C}^1_0 )^2 + \hat{\mu}_0 ( \partial_0 \hat{\Phi}^2 - \hat{\Phi}_0 -\hat{C}^2_0 )^2 \\ 
    & \qquad + \frac{1}{2\hat{\mu}_{12}}( \partial_1 \hat{\Phi}_2 - \partial_2 \hat{\Phi}_1 - \hat{c}_{12} )^2 \\
    & \qquad + \frac{i}{2\pi} \left( C^1_0  \partial_2 \hat{\Phi}^1  - C^2_0  \partial_1 \hat{\Phi}^2 - C^1_2 \partial_0 \hat{\Phi}^1  + C^2_1  \partial_0 \hat{\Phi}^2 \right)  \\
    & \qquad + \frac{i}{2\pi} \left[ h_{01} ( \partial_2 \hat{\Phi}^1 - \hat{\Phi}_2 -\hat{C}^1_2 ) - h_{02} ( \partial_1 \hat{\Phi}^2 - \hat{\Phi}_1 -\hat{C}^2_1 ) \right] \\
    & \qquad + \frac{i}{2\pi} \left[ \hat{\chi}_0 ( C^2_1 - C^1_2 + \partial_1 c_2 - \partial_2 c_1 ) \right. \\
    & \qquad \quad \left. - \hat{\chi}_1 ( C^2_0 + \partial_0 c_2 - \partial_2 c_0) + \hat{\chi}_2 ( C^1_0 + \partial_0 c_1 - \partial_1 c_0 ) \right] \,, \label{genefoliatedphihat3}
}
In this case, integrating out $h_{01}$ and $h_{02}$, we have the constraints
\ali{
    \partial_2 \hat{\Phi}^1 - \hat{\Phi}_2 -\hat{C}^1_2 &= 0 \,, \label{hatconst1} \\
    \partial_1 \hat{\Phi}^2 - \hat{\Phi}_1 -\hat{C}^2_1 &= 0 \,. \label{hatconst2}
}
Using them, we can derive
\ali{
    \frac{1}{2\hat{\mu}_{12}}( \partial_1 \hat{\Phi}_2 - \partial_2 \hat{\Phi}_1 - \hat{c}_{12} )^2 = \frac{1}{2\hat{\mu}_{12}}\left[ \partial_1 ( \partial_2 \hat{\Phi}^1 -\hat{C}^1_2 )  - \partial_2 ( \partial_1 \hat{\Phi}^2 -\hat{C}^2_1 ) - \hat{c}_{12} \right]^2 \,.
}
Next, integrating out $\hat{\Phi}_0$, we have
\ali{
    \hat{\Phi}_0 =  \frac{1}{2} \left( \partial_0 \hat{\Phi}^1 - \hat{C}^1_0 + \partial_0 \hat{\Phi}^2 - \hat{C}^2_0 \right) \,. \label{phihat0formula}
}
Then we can write
\ali{
    \hat{\mu}_0 ( \partial_0 \hat{\Phi}^1 - \hat{\Phi}_0 -\hat{C}^1_0 )^2 + \hat{\mu}_0 ( \partial_0 \hat{\Phi}^2 - \hat{\Phi}_0 -\hat{C}^2_0 )^2 = \frac{\hat{\mu}_0}{2} \left[ ( \partial_0 \hat{\Phi}^1 -\hat{C}^1_0 ) - ( \partial_0 \hat{\Phi}^2 -\hat{C}^2_0  ) \right]^2 \,,
}
and we can obtain the foliated Lagrangian \eqref{foliphihatc}, which is equivalent to the exotic $\hat{\phi}$-theory \eqref{exoticphihatc2}.\footnote{We can change the term $\hat{\mu}_0 ( \partial_0 \hat{\Phi}^1 - \hat{\Phi}_0 -\hat{C}^1_0 )^2 + \hat{\mu}_0 ( \partial_0 \hat{\Phi}^2 - \hat{\Phi}_0 -\hat{C}^2_0 )^2$ to $\frac{\hat{\mu}^1_0}{2} ( \partial_0 \hat{\Phi}^1 - \hat{\Phi}_0 -\hat{C}^1_0 )^2 + \frac{\hat{\mu}^2_0}{2} ( \partial_0 \hat{\Phi}^2 - \hat{\Phi}_0 -\hat{C}^2_0 )^2$, where $\hat{\mu}^1_0$ and $\hat{\mu}^2_0$ are parameters satisfying $1/\hat{\mu}_{0} = 1/\hat{\mu}^1_0 + 1/\hat{\mu}^2_0$. In this case, we have
\alis{
\frac{\hat{\mu}^1_0}{2} ( \partial_0 \hat{\Phi}^1 - \hat{\Phi}_0 -\hat{C}^1_0 )^2 + \frac{\hat{\mu}^2_0}{2} ( \partial_0 \hat{\Phi}^2 - \hat{\Phi}_0 -\hat{C}^2_0 )^2 &= \frac{\hat{\mu}^1_0 \hat{\mu}^2_0}{2 ( \hat{\mu}^1_0 + \hat{\mu}^2_0) } \left[ ( \partial_0 \hat{\Phi}^1 -\hat{C}^1_0 ) - ( \partial_0 \hat{\Phi}^2 -\hat{C}^2_0  ) \right]^2 \\
 &= \frac{\hat{\mu}_0}{2} \left[ ( \partial_0 \hat{\Phi}^1 -\hat{C}^1_0 ) - ( \partial_0 \hat{\Phi}^2 -\hat{C}^2_0  ) \right]^2 \,.
}
This term becomes $\Z_4$ rotational invariant and the parameters $\hat{\mu}^1_0$ and $\hat{\mu}^2_0$ appear only in the form $1/\hat{\mu}_{0} = 1/\hat{\mu}^1_0 + 1/\hat{\mu}^2_0$.
}

If we set the background gauge fields to zero, we can have the foliated $\hat{\phi}$-theory without the background gauge fields
\alis{
    \L_{\hat{\phi},\text{e} \rightarrow \text{f}} &= \hat{\mu}_0 ( \partial_0 \hat{\Phi}^1 - \hat{\Phi}_0 )^2 + \hat{\mu}_0 ( \partial_0 \hat{\Phi}^2 - \hat{\Phi}_0 )^2  + \frac{1}{2\hat{\mu}_{12}}( \partial_1 \hat{\Phi}_2 - \partial_2 \hat{\Phi}_1 )^2 \\
    & + \frac{i}{2\pi} \left[ h_{01} ( \partial_2 \hat{\Phi}^1 - \hat{\Phi}_2 ) - h_{02} ( \partial_1 \hat{\Phi}^2 - \hat{\Phi}_1 ) \right] \,, \label{foliatedphihatczero}
}
or
\alis{
    \L_{\hat{\phi},\text{e} \rightarrow \text{f}} &=
    \frac{\hat{\mu}_0}{2} \left( \partial_0 \hat{\Phi}^1 -  \partial_0 \hat{\Phi}^2  \right)^2 + \frac{1}{2\hat{\mu}_{12}} \left( \partial_1  \partial_2 \hat{\Phi}^1  - \partial_2  \partial_1 \hat{\Phi}^2 \right)^2 \,. 
}

As in the case of the foliated $\phi$-theory, let us consider subsystem symmetry. From the background gauge transformations of $\hat{C}^k$ and $\hat{c}$, which are \eqref{bkggaugetr2} and \eqref{bkggaugetr4}, in the Lagrangian with the background gauge fields \eqref{genefoliatedphihat3}, we have
\begin{gather}
    2 \partial_0 \left\{ i \hat{\mu}_0 ( \partial_0 \hat{\Phi}^1 - \hat{\Phi}_0 ) \right\} - \partial_2 \left( \frac{1}{2\pi} h_{01} \right) = 0 \,, \\
    2 \partial_0 \left\{ i \hat{\mu}_0 ( \partial_0 \hat{\Phi}^2 - \hat{\Phi}_0 ) \right\} + \partial_1 \left( \frac{1}{2\pi} h_{02} \right) = 0 \,, \\
    \hat{\mu}_0 ( \partial_0 \hat{\Phi}^1 - \hat{\Phi}_0  ) + \hat{\mu}_0 ( \partial_0 \hat{\Phi}^2 - \hat{\Phi}_0  ) = 0 \,, \\
    \frac{1}{2\pi} h_{02} - \partial_2 \left\{ \frac{i}{\hat{\mu}_{12}} ( \partial_1 \hat{\Phi}_2 - \partial_2 \hat{\Phi}_1 ) \right\} = 0 \,, \\
    -\frac{1}{2\pi} h_{01} + \partial_1 \left\{ \frac{i}{\hat{\mu}_{12}} ( \partial_1 \hat{\Phi}_2 - \partial_2 \hat{\Phi}_1 ) \right\} = 0 \,.
\end{gather}
These equations can also be obtained as the equations of motion of $\hat{\Phi}^1$, $\hat{\Phi}^2$, $\hat{\Phi}_0$ $\hat{\Phi}_1$ and $\hat{\Phi}_2$ in the theory where the background gauge fields are zero \eqref{foliatedphihatczero}. Combining these equations, we derive a conservation law
\ali{
    \partial_0 \hat{J}^{12}_0 - \partial_1 \partial_2 \hat{J}  = 0 \,,
}
where the currents are
\ali{
    \hat{J}^{12}_0 &= i \hat{\mu}_0 \partial_0 ( \hat{\Phi}^1 - \hat{\Phi}^2 ) \,, \\
    \hat{J} &= -\frac{i}{\hat{\mu}_{12}} ( \partial_1  \hat{\Phi}_2 - \partial_2 \hat{\Phi}_1 ) \,.
}
Using the constraints derived by integrating out the dynamical fields $h_{01}$ and $h_{02}$
\ali{
    \partial_2 \hat{\Phi}^1 - \hat{\Phi}_2 &= 0 \,, \label{donstphinoc1} \\
     \partial_1 \hat{\Phi}^2 - \hat{\Phi}_1 &= 0 \,, \label{donstphinoc2}
}
the $x^1 x^2$-component of the current $\hat{J}_{12}$ becomes
\ali{
    \hat{J} &= \frac{i}{\hat{\mu}_{12}} \partial_1 \partial_2 ( \hat{\Phi}^1 - \hat{\Phi}^2 ) \,.
}
These currents are equivalent to the currents \eqref{hatmomsym1} and \eqref{hatmomsym2} under the correspondence \eqref{corrphihat}, and thus generate the momentum dipole symmetry in the $\hat{\phi}$-theory. We also have conservation laws related to winding symmetry. From the background gauge transformations of $C^k \wedge dx^k$ and $c$, which are \eqref{bkggaugetr1} and \eqref{bkggaugetr3}, we have relations
\begin{gather}
    \frac{1}{2\pi} \left(  \partial_0 \partial_2 \hat{\Phi}^1  - \partial_2 \partial_0 \hat{\Phi}^1 \right) = 0 \,, \\
    \frac{1}{2\pi} \left(  -\partial_0 \partial_1 \hat{\Phi}^2  + \partial_1 \partial_0 \hat{\Phi}^2 \right) = 0 \,, 
\end{gather}
which are locally trivial. Combining these equations, we have a conservation law
\ali{
    \partial_0 J_0 - \partial_1 \partial_2 J_{12} = 0 \,, 
}
where the currents are
\ali{
    J_0 &= - \frac{1}{2\pi} \partial_1 \partial_2 ( \hat{\Phi}^1 - \hat{\Phi}^2 ) \,, \\
    J_{12} &=   \frac{1}{2\pi} \partial_0 (  \hat{\Phi}^1 - \hat{\Phi}^2 ) \,.
}
These currents are equivalent to the currents \eqref{hatwinsym1} and \eqref{hatwinsym2} under the correspondence \eqref{corrphihat}, and thus generate the winding dipole symmetry in the $\hat{\phi}$-theory.

\subsection{Duality in Foliated \texorpdfstring{$\phi$}{phi}-Theory}
\label{section44}

In this section, we directly show the duality between the foliated $\phi$-theory \eqref{etofoliatedphi} and the foliated $\hat{\phi}$-theory \eqref{foliphihatc} in the same way as Section \ref{section41}. In the exotic description, the $\phi$-theory is dual to the $\hat{\phi}$-theory, which is considered as a self-duality, but in the foliated description, the self-duality is non-trivial.

We consider the foliated Lagrangian including the background foliated gauge fields
\alis{
    &\L_{\phi,\text{f}} \left[ C^k \wedge dx^k, c, \hat{C}^k, \hat{c}_{12}\ dx^1 \wedge dx^2 \right] \\
    & \quad = \frac{\mu_0}{2} ( E_0 )^2 + \frac{1}{4\mu_{12}} ( B^1_2 )^2 +  \frac{1}{4\mu_{12}} ( B^2_1 )^2 \\
    & \qquad + \frac{i}{2\pi} \hat{E}^1_0 ( \partial_2 \Phi^1 - C^1_2 - B^1_2 ) + \frac{i}{2\pi} \hat{E}^2_0 ( \partial_1 \Phi^2 - C^2_1 - B^2_1 ) \\
    & \qquad + \frac{i}{2\pi} ( \partial_1 \hat{B}^1_2 - \partial_2 \hat{B}^2_1 + \hat{B}_{12} ) (\partial_0 \Phi - c_0 - E_0 ) \\
    & \qquad + \frac{i}{2\pi} \left[ \hat{C}^1_0 ( 
    \partial_2 \Phi^1 - C^1_2 ) - \hat{C}^1_2 ( 
    \partial_0 \Phi^1 - C^1_0 ) - \hat{C}^2_0 ( 
    \partial_1 \Phi^2 - C^2_1 )  \right. \\
    & \qquad \quad \left. + \hat{C}^2_1 ( 
    \partial_0 \Phi^2 - C^2_0 ) + \hat{c}_{12} ( \partial_0 \Phi - c_0 ) \right] \\
    & \qquad + \frac{i}{2\pi} \left[ -\hat{c}_{02} ( \partial_1 \Phi - \Phi^1 - c_1 )  + \hat{c}_{01} ( \partial_2 \Phi - \Phi^2 - c_2 ) \right] \\
    & \qquad + \frac{i}{2\pi} \left[ \hat{\chi}_0 ( C^2_1 - C^1_2 + \partial_1 c_2 - \partial_2 c_1 ) \right. \\
    & \qquad \quad \left. - \hat{\chi}_1 ( C^2_0 + \partial_0 c_2 - \partial_2 c_0) + \hat{\chi}_2 ( C^1_0 + \partial_0 c_1 - \partial_1 c_0 ) \right] \,,
}
where the fields $E_0$, $B^1_2$, $B^2_1$, $\hat{E}^1_0$, $\hat{E}^2_0$, $\hat{B}^1_2$, $\hat{B}^2_1$ and $\hat{B}_{12}$ are gauge-invariant dynamical fields. Integrating out $\hat{E}^1_0$, $\hat{E}^2_0$ and $\hat{B}_{12}$, we have equations
\ali{
    B^1_2 &= \partial_2 \Phi^1 - C^1_2 \,, \label{dualeq1} \\
    B^2_1 &= \partial_1 \Phi^2 - C^2_1 \,, \label{dualeq2}  \\
    E_0 &= \partial_0 \Phi - c_0 \,,  \label{dualeq3}
}
and then we can obtain the Lagrangian of the foliated $\phi$-theory \eqref{etofoliatedphi}.

On the other hand, integrating out $E_0$, $B^1_2$ and $B^2_1$, we have equations
\begin{gather}
    \mu_0 E_0 = \frac{i}{2\pi}  ( \partial_1 \hat{B}^1_2 - \partial_2 \hat{B}^2_1 + \hat{B}_{12} ) \,, \label{dualeq4}\\
    \frac{1}{2\mu_{12}} B^1_2 = \frac{i}{2\pi} \hat{E}^1_0 \,, \label{dualeq5}\\
    \frac{1}{2\mu_{12}} B^2_1 = \frac{i}{2\pi} \hat{E}^2_0 \,. \label{dualeq6}
\end{gather}
Using them, we have
\alis{
    &\L_{\phi,\text{f}} \left[ C^k \wedge dx^k, c, \hat{C}^k, \hat{c}_{12}\ dx^1 \wedge dx^2 \right] \\
    & \quad = \frac{1}{8\pi^2 \mu_0} ( \partial_1 \hat{B}^1_2 - \partial_2 \hat{B}^2_1 + \hat{B}_{12} )^2 + \frac{\mu_{12}}{4\pi^2} ( \hat{E}^1_0 )^2 +  \frac{\mu_{12}}{4\pi^2} ( \hat{E}^2_0 )^2 \\
    & \qquad + \frac{i}{2\pi} ( \hat{E}^1_0 + \hat{C}^1_0 ) ( \partial_2 \Phi^1 - C^1_2 ) + \frac{i}{2\pi} ( \hat{E}^2_0 + \hat{C}^2_0 ) ( \partial_1 \Phi^2 - C^2_1  ) \\
    & \qquad + \frac{i}{2\pi} ( \partial_1 \hat{B}^1_2 - \partial_2 \hat{B}^2_1 + \hat{B}_{12} + \hat{c}_{12} ) (\partial_0 \Phi - c_0 ) \\
    & \qquad + \frac{i}{2\pi} \left[ - \hat{C}^1_2 ( 
    \partial_0 \Phi^1 - C^1_0 ) + \hat{C}^2_1 ( 
    \partial_0 \Phi^2 - C^2_0 )  \right] \\
    & \qquad + \frac{i}{2\pi} \left[ -\hat{c}_{02} ( \partial_1 \Phi - \Phi^1 - c_1 )  + \hat{c}_{01} ( \partial_2 \Phi - \Phi^2 - c_2 ) \right] \\
    & \qquad + \frac{i}{2\pi} \left[ \hat{\chi}_0 ( C^2_1 - C^1_2 + \partial_1 c_2 - \partial_2 c_1 ) \right. \\
    & \qquad \quad \left. - \hat{\chi}_1 ( C^2_0 + \partial_0 c_2 - \partial_2 c_0) + \hat{\chi}_2 ( C^1_0 + \partial_0 c_1 - \partial_1 c_0 ) \right] \,.
}
Next, integrating out $\Phi$, $\Phi^1$ and $\Phi^2$, we can have equations
\ali{
    \partial_1 \hat{c}_{02} - \partial_2 \hat{c}_{01} &= \partial_0 ( \partial_1 \hat{B}^1_2 - \partial_2 \hat{B}^2_1 + \hat{B}_{12} + \hat{c}_{12} ) \,, \\
    \hat{c}_{02} &= \partial_2 ( \hat{E}^1_0 + \hat{C}^1_0 ) - \partial_0 \hat{C}^1_2 \,, \\
    \hat{c}_{01} &= \partial_1 ( \hat{E}^2_0 + \hat{C}^2_0 ) - \partial_0 \hat{C}^2_1 \,.
}
Combining them, we can obtain
\ali{
    \partial_1 \partial_2 ( \hat{E}^1_0 + \hat{C}^1_0 ) - \partial_2 \partial_1 ( \hat{E}^2_0 + \hat{C}^2_0 ) = \partial_0 \partial_1 ( \hat{B}^1_2 + \hat{C}^1_2 ) - \partial_0 \partial_2 ( \hat{B}^2_1 + \hat{C}^2_1 ) + \partial_0 (\hat{B}_{12} + \hat{c}_{12}) \,.
}
We can locally solve this equation as
\ali{
    \hat{E}^1_0 &= \partial_0 \hat{\Phi}^1 - \hat{\Phi}_0 - \hat{C}^1_0 \,, \label{dualeq7} \\
    \hat{E}^2_0 &= \partial_0 \hat{\Phi}^2 - \hat{\Phi}_0 - \hat{C}^2_0 \,, \label{dualeq8} \\
    \hat{B}^1_2 &= \partial_2 \hat{\Phi}^1 - \hat{\Phi}_2 - \hat{C}^1_2 \,, \label{dualeq9} \\
    \hat{B}^2_1 &= \partial_1 \hat{\Phi}^2 - \hat{\Phi}_1 - \hat{C}^2_1 \,, \label{dualeq10}\\
    \hat{B}_{12} &=  \partial_1 \hat{\Phi}_2 - \partial_2 \hat{\Phi}_1  - \hat{c}_{12} \,,  \label{dualeq11}
}
where $\hat{\Phi}^k\ (k = 1,2)$ is a $U(1)$ type-$B$ foliated zero-form gauge field and $\hat{\Phi}$ is a $U(1)$ type-$B$ bulk one-form gauge field. Their dynamical gauge transformations are
\ali{
    \hat{\Phi}^k  &\sim \hat{\Phi}^k + 2\pi \hat{W}^k + \hat{\xi} \,, \\
    \hat{\Phi} &\sim \hat{\Phi} + d \hat{\xi} \,,
}
where $\hat{W}^k$ is an integer-valued gauge parameter and it can have step function discontinuities in the $x^k$ direction, and $\hat{\xi}$ is a type-$B$ bulk zero-form gauge parameter. Their background gauge transformations are also
\ali{
    \hat{\Phi}^k  &\sim \hat{\Phi}^k + \hat{\lambda}^k \,,  \\
    \hat{\Phi} &\sim \hat{\Phi} + \hat{\kappa} \,.
}
The dynamical gauge fields $\hat{c}_{01}$ and $\hat{c}_{02}$ are written as
\ali{
    \hat{c}_{01} &= \partial_1 ( \partial_0 \hat{\Phi}^2 - \hat{\Phi}_0 ) - \partial_0 \hat{C}^2_1 \,, \\
    \hat{c}_{02} &= \partial_2 ( \partial_0 \hat{\Phi}^1 - \hat{\Phi}_0 ) - \partial_0 \hat{C}^1_2 \,.
}

Then, we can derive
\alis{
    &\L_{\phi,\text{f}} \left[ C^k \wedge dx^k, c, \hat{C}^k, \hat{c}_{12}\ dx^1 \wedge dx^2 \right] \\
    & \quad = \hat{\mu}_{0} ( \partial_0 \hat{\Phi}^1 - \hat{\Phi}_0 - \hat{C}^1_0 )^2 +  \hat{\mu}_{0} ( \partial_0 \hat{\Phi}^2 - \hat{\Phi}_0 - \hat{C}^2_0 )^2 \\
    & \qquad + \frac{1}{2\hat{\mu}_{12}} \left[ \partial_1 ( \partial_2 \hat{\Phi}^1 - \hat{C}^1_2 ) - \partial_2 ( \partial_1 \hat{\Phi}^2 - \hat{C}^2_1)  - \hat{c}_{12} \right]^2 \\
    & \qquad - \frac{i}{2\pi} C^1_2 ( \partial_0 \hat{\Phi}^1 - \hat{\Phi}_0  ) + \frac{i}{2\pi}  C^2_1 ( \partial_0 \hat{\Phi}^2 - \hat{\Phi}_0 )  \\
    & \qquad - \frac{i}{2\pi} c_0 \left[ \partial_1 ( \partial_2 \hat{\Phi}^1 - \hat{C}^1_2 ) - \partial_2 ( \partial_1 \hat{\Phi}^2 - \hat{C}^2_1 ) \right] + \frac{i}{2\pi} \left[ C^1_0 \hat{C}^1_2  - C^2_0 \hat{C}^2_1  \right] \\
    & \qquad + \frac{i}{2\pi} \left\{ c_1 \left[ \partial_2 ( \partial_0 \hat{\Phi}^1 - \hat{\Phi}_0 ) - \partial_0 \hat{C}^1_2 \right]  - c_2 \left[ \partial_1 ( \partial_0 \hat{\Phi}^2 - \hat{\Phi}_0 ) - \partial_0 \hat{C}^2_1 \right] \right\} \\
    & \qquad + \frac{i}{2\pi} \left[ \hat{\chi}_0 ( C^2_1 - C^1_2 + \partial_1 c_2 - \partial_2 c_1 ) \right. \\
    & \qquad \quad \left. - \hat{\chi}_1 ( C^2_0 + \partial_0 c_2 - \partial_2 c_0) + \hat{\chi}_2 ( C^1_0 + \partial_0 c_1 - \partial_1 c_0 ) \right] \,,
}
where $\hat{\mu}_0 = \mu_{12}/(4\pi^2)$ and $\hat{\mu}_{12} = 4\pi^2 \mu_0$. We consider changes of variables
\ali{
    \hat{\chi}_0 &\rightarrow \hat{\chi}_0 + \hat{\Phi}_0 \,, \\
    \hat{\chi}_1 &\rightarrow \hat{\chi}_1 + \partial_1 \hat{\Phi}^2 - \hat{C}^2_1 \,, \\
    \hat{\chi}_2 &\rightarrow \hat{\chi}_2 + \partial_2 \hat{\Phi}^1 - \hat{C}^1_2 \,, 
}
and then, the field $\chi$ has a dynamical gauge transformation
\ali{
    \hat{\chi} \sim \hat{\chi} - d\hat{\xi} \,,
}
and a background gauge transformation
\ali{
    \hat{\chi} \sim \hat{\chi} - \hat{\kappa} \,,
}
as in the previous section. The result is
\alis{
    &- \frac{i}{2\pi} C^1_2 ( \partial_0 \hat{\Phi}^1 - \hat{\Phi}_0  ) + \frac{i}{2\pi}  C^2_1 ( \partial_0 \hat{\Phi}^2 - \hat{\Phi}_0 )  \\
    & \quad - \frac{i}{2\pi} c_0 \left[ \partial_1 ( \partial_2 \hat{\Phi}^1 - \hat{C}^1_2 ) - \partial_2 ( \partial_1 \hat{\Phi}^2 - \hat{C}^2_1 ) \right] + \frac{i}{2\pi} \left[ C^1_0 \hat{C}^1_2  - C^2_0 \hat{C}^2_1  \right] \\
    &  \quad + \frac{i}{2\pi} \left\{ c_1 \left[ \partial_2 ( \partial_0 \hat{\Phi}^1 - \hat{\Phi}_0 ) - \partial_0 \hat{C}^1_2 \right]  - c_2 \left[ \partial_1 ( \partial_0 \hat{\Phi}^2 - \hat{\Phi}_0 ) - \partial_0 \hat{C}^2_1 \right] \right\} \\
    & \rightarrow  \frac{i}{2\pi} \left(  C^1_0  \partial_2 \hat{\Phi}^1 -  C^2_0  \partial_1 \hat{\Phi}^2 -  C^1_2  \partial_0 \hat{\Phi}^1 +  C^2_1  \partial_0 \hat{\Phi}^2 \right) \,,
}
and finally we obtain the foliated $\hat{\phi}$-theory
\alis{
    &\L_{\hat{\phi},\text{e} \rightarrow \text{f}} \left[ C^k \wedge dx^k, c, \hat{C}^k, \hat{c}_{12}\ dx^1 \wedge dx^2 \right] \\
    & = \frac{\hat{\mu}_0}{2} \left[ ( \partial_0 \hat{\Phi}^1 - \hat{C}^1_0 ) - ( \partial_0 \hat{\Phi}^2 - \hat{C}^2_0 ) \right]^2 \\
    & \quad + \frac{1}{2\hat{\mu}_{12}} \left[ \partial_1 ( \partial_2 \hat{\Phi}^1 - \hat{C}^1_2 ) - \partial_2 ( \partial_1 \hat{\Phi}^2 - \hat{C}^2_1 ) - \hat{c}_{12} \right]^2 \\
    & \quad + \frac{i}{2\pi} \left(  C^1_0  \partial_2 \hat{\Phi}^1 -  C^2_0  \partial_1 \hat{\Phi}^2 -  C^1_2  \partial_0 \hat{\Phi}^1 +  C^2_1  \partial_0 \hat{\Phi}^2 \right) \\
    & \quad + \frac{i}{2\pi} \left[ \hat{\chi}_0 ( C^2_1 - C^1_2 + \partial_1 c_2 - \partial_2 c_1 ) \right. \\
    & \qquad \left. - \hat{\chi}_1 ( C^2_0 + \partial_0 c_2 - \partial_2 c_0) + \hat{\chi}_2 ( C^1_0 + \partial_0 c_1 - \partial_1 c_0 ) \right] \,,
}
which is the same as \eqref{foliphihatc}. We can find relations among the fields using the equations \eqref{dualeq1}--\eqref{dualeq3},\eqref{dualeq4}--\eqref{dualeq6} and \eqref{dualeq7}--\eqref{dualeq11}. The relations are
\begin{gather}
    \mu_0 (\partial_0 \Phi - c_0 ) = \frac{i}{2\pi} \left\{ \partial_1 \partial_2 (\hat\Phi^1 - \hat\Phi^2 ) - (\partial_1 \hat{C}^1_2 - \partial_2 \hat{C}^2_1 + \hat{c}_{12}  )  \right\} \,, \\
    \frac{1}{2\mu_{12}} ( \partial_2 \Phi^1 - C^1_2 ) = \frac{i}{2\pi} ( \partial_0 \hat\Phi^1 - \hat\Phi_0 - \hat{C}^1_0 ) \,, \\
    \frac{1}{2\mu_{12}} ( \partial_1 \Phi^2 - C^2_1 ) = \frac{i}{2\pi} ( \partial_0 \hat\Phi^2 - \hat\Phi_0 - \hat{C}^2_0 ) \,.
\end{gather}
Turning off the background gauge fields and using the relations  \eqref{phihat0formula}, \eqref{donstphinoc1} and \eqref{donstphinoc2}, these relations become
\begin{gather}
    \mu_0 \partial_0 \Phi  = \frac{i}{2\pi} \left( \partial_1  \hat\Phi_2 -  \partial_2 \hat\Phi_1  \right) \,, \\
    \frac{1}{2\mu_{12}} \partial_2 \Phi^1  =  \frac{i}{2\pi} ( \partial_0 \hat\Phi^1 - \hat\Phi_0 ) \,,  \\
    \frac{1}{2\mu_{12}} \partial_1 \Phi^2  =-  \frac{i}{2\pi} ( \partial_0 \hat\Phi^2 - \hat\Phi_0 ) \,,
\end{gather}
which indicates that $\hat\Phi$ is the one-form $T$-dual field of the zero-form field $\Phi$ in 2+1 dimensions, and $\hat\Phi^k$ is the zero-form $T$-dual field of the zero-form field $\Phi^k$ on the (1+1)-dimensional layers.

\subsection*{Acknowledgment}
KO is supported by JSPS KAKENHI Grant-in-Aid No.22K13969 and No.24K00522. KO also acknowledge support by the Simons Foundation Grant \#888984 (Simons Collaboration on Global Categorical Symmetries). SS is supported by the World-leading INnovative Graduate Study Program for Frontiers of Mathematical Sciences and Physics (WINGS-FMSP), The University of Tokyo.

\appendix
\section{Exotic QFT and Tensor Gauge Fields}
\label{appendix1}

Fractonic theories do not have the full continuous rotational symmetry, but instead, have a discrete spatial rotational symmetry in particular directions.
Some fractonic QFTs admit an alternative description by exotic QFTs with exotic tensor gauge fields \cite{Pretko:2018jbi,You:2019ciz,Slagle:2017wrc,Seiberg:2020bhn,Seiberg:2020wsg,Seiberg:2020cxy,Gorantla:2020xap,Gorantla:2021bda,Gorantla:2020jpy,Geng:2021cmq,Yamaguchi:2021qrx,Yamaguchi:2021xeq,Burnell:2021reh,Gorantla:2022eem,Gorantla:2022ssr,Honda:2022shd}.
The tensor gauge fields are the gauge fields in the representations of the discrete symmetry.
In this paper, we deal with theories in 2+1 dimensions with the 90-degree rotational symmetry for $(x^1,x^2)$ whose symmetry group is $\Z_4$, and theories in 3+1 dimensions with the 90-degree rotational symmetry for $(x^1,x^2)$ and the continuous rotational symmetry for $(x^0,x^3)$ whose symmetry group is $\Z_4 \times SO(2)$.\footnote{In this paper, we take Euclidean spacetime and the coordinate is written as $(x^0, x^k)$, where $x^0$ is the time component and $x^k \ (k = 1,2,...)$ are spatial components.}
Therefore, we can consider the representation of $\Z_4$.
The irreducible representations of $\Z_4$ are one-dimensional representations $\bm{1}_n \ (n = 0,\pm1,2)$.
The 90-degree rotation $e^{i\frac{\pi}{2}}$ acts on the representations in $\bm{1}_n$ as $e^{i\frac{n\pi}{2}}$.
We write the tensor using $SO(2)$ vector indices as $F$ for $\bm{1}_0$, $F_{12}=F_{21}$ for $\bm{1}_2$. Under the 90-degree rotation, the tensors transform as
\ali{
    F &\rightarrow F \,, \\
    F_{12} &\rightarrow - F_{12} \,.
}
These transformations can be understood %
as follows.
$F_{12}$ is the component of a hollow symmetric tensor $F_{12}\, dx^1 \odot dx^2$,\footnote{The symbol $\odot$ represents the symmetric tensor $dx^1 \odot dx^2 = \frac{1}{2} (dx^1 \otimes dx^2 + dx^2 \otimes dx^1)$, while the wedge product $\wedge$ means $dx^1 \wedge dx^2 = \frac{1}{2} (dx^1 \otimes dx^2 - dx^2 \otimes dx^1)$.}
so under the rotation $x'^1 = x^2$, $x'^2 = -x^1$, it transforms as
\alis{
    F_{12} \, dx^1 \odot dx^2 &= F'_{12} \, dx'^1 \odot dx'^2 \\
    &= - F'_{12} \, dx^2 \odot dx^1 \,,
}
and thus, we have $F'_{12} = - F_{21} = -F_{12}$.

We consider $U(1)$ tensor gauge fields.
A feature of the tensor gauge fields is that they can have some delta function singularities and step function discontinuities, which is a consequence of the UV/IR mixing in fracton phases \cite{Seiberg:2020bhn,Seiberg:2020wsg,Seiberg:2020cxy}.\footnote{The allowed singularities and discontinuities are determined so that the action has up to a $O(a^{-1})$ dependence where $a$ is the UV cutoff length.}
As examples of the tensor gauge fields, we describe the fields in a (2+1)-dimensional $\Z_4$-invariant fractonic theory.
The (3+1)-dimensional case, which we also consider, is similar to the (2+1)-dimensional case.

\subsection{Tensor Gauge Fields in 2+1 Dimensions}

\subsubsection*{$U(1)$ tensor zero-form gauge fields}

We consider a compact scalar field $\phi$ in the representation $\bm{1}_0$ with the periodicity $\phi \sim \phi + 2\pi$.
Unlike ordinary compact scalar fields, the field $\phi$ has a gauge transformation
\ali{
    \phi \sim \phi + 2\pi w^1 + 2\pi w^2 \,,
}
where $w^1$ and $w^2$ are integer-valued gauge parameters.
Under the 90-degree rotation, they transform as $w^1 \rightarrow w^2$ and $w^2 \rightarrow w^1$.
We let %
$w^k \ (k=1,2)$ have step function discontinuities in the $x^k$ direction.
This compact scalar $\phi$ is similar to the $U(1)$ zero-form gauge field, so we call $\phi$ a $U(1)$ tensor zero-form gauge field $\bm{\phi}$.
Another tensor zero-form gauge field is a compact scalar $\bm{\hat\phi} = \hat{\phi}^{12}$ in the representation $\bm{1}_2$.
The field $\hat{\phi}^{12}$ has a gauge transformation
\ali{
    \hat{\phi}^{12} \sim \hat{\phi}^{12} + 2\pi \hat{w}^1 - 2\pi \hat{w}^2 \,,
}
where $\hat{w}^k\ (k=1,2)$ has the same properties as $w^k$.
We define the exterior derivatives of $\bm
{\phi} = \phi$ and $\bm{\hat\phi} = \hat\phi^{12}$ as
\ali{
    d_\text{e} \bm{\phi} &= (\partial_0 \phi, \partial_1 \partial_2 \phi ) \,, \\
    d_\text{e} \bm{\hat\phi} &= (\partial_0 \hat\phi^{12}, \partial_1 \partial_2 \hat\phi^{12} ) 
}
in the representation $(\bm{1}_0,\bm{1}_2)$ and $(\bm{1}_2,\bm{1}_0)$, which are gauge invariant.

\subsubsection*{$U(1)$ tensor one-form gauge fields}

Next, we consider $U(1)$ tensor one-form gauge fields.
These fields are $\bm{C}=(C_0,C_{12})$ in $(\bm{1}_0,\bm{1}_2)$ and $\bm{\hat{C}}=(\hat{C}^{12}_0, \hat{C})$ in $(\bm{1}_2,\bm{1}_0)$.
Their gauge transformations are 
\ali{
    C_0 &\sim C_0 + \partial_0 \Lambda \,,  \\
    C_{12} &\sim C_{12} + \partial_1\partial_2 \Lambda \,. 
}
and
\ali{
    \hat{C}^{12}_0 &\sim \hat{C}^{12}_0 + \partial_0  \hat{\Lambda}^{12} \,,  \\
    \hat{C} &\sim \hat{C} + \partial_1 \partial_2  \hat{\Lambda}^{12} \,,
}
where $\bm{\Lambda}= \Lambda$ and $\bm{\hat\Lambda} = \hat{\Lambda}^{12}$ are $U(1)$ tensor zero-form gauge parameters in the $\bm{1}_0$ and $\bm{1}_2$ respectively, and they have their own gauge transformations.
We define the exterior derivatives of $\bm{C}$ and $\bm{\hat{C}}$ as
\ali{
    d_\text{e} \bm{C} &= \partial_0 C_{12} - \partial_1 \partial_2 C_0 \,, \\
    d_\text{e} \bm{\hat{C}} &= \partial_0 \hat C  - \partial_1 \partial_2 \hat C^{12}_0  \,,
}
in $\bm{1}_2$ and $\bm{1}_0$, which are gauge invariant.
We can rewrite the gauge transformations as
\ali{
    \bm{C} &\sim \bm{C} + d_\text{e} \bm{\Lambda} \,, \\
    \bm{\hat{C}} &\sim \bm{\hat{C}} + d_\text{e} \bm{\hat{\Lambda}} \,,
}
and we can check that $(d_\text{e})^2=0$.

\subsubsection*{$U(1)$ tensor two-form gauge fields}

We can also consider $U(1)$ tensor two-form gauge fields in 2+1 dimensions, although we do not deal with them in this paper.
These fields are $B_{012}$ in $\bm{1}_2$ and $\hat{B}_0$ in $\bm{1}_0$.
Their gauge transformations are
\ali{
    B_{012} &\sim B_{012} + \partial_0 \Gamma_{12} - \partial_1 \partial_2 \Gamma_0 \,, \\
    \hat{B}_0 &\sim \hat{B}_0 + \partial_0 \hat{\Gamma} - \partial_1 \partial_2 \hat{\Gamma}^{12}_0 \,,
}
where $\bm{\Gamma}=(\Gamma_0, \Gamma_{12})$ and $\bm{\hat{\Gamma}}= (\hat{\Gamma}^{12}_0,\hat{\Gamma})$ are $U(1)$ tensor one-form gauge parameters in the $(\bm{1}_0,\bm{1}_{2})$ and $(\bm{1}_{2},\bm{1}_0)$ respectively, and they have their own gauge transformations.
Exterior derivatives of $B_{012}$ and $\hat{B}_0$ are assumed to be zero.
We can also rewrite the gauge transformations in terms of the exterior derivatives as
\ali{
    B_{012} &\sim B_{012} + d_\text{e} \bm{\Gamma} \,, \\
    \hat{B}_0 &\sim \hat{B}_0 + d_\text{e} \bm{\hat{\Gamma}} \,. 
}

\subsection{Exotic de Rham Complex in 2+1 Dimensions}

To summarize this appendix, here we compose the structure so far discussed into a mathematical system analogous to the de Rham complex, which we call the exotic de Rham complex.
Here we assume the 2+1-dimensional spacetime with flat foliations along every spatial dimension.
It would be interesting to develop a more robust mathematical theory working in more general cases, as well as the differential enrichment (that is, to take the patch-to-patch gluing data on top of the local degrees of freedom) of the generalized de Rham complex.

As the underling algebra of the exotic de Rham complex on a spacetime manifold $M = M_\text{t} \times M_\text{s}$ that is the product of a time $M_\text{t}$ and a spatial manifold $M_\text{s}$, we take the exterior algebra\footnote{\ytableausetup{boxsize=0.4em}
    The higher-dimensional generalization depends on the spatial kinetic term assumed for a scalar field theory. For a kinetic term of the form $\sum_{i,j}(\partial_i\partial_j \phi)^2$ in $2+1$ dimensions, the relevant bundle of degree 1 over $M_\text{s}$ is expected to be generated by $dx^i \odot dx^j = dx^j \odot dx^i$, where $i$ and $j$ are distinct indices. Additionally, we consider another bundle of rank 2, assigned a degree of 2, over $M_\text{s}$. This bundle is generated by $dx^i \otimes dx^j \otimes dx^k$ with distinct indices $i, j, k$, projected down by the Young symmetrizers associated with the Young diagram $\ydiagram{2,1}$. To make sense of this bundle, $M_\text{s}$ has to be equipped with a structure reducing its tangent $SO(3)$ bundle into a $S_4$ bundle. The exterior derivative also includes a term $dx^i \partial_i$, which sends a section of the degree 1 bundle to a section of the degree 2 bundle.
}
\begin{equation}
    \Omega^p_\text{e}(M) := \Gamma(M,\Lambda^p(T^\ast M_\text{t} \oplus S_{\odot}T^\ast M_\text{s})),
\end{equation}
where $S_{\odot}T^\ast M_\text{s}$ is the bundle locally generated by $dx^1\odot dx^2$, and we assign the degrees so that both $T^\ast M_\text{t}$ and $S_{\odot}T^\ast M_\text{s}$ are of degree 1.
This bundle is not closed under the local $SO(2)$ rotation, and thus the $M_\text{s}$ should be equipped with an additional structure reducing its tangential $SO(2)$ bundle to a $\mathbb{Z}_4$ bundle.
 In addition, $\Gamma(M,E)$ means the set of sections (allowing appropriate singularities and discontinuities) of a vector bundle $E$ over $M$, and $\Lambda^\ast$ represents the exterior algebra bundle over the trivial vector bundle.
Explicitly in the coordinate system $(x^0,x^1,x^2)$, the algebra $\Omega^{p}_\text{e}(M)$ is generated by $dx^0$ of degree 1 and $dx^1\odot dx^2$ of degree 1, which anticommutes:
\begin{equation}
    dx^0 \wedge (dx^1 \odot dx^2) = - (dx^1 \odot dx^2) \wedge dx^0,
\end{equation}
and also are nilpotent:
\begin{equation}
    dx^0 \wedge dx^0 = 0, \quad (dx^1 \odot dx^2) \wedge (dx^1 \odot dx^2) = 0.
\end{equation}
Note that $\Omega^p_\text{e}(M)$ does not include elements $dx^1$ or $dx^2$, as opposed to the usual de Rham complex.
On this algebra, we define the exterior derivative $d_\text{e}$ as
\begin{equation}
    d_\text{e} = dx^0 \partial_0 + dx^1 \odot dx^2 \partial_1 \partial_2,
    \label{eq:exotic_derivative}
\end{equation}
which is nilpotent: $d_\text{e}^2 = 0$.
The tensor gauge fields with $p=0,1,2$ are expanded as
 \begin{align}
    \bm{\phi} &= \phi\\
    \bm{C} &= C_0\, dx^0 + C_{12}\, dx^{1}\odot dx^2\\ 
    \bm{B} &= B_{012}\, dx^0 \wedge (dx^1 \odot dx^2),
 \end{align}
  and their local gauge transformations can be described in terms of the exterior derivative $d_\text{e}$.
To match the $\mathbb{Z}_4$ representations of each component, one notices that $dx^0$ is in the $\bm{1}_0$ representation and $dx^1\odot dx^2$ is in the $\bm{1}_2$ representation of $\mathbb{Z}_4$.

Note that, unlike the usual de Rham complex, the complex $\Omega^p_\text{e}(M)$ does not contain the local volume form $dx^0\wedge dx^1 \wedge dx^2$.
The element $dx^0\wedge (dx^1 \odot dx^2)$ does not have the correct transformation property under the spatial rotation and thus it is not suitable for the volume form.
As a Lagrangian density has to be locally proportional to the volume form, we need to introduce another dual de Rham complex $\widehat{\Omega}_\text{e}^p(M)$ by 
\begin{align}
    \widehat{\Omega}_\text{e}^0(M) &:= \Gamma(M,\Lambda^2T^\ast M_\text{s} \otimes S_{\odot}TM_\text{s}),\\
    \widehat{\Omega}_\text{e}^1(M) &:= \Gamma(M,(\Lambda^3T^\ast M\otimes S_{\odot}TM_\text{s}) \oplus \Lambda^2T^\ast M_\text{s}),\\
    \widehat{\Omega}_\text{e}^2(M) &:= \Gamma(M,\Lambda^3T^\ast M).
\end{align}
Here, $S_{\odot}TM_s$ is the bundle locally generated by $\partial_1\odot \partial_2$, which has a natural pairing with $S_{\odot}T^\ast M_s$ defined by
\begin{equation}
    \langle dx^1\odot dx^2, \partial_1\odot\partial_2 \rangle = 1. \\ 
\end{equation}

These bundles are determined so that we have the pairing valued in a volume form:
\begin{equation}
    \langle\cdot,\cdot\rangle: \Omega_\text{e}^p(M) \times \widehat{\Omega}_\text{e}^{2-p}(M) \to \Gamma(M,\Lambda^3T^\ast M),
    \label{eq:exotic_pairing}
\end{equation}
induced from the pairing of $S_{\odot}TM_\text{s}$ and $S_{\odot}T^\ast M_\text{s}$. For the $\Lambda^pT^\ast M_\text{t}$ and $\Lambda^pT^\ast M_\text{s}$ components, we simply take the wedge product, as we are assuming the Euclidean flat metric.

The dual $p=0,1,2$ forms are expanded as
\begin{align}
    \bm{\hat{\phi}} &= \hat{\phi}^{12} \, (dx^1 \wedge dx^2) \otimes (\partial_1\odot\partial_2) ,\\
    \bm{\hat{C}} &= \hat{C}_0^{12} \, (dx^0 \wedge dx^1 \wedge dx^2)\otimes ( \partial_1\odot\partial_2 ) + \hat{C} \,dx^1 \wedge dx^2 , \\
    \bm{\hat{B}} &= \hat{B}_0 \, dx^0 \wedge dx^1 \wedge dx^2 ,
\end{align}
where in these expressions $\partial_1\partial_2$ is the formal basis of $S_{\odot}TM_\text{s}$.
The exterior derivative on $\widehat{\Omega}_\text{e}^p(M)$ is again expressed as \eqref{eq:exotic_derivative}, while now $dx^1\odot dx^2$ is understood to be paired with the $S_{\odot}TM_s$ component of the bundle.

The pairing \eqref{eq:exotic_pairing} induces the Hodge star operator:
\begin{equation}
    \ast_\text{e}: \Omega_\text{e}^p(M) \to \widehat{\Omega}_\text{e}^{2-p}(M).
\end{equation}
For example, one calculates that
\begin{align}
    \ast_\text{e} (dx^0) &= dx^1 \wedge dx^2, \\ 
    \ast_\text{e} (dx^1 \odot dx^2) &= -\, (dx^0 \wedge dx^1 \wedge dx^2) \otimes(\partial_1\odot\partial_2).
\end{align}
The sign in the latter equation is because $dx^1\odot dx^2$ has degree 1 and thus it anticommutes with $dx^0\wedge dx^1\wedge dx^2$.
With this operator, the Lagrangian density of the exotic $\phi$-theory \eqref{exoticphi}, when $
\frac1{\mu_{12}}=\mu_0$, can be written as
\begin{equation}
    \mathcal{L}_{\phi,\text{e}} = \frac{1}{2\mu_0}\langle d_\text{e} \phi, \ast_\text{e} d_\text{e}\phi\rangle.
\end{equation}
Lagrangian density with a general $\mu_{12}$ can be obtained by considering a more general (diagonal) metric, which modifies the pairing \eqref{eq:exotic_pairing} and accordingly the Hodge star operator $\ast_\text{e}$.

\section{Foliated QFT and Foliated Gauge Fields}
\label{appendix2}

Although this section is based on \cite{Slagle:2020ugk,Hsin:2021mjn, Ohmori:2022rzz, Shimamura:2024kwf}, we present a reorganized formulation that provides a new perspective.

We can construct fractonic QFTs by using the foliation structure.
A foliation is a decomposition of a manifold into a stack of an infinite number of submanifolds, each of which we call a \textit{leaf}.
We consider codimension-one foliations specified by the one-form foliation field $e$.
The foliation field $e$ is nonzero everywhere and must satisfy $e \wedge de = 0$.
Given a foliation field $e$, we can define the foliation structure as the set of the codimension-one submanifolds orthogonal to $e$.
The foliation field $e$ has a gauge transformation $e \rightarrow f e$, where $f$ is a zero-form parameter.
QFTs coupled to the foliation structures on the spacetime are called foliated QFTs.
A foliation is called flat when $de = 0$.
In this paper, we focus on the particular flat foliation fields that are aligned with the coordinate axes and fix the gauge by setting $e^k = d x^k\ (k =1,2,...)$, where $x^k$ donates a spatial coordinate.
In 2+1 dimensions, we consider the two simultaneous foliations $e^1 = dx^1$ and $e^2 = dx^2$, and the leaves are the lines orthogonal to the $x^1$ direction and the lines orthogonal to the $x^2$ direction.

Next, we introduce $U(1)$ foliated gauge fields indexed by $k$.
They can be considered as $U(1)$ gauge fields on the leaves orthogonal to the $x^k$ direction.
We also introduce $U(1)$ gauge fields that mediate foliated gauge fields, which are called \textit{bulk} gauge fields.
We also refer to a pair of a foliated gauge field and a bulk gauge field as a foliated gauge field as well.
As in the case of the tensor gauge fields, the foliated gauge fields can have delta function singularities and step function discontinuities.
In the following, we consider a foliated QFT in 2+1 dimensions with two flat foliations $dx^1$ and $dx^2$. The (3+1)-dimensional case is similar to the (2+1)-dimensional case.
We have two types of foliated gauge fields: type-$A$ foliated gauge fields and type-$B$ foliated gauge fields.
Type-$A$ foliated gauge fields are ($n$+1)-form gauge fields $\tilde{A}^k$ satisfying $\tilde{A}^k \wedge e^k = 0$, and then we 
write $\tilde{A}^k = A^k \wedge e^k$ locally, where $A^k$ is some $n$-form gauge field.
In addition to the gauge transformations $A^k$ inherits from $\tilde{A}^k$, it has an additional gauge transformation $A^k \sim A^k + \sigma^k\wedge e^k$, with which we set the component $A^k_{i_1,\cdots,i_n}$ to zero when $e^k = dx^k$ and one of $i_p$ is $k$.

\subsection{Foliated Gauge fields in 2+1 Dimensions}

\subsubsection*{$U(1)$ type-$A$ foliated (0+1)-form gauge field}

The gauge transformation of a one-form gauge field $\Phi^k\, dx^k\  (k =1,2)$ is
\ali{
    \Phi^k \, dx^k \sim \Phi^k \, dx^k + 2\pi d W^k \,, 
}
where $W^k$ is an integer-valued gauge parameter and it can have step function discontinuities in the $x^k$ direction.
We call this field a $U(1)$ type-$A$ foliated (0+1)-form gauge field.
In addition, we consider zero-form gauge field $\Phi$ whose gauge transformation is
\ali{
    \Phi \sim \Phi  + 2\pi W^1 + 2\pi W^2 \,. 
}
We call this field a $U(1)$ type-$A$ bulk zero-form gauge field.
Then, the tuple $\bm{\Phi} = (\Phi^1 \, dx^1, \Phi^2 \, dx^2  ;\, \Phi)$ is also called $U(1)$  type-$A$ foliated (0+1)-form gauge field.
We define the exterior derivatives of $\bm{\Phi}$ as
\ali{
    d_\text{f} \bm{\Phi} = \left(d\Phi^1 \wedge dx^1, d\Phi^2 \wedge dx^2;\,   d \Phi - \sum^{2}_{k=1} \Phi^k \, dx^k \right) \,, 
}
which is gauge invariant.\footnote{
Mathematically, the resulting complex is the mapping cone construction known in the context of homological algebra. About the mapping cone, see a standard textbook, e.g.\ \cite{Weibel_1994}.
}

\subsubsection*{$U(1)$ type-$A$ foliated (1+1)-form gauge field}

Similarly, we can consider $U(1)$ type-$A$ foliated (1+1)-form gauge field. The foliated fields are the two-form gauge fields $C^k\wedge dx^k\  (k =1,2)$ with gauge transformations
\ali{
    C^k\wedge dx^k \sim C^k \wedge dx^k + d\lambda^k \wedge dx^k\,, 
}
where $\lambda^k\, dx^k$ is a $U(1)$ type-$A$ foliated (0+1)-form gauge parameter and has its own gauge transformation. 
The bulk field is the one-form gauge field $c$ with gauge transformations
\ali{
    c \sim c + d\lambda - \sum^{2}_{k=1} \lambda^k\, dx^k \,,
}
where $\lambda$ is a $U(1)$ type-$A$ bulk zero-form gauge parameter and has its own gauge transformation.  
We define the exterior derivatives of $\bm{C} = (C^1 \wedge dx^1, C^2 \wedge dx^2;\, c)$ as
\ali{
    d_\text{f} \bm{C} = (d C^1 \wedge dx^1,d C^2 \wedge dx^2;\, dc + \sum^2_{k=1} C^k \wedge dx^k ) \,,
}
which is gauge invariant. The gauge transformation of $\bm{C}$ can be written as
\ali{
    \bm{C} \sim \bm{C} + d_\text{f} \bm{\lambda} \,,
}
where $\bm{\lambda} = (\lambda^1 \, dx^1, \lambda^2 \, dx^2;\, \lambda)$. Therefore, we can derive $(d_\text{f})^2 = 0$.

\subsubsection*{$U(1)$ type-$A$ foliated (2+1)-form gauge field}

We can also consider $U(1)$  type-$A$ foliated (2+1)-form gauge field $\bm{B}= ( B^1 \wedge dx^1, B^2 \wedge dx^2;\, b)$, although we do not deal with them in this paper. $B^k \wedge dx^k \ (k=1,2)$ is the (2+1)-form foliated gauge fields and $b$ is the two-form bulk gauge field. Their gauge transformations are
\ali{
    B^k \wedge dx^k &\sim B^k \wedge dx^k + d\gamma^k \wedge dx^k \,, \\
    b &\sim b + d\gamma + \sum^{2}_{k=1} \gamma^k \wedge dx^k  \,,
}
or
\ali{
    \bm{B} \sim \bm{B} + d_\text{f} \bm{\gamma} \,,
}
where $\bm{\gamma} = (\gamma^1 \wedge dx^1, \gamma^2 \wedge dx^2;\, \gamma)$ is a $U(1)$ type-$A$ foliated (1+1)-form gauge parameter. We can define the exterior derivative of $\bm{B}$ as
\ali{
    d_\text{f} \bm{B} = \left(0,0;\, db - \sum^{2}_{k=1} B^k \wedge dx^k \right) \,.
}
We set the exterior derivative of any three-form gauge fields to zero in 2+1 dimensions, and then $(d_\text{f})^2 = 0$ always holds.

\subsubsection*{$U(1)$ type-$B$ foliated zero-form gauge field}

The other type of foliated gauge fields are type-$B$ foliated gauge fields. We consider zero-form gauge fields $\hat\Phi^k \ (k = 1,2)$ whose gauge transformations are
\ali{
    \hat\Phi^k \sim \hat\Phi^k + 2\pi \hat{W}^k - \hat{\xi} \,,
}
where $\hat{W}^k$ is an integer-valued gauge parameter and it can have step function discontinuities in the $x^k$ direction, and $\hat{\xi}$ is a zero-form gauge parameter. 
The bulk gauge field is the one-form gauge field $\hat\Phi$ whose gauge transformation
\ali{
    \hat\Phi \sim \hat\Phi + d\hat\xi \,.
}
In addition, we must consider the auxiliary fields $\alpha^k\, dx^k \ (k =1,2)$, which is a $U(1)$ type-$A$ foliated (0+1)-form gauge field whose gauge transformations are
\ali{
    \alpha^k\, dx^k \sim \alpha^k\, dx^k + 2\pi d\hat{W}^k \,.
}
Then, the $U(1)$ type-$B$ foliated zero-form gauge field is the tuple $\bm{\hat{\Phi}} = (\hat\Phi^1, \hat\Phi^2;\, \hat\Phi;\, \alpha^1 \, dx^1, \alpha^2 \, dx^2 )$, and we define the exterior derivative as
\ali{
    d_\text{f}\bm{\hat{\Phi}} = (d\hat\Phi^1 -\hat\Phi-\alpha^1 \, dx^1 , d\hat\Phi^2 -\hat\Phi- \alpha^2 \, dx^2  ;\, d\hat\Phi;\, d\alpha^1 \wedge dx^1, d\alpha^2 \wedge dx^2 ) \,,
}
which is gauge invariant. However, we always couple $d\hat\Phi^k -\hat\Phi-\alpha^k \, dx^k$ to type-$A$ foliated gauge field with $dx^k$, so we can ignore the auxiliary fields $\alpha^k\, dx^k$.

\subsubsection*{$U(1)$ type-$B$ foliated one-form gauge field}

Next, we consider the $U(1)$ type-$B$ foliated one-form gauge field. The foliated fields are one-form gauge fields $\hat{C}^k\ (k=1,2)$, the bulk field is a two-form gauge field $\hat{c}$, and the auxiliary fields $\zeta^k\wedge dx^k$ are type-$A$ foliated (1+1)-form gauge field. The gauge transformations of $\bm{\hat{C}} = (\hat{C}^1, \hat{C}^2;\, \hat{c};\,\zeta^1\wedge dx^1,\zeta^2\wedge dx^2  )$ are
\ali{
    \hat{C}^k &\sim \hat{C}^k + d \hat\lambda^k - \hat{\kappa} - \beta^k \, dx^k \,, \\
    \hat{c} &\sim \hat{c} + d\hat\kappa \,, \\
    \zeta^k\wedge dx^k &\sim \zeta^k\wedge dx^k + d\beta^k \wedge dx^k \,,
}
or
\ali{
    \bm{\hat{C}} \sim \bm{\hat{C}} + d_\text{f} \bm{\hat\lambda} \,, 
}
where $\bm{\hat\lambda}= (\hat\lambda^1, \hat\lambda^2;\, \hat\kappa;\, \beta^1 \, dx^1, \beta^2 \, dx^2 )$ is a $U(1)$ type-$B$ foliated zero-form gauge parameter. %
We define the exterior derivative of $\bm{\hat{C}}$ as
\ali{
    d_\text{f} \bm{\hat{C}} = (d\hat{C}^1 + \hat{c} + \zeta^1 \wedge dx^1, d\hat{C}^2 + \hat{c} + \zeta^2 \wedge dx^2;\, d\hat{c};\, d\zeta^1 \wedge dx^1, d\zeta^2 \wedge dx^2 ) \,,
}
which is gauge invariant, and we also have $(d_\text{f})^2 = 0$. As in the case of type-$B$ foliated zero-form gauge fields, we can ignore the auxiliary fields $\zeta^k\wedge dx^k$ and the gauge parameter $\beta^k\, dx^k$. Furthermore, due to the gauge  parameter $\beta^k \, dx^k$, we can set $\hat{C}^k_k$ to zero. Thus, we have only to consider $\hat{C}^1_0, \hat{C}^1_2, \hat{C}^2_0,$ and $\hat{C}^2_1$ as the components of $\hat{C}^k$.

\subsubsection*{$U(1)$ type-$B$ foliated two-form gauge field}

Although we do not deal with the $U(1)$ type-$B$ foliated two-form gauge field in this paper, we can consider it in the same way. The foliated fields are two-form gauge fields $\hat{B}^k\ (k=1,2)$, the bulk field is a three-form gauge field $\hat{b}$, and the auxiliary fields $\rho^k\wedge dx^k$ are type-$A$ foliated (2+1)-form gauge field. The gauge transformations of $\bm{\hat{B}} = (\hat{B}^1, \hat{B}^2;\, \hat{b};\,\rho^1\wedge dx^1,\rho^2\wedge dx^2  )$ are
\ali{
    \hat{B}^k &\sim \hat{B}^k + d \hat\gamma^k + \hat{\tau} + \psi^k \wedge dx^k \,, \\
    \hat{b} &\sim \hat{b} + d\hat\tau \,, \\
    \rho^k\wedge dx^k &\sim \rho^k\wedge dx^k + d\psi^k \wedge dx^k \,,
}
or
\ali{
    \bm{\hat{B}} \sim \bm{\hat{B}} + d_\text{f} \bm{\hat\gamma} \,, 
}
where $\bm{\hat\gamma}= (\hat\gamma^1, \hat\gamma^2;\, \hat\tau;\, \psi^1 \wedge dx^1, \psi^2 \wedge dx^2 )$ is a $U(1)$ type-$B$ foliated one-form gauge parameter. %
We define the exterior derivative of $\bm{\hat{B}}$ as
\ali{
    d_\text{f} \bm{\hat{B}} = (d\hat{B}^1 - \hat{b} - \rho^1 \wedge dx^1, d\hat{B}^2 - \hat{b} - \rho^2 \wedge dx^2;\, 0 ;\, 0, 0) \,,
}
which is gauge invariant, and we also have $(d_\text{f})^2 = 0$. We can also ignore the auxiliary fields $\rho^k\wedge dx^k$ and the gauge parameter $\psi^k \wedge dx^k$. Furthermore, due to the gauge  parameter $\psi^k \wedge dx^k$, we can set $\hat{B}^k_{ki}$ to zero. Thus, we have only to consider $\hat{B}^1_{02}$ and $\hat{B}^2_{01}$ as the components of $\hat{B}^k$.

\subsection{Foliated Lagrangians}

Using these notation, we can rewrite the foliated Lagrangians. We define the wedge product of a type-$A$ and a type-$B$ foliated gauge fields as follows. We let a type-$A$ foliated ($n$+1)-form gauge field be $\bm{A} = (A^k \wedge dx^k;\, a)$ and a type-$B$ foliated $m$-form gauge field be $\bm{\hat{B}} = (\hat{B}^k ;\, \hat{b};\, \beta^k \wedge dx^k)$, and we define the wedge product of these as
\ali{
    \bm{A} \wedge_\text{f} \bm{\hat{B}} &= \sum_k A^k \wedge \hat{B}^k \wedge dx^k   + a \wedge \hat{b} \,, \\
    \bm{\hat{B}} \wedge_\text{f} \bm{A} &= \sum_k  \hat{B}^k \wedge A^k \wedge dx^k + (-1)^n \, \hat{b} \wedge a \,,
}
where the auxiliary fields $\beta^k \wedge dx^k$ do not appear, and then we have the relation
\ali{
    \bm{A} \wedge_\text{f} \bm{\hat{B}} = (-1)^{n m} \bm{\hat{B}} \wedge_\text{f} \bm{A} \,.
}
For example, the wedge product of the type-$B$ foliated one-form gauge field $\bm{\hat{C}} = (\hat{C}^1, \hat{C}^2;\, \hat{c};\,\zeta^1\wedge dx^1,\zeta^2\wedge dx^2  )$ and the type-$A$ foliated (2+1)-form gauge field $d_\text{f} \bm{C} = (d C^1 \wedge dx^1,d C^2 \wedge dx^2;\, dc + \sum^2_{k=1} C^k \wedge dx^k )$ is
\alis{
    \bm{\hat{C}} \wedge_\text{f} d_\text{f} \bm{C} &=  \hat{C}^1 \wedge d C^1 \wedge dx^1 + \hat{C}^2 \wedge d C^2 \wedge dx^2 + \hat{c} \wedge \left( dc + \sum^2_{k=1} C^k \wedge dx^k \right) \\
    &= \sum^2_{k = 1} \hat{C}^k \wedge d C^k \wedge dx^k + \hat{c} \wedge \left( \sum^2_{k = 1} C^k \wedge dx^k  + dc \right) \,.
}
Therefore, the foliated Lagrangian of the SSPT phase \eqref{folissptlag} can be written as
\ali{
    L_{\text{SSPT,f}} \left[ \bm{C}, \bm{\hat{C}} \right] = \frac{i}{2\pi} \bm{\hat{C}} \wedge_\text{f} d_\text{f} \bm{C} \,.
}
In the same way, we can rewrite the foliated Lagrangian of the $\phi$-theory \eqref{foliphideff} as
\alis{
    L_{\phi,\text{f}} \left[ \bm{C}, \bm{\hat{C}} \right] = \frac{1}{2} \left( d \bm{\Phi}  - \bm{C} \right)^2 - \frac{i}{2\pi} \bm{\hat{C}} \wedge_\text{f} ( d\bm{\Phi} - \bm{C} ) + \bm{\hat\chi} \wedge_\text{f} d_\text{f} \bm{C} \,, 
}
where the $\left( d \bm{\Phi}  - \bm{C} \right)^2$ is defined by
\ali{
    ( \bm{A} )^2 = \sum_k ( A^k \wedge dx^k ) \wedge \ast ( A^k \wedge dx^k ) + a \wedge \ast a \,,
}
for $\bm{A} = ( A^k \wedge dx^k ;\, a)$, and $\bm{\hat\chi}$ is formally $(0,0;\,\hat\chi;\,0,0)$, 
We can also rewrite the foliated Lagrangian of the $\hat\phi$-theory \eqref{foliatedphihatdeff} as
\ali{
   L_{\hat{\phi},\text{f}} \left[ \bm{C}, \bm{\hat{C}} \right] = \frac{1}{2} \left( d_\text{f} \bm{\hat\Phi}  - \bm{\hat{C}} \right)^2 + \frac{i}{2\pi}  d_\text{f} \bm{\hat\Phi} \wedge_\text{f} \bm{C} + \bm{\hat\chi} \wedge_\text{f} d_\text{f} \bm{C} \,,
}
where the $\left( d_\text{f} \bm{\hat\Phi}  - \bm{\hat{C}} \right)^2$ is defined by
\ali{
    ( \bm{\hat{B}} )^2 = \sum_k ( \hat{B}^k \wedge dx^k ) \wedge \ast ( \hat{B}^k \wedge dx^k ) + \hat{b} \wedge \ast \hat{b} \,,
}
for $\bm{\hat{B}} = ( \hat{B}^k ;\, \hat{b} ;\, \beta^k\wedge dx^k )$.

\subsection{\texorpdfstring{$\Z_4$}{Z4} Rotation of Foliated Gauge Fields}
\label{appendix23}

Here we collect how the foliated gauge fields transform under the 90-degree rotation $\Z_4$. We consider the rotation $x'^1 = x^2$, $x'^2 = -x^1$. From the transformations of $dx^k$, we determine the transformation rules for the type-$A$ foliated gauge field as
\ali{
    \Phi^1 &\rightarrow \Phi^2 \,, \\
    \Phi^2 &\rightarrow - \Phi^1 \,, \\
    C^1_2 &\rightarrow - C^2_1 \,, \\
    C^2_1 &\rightarrow - C^1_2 \,, \\
    C^1_0 &\rightarrow  C^2_0 \,, \\
    C^2_0 &\rightarrow - C^1_0 \,, 
}
and so forth. As for the type-$B$ foliated gauge field, the rules are
\ali{
    \hat{\Phi}^1 &\rightarrow \hat{\Phi}^2 \,, \\
    \hat\Phi^2 &\rightarrow  \hat\Phi^1 \,, \\
    \hat{C}^1_2 &\rightarrow -\hat{C}^2_1 \,, \\
    \hat{C}^2_1 &\rightarrow \hat{C}^1_2 \,, \\
    \hat{C}^1_0 &\rightarrow  \hat{C}^2_0 \,, \\
    \hat{C}^2_0 &\rightarrow \hat{C}^1_0 \,, 
}
and so forth. The bulk fields obey the ordinary rules.

\bibliography{ref2.bib}
\bibliographystyle{utphys}

\end{document}